\documentclass[fleqn,usenatbib]{mnras}

\usepackage{newtxtext,newtxmath,booktabs}

\usepackage[T1]{fontenc}

\DeclareRobustCommand{\VAN}[3]{#2}
\let\VANthebibliography\thebibliography
\def\thebibliography{\DeclareRobustCommand{\VAN}[3]{##3}\VANthebibliography}

\usepackage{graphicx}	
\usepackage{amsmath}	
\usepackage{placeins}
\usepackage{xcolor}
\usepackage{lipsum}

\title[Sub-arcsecond metre-wave polarisation imaging]{Polarisation and Faraday rotation measure imaging at metre wavelengths with sub-arcsecond resolution: a foundational calibration strategy}

\author[R. J. van Weeren et al.]{
R. J. van Weeren$^{1}$\thanks{E-mail: rvweeren@strw.leidenuniv.nl (RJvW)},
           J.~M.~G.~H.~J.~de~Jong$^{1,2}$,
           X.~K Le~Saux$^{1}$,
           V.~A. Chakawri$^{1}$,
           Q.~W.~E.~van~Zegveld$^{1}$,\newauthor
           D.~de~Jong$^{1}$,
           S.~P.~O'Sullivan$^{3}$,
           F.~Sweijen$^{4}$,
           V.~H.~Mahatma$^{5,6}$,
           E.~De Rubeis$^{7,8}$,
           L.K. Morabito$^{4,9}$,\newauthor
           D. Alonso-L\'opez$^{3}$,
           A.~Bonafede$^{10,8}$,
           C.~Horellou$^{11}$,
           M.~van~der~Wild$^{4}$
\\ \\
$^{1}$Leiden Observatory, Leiden University, PO Box 9513, 2300 RA Leiden, The Netherlands\\
$^{2}$ASTRON, The Netherlands Institute for Radio Astronomy, Postbus 2, 7990 AA Dwingeloo, The Netherlands\\
$^{3}$Departamento de F\'isica de la Tierra y Astrof\'isica \& IPARCOS-UCM, Universidad Complutense de Madrid, 28040 Madrid, Spain\\
$^{4}$Centre for Extragalactic Astronomy, Department of Physics, Durham University, Durham, DH1 3LE, UK\\ 
$^{5}$Cavendish Laboratory -- Astrophysics Group, University of Cambridge, 19 JJ Thomson Avenue, Cambridge CB3 0HE, UK\\
$^{6}$Kavli Institute for Cosmology, University of Cambridge, Madingley Road, Cambridge CB3 0HA, UK\\
$^{7}$Hamburger Sternwarte, Universit\"at Hamburg, Gojenbergsweg 112, 21029 Hamburg, Germany\\
$^{8}$INAF -- Istituto di Radioastronomia di Bologna, Via Gobetti 101, I-40129 Bologna, Italy\\ 
$^{9}$Institute for Computational Cosmology, Department of Physics, Durham University, South Road, Durham DH1 3LE, UK\\ 
$^{10}$Dipartimento di Fisica e Astronomia, Universit\`a di Bologna, via P. Gobetti 93/2, I-40129 Bologna, Italy\\
$^{11}$Department of Physics and Astronomy, Chalmers University of Technology, Onsala Space Observatory, 43992 Onsala, Sweden
}

\date{Accepted XXX. Received YYY; in original form ZZZ}

\pubyear{\the\year{}}

\begin{document}
\label{firstpage}
\pagerange{\pageref{firstpage}--\pageref{lastpage}}
\maketitle

\begin{abstract}
Low-frequency radio polarimetric observations provide a powerful probe of magnetic fields in astrophysical sources and the intervening medium, as well as magnetospheric emission from compact objects such as pulsars, magnetically active stars, brown dwarfs, and planetary aurorae. With baselines of up to 2000~km, LOFAR offers a unique opportunity to study the low-frequency polarised Universe at sub-arcsecond resolution. However, polarimetric studies with LOFAR have so far been limited to angular resolutions of ${\sim}6\arcsec$, resulting in stronger beam depolarisation.
Here we present a calibration strategy that enables full-resolution polarimetric imaging with the LOFAR pan-European array. Our method applies full-Jones corrections to the international stations using an in-field unpolarised calibrator. In addition, when a sufficiently bright polarised source is present in the field, multi-epoch observations can be aligned in Faraday depth using a visibility-based correction that accounts for polarisation angle and rotation measure offsets. This approach enables deeper combined imaging and deconvolution. We apply this strategy to the LOFAR ELAIS-N1 field, combining four 8-h observations for a total integration time of 32~h. At $0.3\arcsec$ resolution, we detect two previously known polarised sources identified in lower-resolution studies, resolve additional polarised components, and localise emission regions with sub-arcsecond precision. We also identify a new polarised source and detect circularly polarised emission from the binary M-dwarf system CR~Draconis, measuring its proper motion across epochs. These results demonstrate that sub-arcsecond polarimetry at metre wavelengths is now feasible with LOFAR, opening new science opportunities in the LOFAR2.0 era.

\end{abstract}

\begin{keywords}

polarisation -- techniques: polarimetric -- techniques: interferometric -- surveys -- galaxies: active
\end{keywords}



\section{Introduction}

Radio waves arriving from astronomical sources carry an electric field that may exhibit oscillations with a preferred direction or rotational pattern; a phenomenon called polarisation. This behaviour is described using the Stokes parameters: total intensity ($I$), linear polarisation ($Q$ and $U$), and circular polarisation ($V$). Analysing these polarised components of the signal and their wavelength dependence provides insight into the underlying emission mechanisms, the properties of the intervening medium, and the intrinsic characteristics of the source. This is information that cannot be obtained from total-intensity measurements alone.

Radio polarisation observations provide a unique probe of magnetic fields, both within astrophysical sources and in the intervening medium through which the polarised emission propagates \citep[e.g.,][]{1966MNRAS.133...67B,2005A&A...441.1217B,2020NatAs...4..577V,2023SciA....9E7233V,2024NatAs...8.1359C,2024A&A...690A.314H,2024AJ....167..226V}. Polarisation also helps to distinguish between different emission mechanisms in galaxies, such as contributions from star formation, AGN accretion, and jets \citep[e.g.,][]{2020MNRAS.499..334S}, and provides insight into the orientation of radio galaxies and the structure of their hotspots \citep[e.g.,][]{1989ApJ...341...68H,2021MNRAS.502..273M}. Large samples of linearly polarised sources, combined with measurements of Faraday rotation, serve as powerful statistical tools to study magnetic field properties in large-scale structures such as galaxy clusters \citep[e.g.,][]{2004A&A...424..429M,2010A&A...513A..30B,2022A&A...665A..71O,2025A&A...694A..44O,2024A&A...691A..23D,2025A&A...694A.125L}, filaments and the cosmic web \citep[e.g.,][]{2022MNRAS.512..945C,2025A&A...693A.208C,2025A&A...696A.203P}, as well as in our own Milky Way’s magnetoionic environment \citep[e.g.,][]{2007ApJ...663..258B,2009ApJ...702.1230T,2022A&A...657A..43H,2024ApJ...970...95U}. Compared to centimetre-wavelength observations, metre-wavelength measurements can yield Faraday rotation estimates that are up to two orders of magnitude more precise, owing to the $\lambda^{2}$ dependence of Faraday rotation. This allows probing weak magnetic fields and studying the magnetoionic structure of diffuse cosmic plasmas. Additionally, these low-frequency observations are highly sensitive to depolarisation effects, providing further constraints on the properties of intervening media in the low-density regime.

Circularly polarised emission at low radio frequencies typically originates from compact objects such as pulsars, magnetically active stars, brown dwarfs, and aurorae from planets, including potentially extrasolar planets \citep[for an overview see][]{2024NatAs...8.1359C}. This emission is often generated by coherent mechanisms, such as the electron cyclotron maser instability (ECMI), and can serve as a diagnostic of star--planet interactions or extrasolar coronal mass ejections. In the context of star--planet interactions, ECMI emission is expected to peak in the MHz regime, with the cutoff frequency directly tracing the magnetic field strength of the planet \citep[e.g.,][]{2007P&SS...55..598Z,2011A&A...531A..29H}. As such, low-frequency polarimetric observations are crucial for detecting and characterising this emission. They provide a powerful pathway to study and constrain exoplanetary magnetospheres, offering insights into magnetic field strengths, rotation, and the surrounding plasma environment \citep[e.g.,][]{2023A&A...675L...6V}.

In recent years, the number of polarisation studies of the low-frequency sky has steadily increased. For example, using the Murchison Widefield Array \citep[MWA;][]{2013PASA...30....7T}, \cite{2020PASA...37...29R,2018PASA...35...43R} reported 512 linearly polarised sources in the sky south of declination $+30^\circ$, at a resolution of a few arcminutes. With LOFAR \citep{haarlem2013}, several polarisation studies have been conducted at similar arcminute-scale resolution, primarily focusing on Galactic foreground emission \citep[e.g.,][]{2013A&A...558A..72I,2015A&A...583A.137J,2017A&A...597A..98V,2018A&A...615L...3J,2019A&A...623A..71V,2022A&A...663A...7E,2024A&A...688A.200E}. An overview of some polarisation-related aspects of LOFAR is provided by \citet{2018ASSL..426..159B}.

However, LOFAR also offers significantly higher angular resolution at metre wavelengths. A systematic search for linearly polarised sources in the LOFAR Two-metre Sky Survey Data Release 2 \citep[LoTSS DR2;][]{shimwell2022} was performed by \cite{2023MNRAS.519.5723O}, detecting 2461 sources at 20\arcsec{} resolution across 5720 deg$^2$, with a median noise of 0.08~mJy~beam$^{-1}$ in the Faraday spectra. This represents the deepest large-area low-frequency polarisation survey to date. Additional studies based on LoTSS data have also been carried out \citep[e.g.,][]{2020MNRAS.495.2607O,2020A&A...638A..48S,2021MNRAS.502..273M,2023A&A...670L..23H}.
A large-area search for circularly polarised sources was carried out at 20\arcsec{} resolution at 144~MHz by \cite{callingham2023}, resulting in the detection of 68 sources within the LoTSS DR2 footprint. The median noise level in the Stokes~V images was approximately 140~\textmu Jy~beam$^{-1}$.

The ELAIS-N1 field has become one of the best-studied regions in low-frequency polarisation, as it is one of LOFAR’s deep extragalactic survey fields \citep[][]{shimwell2025,tasse2021}. Galactic polarised emission in this field has been explored at 3-4\arcmin{} resolution by \cite{2014A&A...568A.101J,2023A&A...674A.119S}. Discrete sources were investigated by \cite{ruiz2021} at 20\arcsec{} resolution, achieving a median noise of 26~\textmu Jy~beam$^{-1}$ and detecting 10 sources, corresponding to a polarised source density of 0.8~deg$^{-2}$. More recently, \cite{2024A&A...687A.267P,2025A&A...693A.100P} probed the field at the full angular resolution of LoTSS (6\arcsec) using 152~hrs of LoTSS deep-field data, detecting 31 sources with a median noise of 19~\textmu Jy~beam$^{-1}$, yielding a polarised source density of 1.24~deg$^{-2}$.

Since the first LoTSS data release \citep[][]{shimwell2017}, we have known that the majority of radio sources detected by LOFAR at 6\arcsec{ }resolution remain unresolved \citep[see also][]{2025MNRAS.540..416S}. This lack of spatial resolution limits our ability to study the polarisation structure across sources and may result in significant beam depolarisation, thereby reducing the number of polarised sources that can be detected \citep[e.g.,][]{2018MNRAS.475.4263O}. LOFAR's international (European) baselines, extending up to 2000~km, provide significantly higher angular resolution, reaching $\sim$0.3\arcsec{} within the typical LoTSS observing band of 120--168~MHz \citep[e.g.,][]{morabito2022,sweijen2022,2025Ap&SS.370...19M}. This enables sub-arcsecond resolution polarimetric studies and, for example, allows structures on sub-kpc scales to be probed at $z=0.1$, typically resolving the hotspots of radio galaxies into complex morphologies. However, until now, no comprehensive strategy for full-polarisation calibration and imaging at this resolution has been developed or demonstrated.

In this work, we present a strategy to perform full-polarisation calibration using LOFAR’s international baselines. We apply and validate this approach using observations of the ELAIS-N1 field. This field was previously imaged by \cite{dejong2024} at resolutions of 1.2\arcsec, 0.6\arcsec, and 0.3\arcsec{} over the full LOFAR international station's field of view ($2.5\degr \times 2.5\degr$), reaching a central Stokes~$I$ map noise of 14~\textmu Jy~beam$^{-1}$ at 0.3\arcsec{} resolution.

The paper is structured as follows. Sect~\ref{sec:data} describes the ELAIS-N1 observations used in this study.
In Sect.~\ref{sec:calibration}, we outline the calibration steps required, building upon the framework developed by \citet{dejong2024} and \citet{dejong2025b}. The results are presented in Sect.~\ref{sec:results}, followed by discussion and conclusions in Sects.~\ref{sec:discussion} and~\ref{sec:conclusions}, respectively.

\section{Data}
\label{sec:data}

We make use of four calibrated 8-hour LOFAR observations from the ELAIS-N1 Deep Field, originally analysed by \citet{dejong2024}. These observations, L686962, L769393, L798074, and L816272, were taken between 2018 and 2021 and selected based on having international LOFAR stations and good ionospheric conditions, as detailed in \cite{dejong2024}'s section~2. The observations include baselines up to 1980~km in length and frequencies from 115 to 166~MHz. All four observations use 3C~295 as the primary calibrator. Two of the datasets (L798074 and
L816272) were originally archived with a 2~s time resolution, which is twice the typical 1~s resolution for LOFAR, introducing additional time smearing effects at larger angular distances from the pointing centre \citep[see fig.~1 in][]{dejong2024}. Due to the calibration on the in-field calibrator (see Section \ref{sec:maincalibrator}), the astrometric accuracy achieves median offsets in right ascension and declination of less than 0.1\arcsec. The flux density is also consistent with radio maps made at the same frequencies with LOFAR at 6\arcsec{} resolution by \cite{sabater2021} and \cite{shimwell2025}. With the calibrated visibility data retained, we can re-image the observations at different resolutions, explore different Briggs weightings, and adjust other imaging parameters using the \texttt{WSClean} imager with its \texttt{wgridder} module \citep{wsclean,arras2021,ye2022}.

\section{Calibration}
\label{sec:calibration}
The polarisation calibration of LOFAR international baseline data consists of several steps. These include a station-based correction for ionospheric Faraday rotation, based on a global ionospheric model using \texttt{RMextract} \citep{mevius2018}\footnote{\url{ https://pypi.org/project/RMextract/}; Predecessor of \texttt{spinifex}{ } (\url{https://pypi.org/project/spinifex/}).}, which is a standard part of LOFAR processing carried out by \texttt{LINC}\footnote{\url{https://git.astron.nl/RD/LINC/}} and is not specific to international baseline observations. 
A subsequent step is a full-Jones polarisation calibration for the Dutch LOFAR stations, performed by the LoTSS DDF-pipeline\footnote{\url{https://github.com/mhardcastle/ddf-pipeline}} \citep{tasse2014, tasse2014b, smirnov2015, shimwell2019, tasse2021}. The final steps are specific to the international baselines and are discussed below for both an unpolarised and a polarised calibrator.

\subsection{Unpolarised calibrator}\label{sec:maincalibrator}

\begin{figure}
    \centering
    \includegraphics[width=0.45\textwidth]{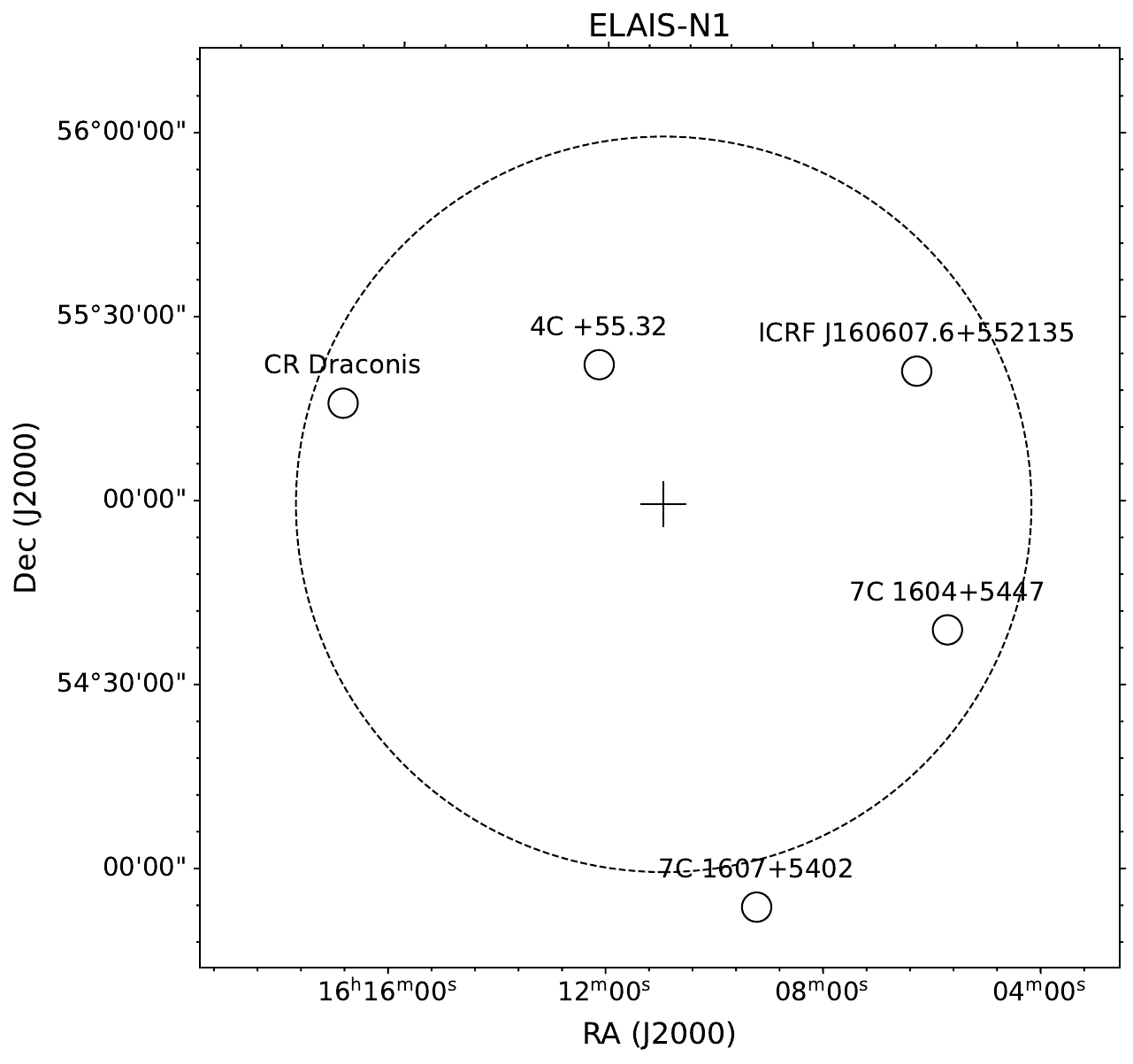}
    \caption{Location of the sources discussed in this work within the ELAIS-N1 field. The unpolarised calibrator is ICRF\,J160607.6+552135, and the polarised calibrator is 7C\,1604+5447. The cross marks the field centre, while the dashed circle indicates a radius of $1^\circ$.}
    \label{fig:sourcelocations}
\end{figure}

As part of the LOFAR calibration strategy for observations including international stations, a high--signal-to-noise-ratio (S/N) in-field calibrator is selected from the Long-Baseline Calibrator Survey \citep[LBCS;][]{jackson2022}. This source is typically among the brightest objects in the field, providing sufficient S/N on the international baselines. Given the high-S/N requirement for calibrating at a fine time and frequency resolution, driven by the rapidly varying ionosphere, such a source is well suited not only for correcting scalar Stokes~$I$ effects but also for solving for polarisation leakage terms.

For polarisation calibration, the in-field calibrator must be unpolarised  ($\ll 1\%$) in the HBA band (about 115--170~MHz). For the ELAIS-N1 observations, the Seyfert~2 galaxy ICRF\,J160607.6+552135 was selected as the calibrator \citep{charlot2020, sexton2022}, and has previously been shown to be unpolarised at LOFAR frequencies in linear and circular polarisation \citep{ruiz2021, callingham2023}. In addition, the source remains unpolarised in higher frequency Very Long Baseline Array (VLBA) observations \citep{tremblay2016}. Furthermore, the source is unresolved at the longest LOFAR baselines and has a flux density of about 0.3~Jy in the LOFAR HBA band.

We now highlight a few key calibration steps that are discussed in detail by \citet{dejong2024} and are specific to polarisation calibration. The direction-independent full-Jones solutions from the DDF-pipeline were carried over as a starting point for the calibration of the other international LOFAR stations. These stations were calibrated with \texttt{facetselfcal} \citep{vanweeren2021}\footnote{\url{github.com/rvweeren/lofar_facet_selfcal}}, using a point-source sky model of ICRF\,J160607.6+552135. The model incorporates a well-constrained radio spectrum derived from multiple radio surveys, together with a precise astrometric position from \citet{charlot2020}, accurate to approximately 2~mas in both right ascension and declination. The location of this source within the $2.5^\circ \times 2.5^\circ$ ELAIS-N1 field is indicated in Fig.~\ref{fig:sourcelocations}. The first calibration step corrects for differences between right-handed (RR) and left-handed (LL) circular polarisations by solving for their relative phase offset, which can include contributions from differential Faraday rotation. In a circular polarisation basis, differential Faraday rotation reduces to a simple phase difference between the two circular hands. For the ELAIS-N1 field, this calibration was applied using 8-minute solution intervals with a frequency smoothness kernel of 10~MHz. After polarisation-independent phase and amplitude corrections, a longer timescale full-Jones correction for instrumental leakage effects was applied to also correct the RL and LR cross-hands. This was done with a solution interval of 20 minutes and a frequency smoothness kernel of 5~MHz. Since full-Jones corrections for the Dutch LOFAR stations from the \texttt{DDF}-pipeline were already carried over, these stations were constrained to have the same solutions for the full-Jones solve. During calibration, short baselines were ignored by using a \textit{uv}-limit of 20~k$\lambda$, which corresponds to a largest angular scale of about 10\arcsec{} at 140~MHz. In addition, for these steps the data are phase-shifted and averaged to a frequency resolution of 488~kHz and a time resolution of 32~s. Both this averaging and the exclusion of short baselines during the calibration solve help mitigate the effects of incomplete sky models arising from other sources within the station's primary beam, as these sources are suppressed through fringe washing. This includes potentially polarised sources. The core stations can also be phased up into a superstation to further suppress contributions from sources elsewhere in the primary beam and to reduce processing time \citep{morabito2022}; however, this was not necessary for the ELAIS-N1 field observations. 

\begin{figure}
    \centering
    \includegraphics[width=0.5\textwidth]{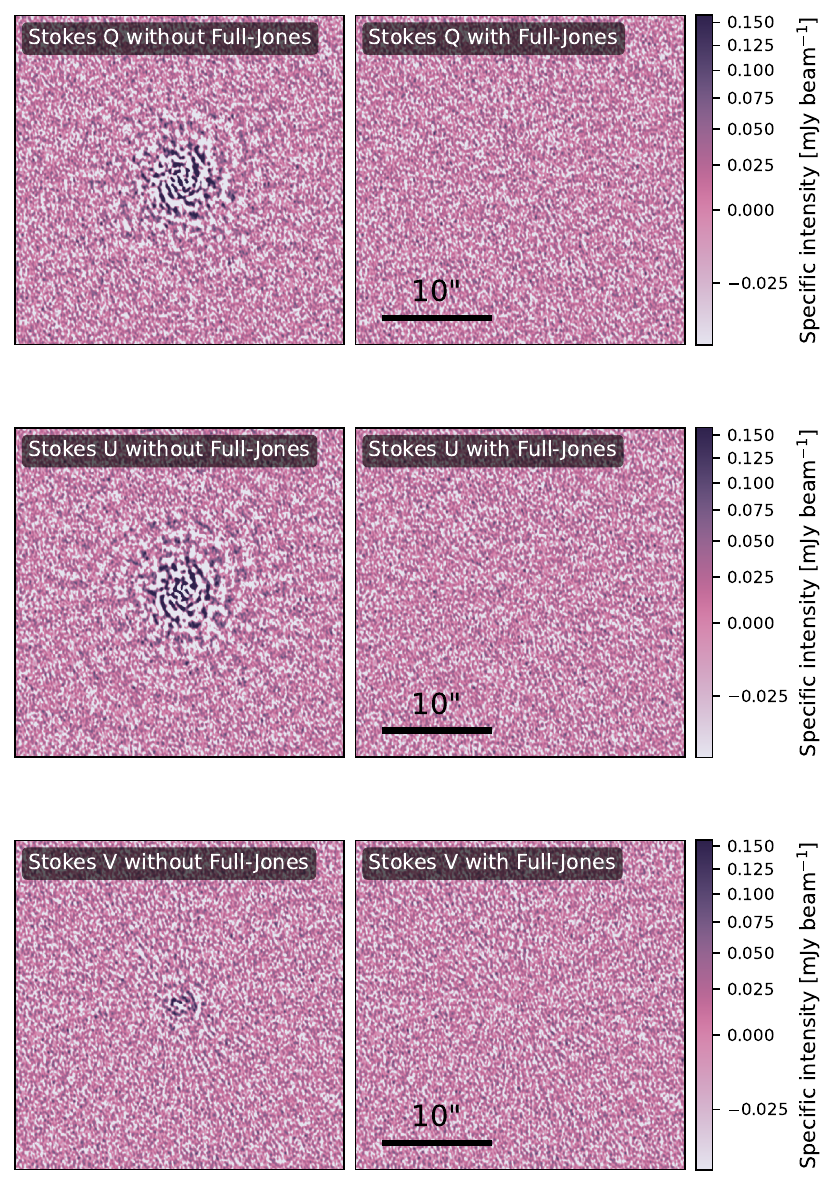}
    \caption{Stokes~$Q$, $U$, and~$V$ images of the unpolarized infield calibrator ICRF\,J160607.6+552135 with full-Jones calibration applied (\textit{right panels}) and without (\textit{left panels}). A 10\arcsec~scale bar is added in the right panels.}
    \label{fig:stokes_calibration}
\end{figure}

By creating Stokes $Q$, $U$, and $V$ images before and after applying the full-Jones calibration, we confirm that this step successfully corrects the data, forcing the Stokes $Q$, $U$, and $V$ emission of the calibrator source to zero, as intended. Fig. \ref{fig:stokes_calibration} shows the corresponding images with and without the full-Jones calibration, clearly demonstrating the removal of calibration artefacts. In Sect.~\ref{sec:4C+55.32}, we assess the sensitivity of the Stokes~$I$ to the linear and circular polarisation leakage using the source 4C\,+55.32, which has the highest integrated Stokes~$I$ flux density in the field. 

When working with sources other than the in-field (unpolarised) calibrator, such as the sources discussed in Sects.~\ref{sec:polanglecalibrator} and~\ref{sec:results}, we always apply the additional direction-dependent corrections derived in \citet{dejong2024}. These are scalar corrections and are therefore not polarisation specific. We do not apply direction-dependent differential Faraday rotation corrections, as the bulk of this effect is already removed through the in-field calibrator. The remaining direction-dependent differential Faraday rotation terms are expected to be very small, with magnitudes $\ll 1$~rad, and therefore do not require additional correction.

\subsection{Polarised calibrator}
\label{sec:polanglecalibrator}
Although each of the four LOFAR observing epochs was corrected for the bulk ionospheric Faraday rotation, residual offsets in the polarisation angles between epochs may remain, at the level of $\sim$0.1--$0.3$~rad~m$^{-2}$ \citep[e.g.,][]{2013A&A...552A..58S}, larger than the typical Rotation Measure (RM) precision obtained with these data.

To correct for these residual offsets, the observations must be aligned in Faraday depth using a linearly polarised reference source \citep[e.g.,][]{2017AJ....154...54H, 2024A&A...687A.267P}. This polarised source needs to have an  $\left|\textrm{RM}\right|\gtrsim 1 $~rad~m$^{-2}$ to avoid potential confusion with any leftover polarisation leakage from imperfect calibration contaminating the region around zero Faraday
depth.

We selected the source 7C~1604+5447 to carry out the Faraday depth alignment. This was the source with the highest S/N in polarised intensity previously detected with LOFAR in this field by \cite{2024A&A...687A.267P}. The location of 7C~1604+5447 within the ELAIS-N1 field is indicated in Fig.~\ref{fig:sourcelocations}. For this source, \citeauthor{2024A&A...687A.267P} reported an RM of $6.062 \pm 0.003$~rad~m$^{-2}$. We applied the solutions derived from the unpolarised in-field calibrator, including the necessary correction using the full-Jones station beam model. We then phase-shifted the data to the position of 7C~1604+5447, again applying the station beam correction, now evaluated in the direction of 7C~1604+5447. Since the full-Jones correction does not commute with the station beam correction, these terms must be applied in the correct order \citep[for details, see sect.~3.3.1 of][]{dejong2024}.

As 7C~1604+5447 has an angular extent of approximately 15\arcsec{} and is well-resolved on the long baselines of LOFAR, we first performed RM-synthesis to identify more precisely the region from which the polarised emission originates. From the resulting maximum linearly polarised intensity map, we find that the vast majority of the polarised emission arises from the southern hotspot. For further details on the RM-synthesis procedure and the scientific results for this source, we refer the reader to Sects.~\ref{sec:RMsynthesis} and~\ref{sec:7C1604+5447}, respectively.

\begin{figure*}
    \centering
    \includegraphics[width=0.95\textwidth]{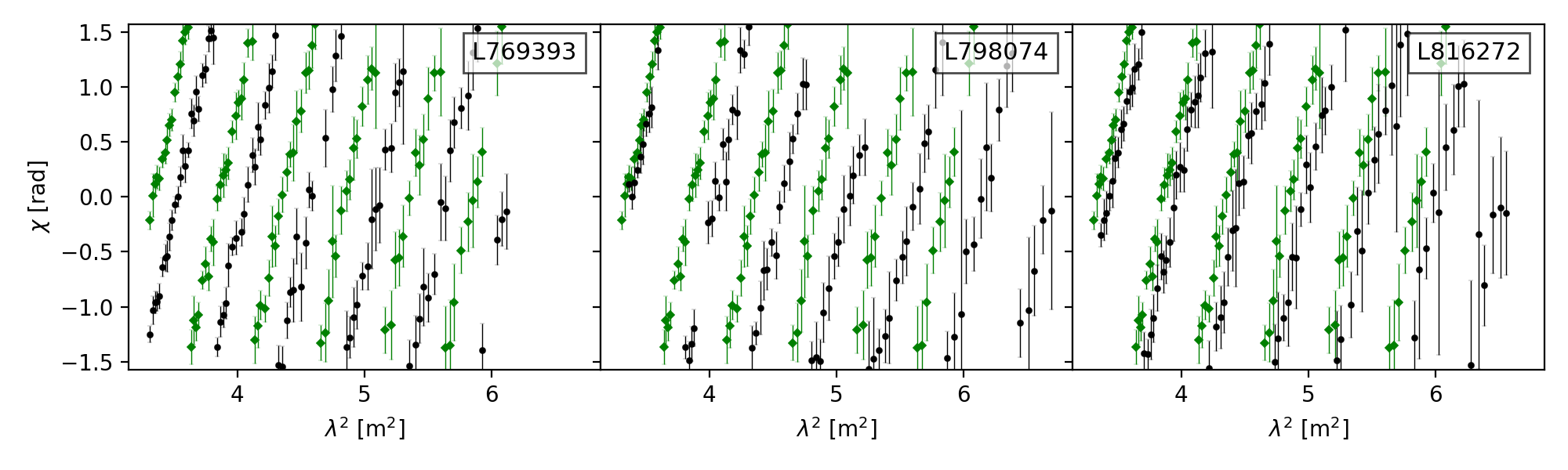}
    \caption{The polarisation angle ($\chi$) as a function of the wavelength squared ($\lambda^{2}$) for our reference observation L686962 in green and the three other observations in black.}
    \label{fig:polangle}
\end{figure*}

We extracted integrated Stokes $Q$ and $U$ flux densities from 125 channel maps by summing the flux within the southern hotspot region for each channel. Using these measurements at 125 frequencies, we plot in Fig.~\ref{fig:polangle} the polarisation angle $\chi$,
\begin{equation}
\chi = \frac{1}{2} \arctan\left(\frac{U}{Q}\right),
\end{equation}
as a function of wavelength squared ($\lambda^{2}$) for the four observing epochs. We adopt observation L686962 as the reference epoch, as it provided the highest data quality and the lowest noise levels. 

For a single Faraday-rotating screen, the polarisation angle varies with wavelength squared as
\begin{equation}
\chi(\lambda^{2}) = \chi_{0} + \mathrm{RM}\,\lambda^{2},
\end{equation}
where $\chi_{0}$ is the intrinsic polarisation angle in the absence of Faraday rotation.

As can be seen in Fig.~\ref{fig:polangle}, the polarisation angles of the four observing epochs are not aligned with respect to the reference observation, exhibiting systematic offsets. To quantify these offsets\footnote{these are also called Xf-terms in \texttt{CASA} \citep{2007ASPC..376..127M}}, we fit the following parametrisation to the observed $\chi(\lambda^{2})$ for each epoch:
\begin{equation}
\chi(\lambda^{2}) = \chi_{\lambda_{0}^{2}} + \mathrm{RM}\,(\lambda^{2} - \lambda_{0}^{2}),
\end{equation}
where $\chi_{\lambda_{0}^{2}}$ is the polarisation angle at a central wavelength squared $\lambda_{0}^{2}$.

The fit is performed using \texttt{scipy.optimize.curve\_fit} by modelling the fractional Stokes parameters
\begin{eqnarray}
\label{eq:qu1}
q(\lambda^{2}) &=& p\cos\left\{2\left[\mathrm{RM}(\lambda^{2}-\lambda_{0}^{2}) + \chi_{\lambda_{0}^{2}}\right]\right\}, \\
u(\lambda^{2}) &=& p\sin\left\{2\left[\mathrm{RM}(\lambda^{2}-\lambda_{0}^{2}) + \chi_{\lambda_{0}^{2}}\right]\right\},
\label{eq:qu2}
\end{eqnarray}
and minimising the combined residuals of $q$ and $u$, where $p$ is the polarisation fraction with $p=\sqrt{Q^2+U^2}/I$. The fractional polarisations  $q=Q/I$ and $u=U/I$ are therefore insensitive to the absolute Stokes~$I$ intensity. The fractional polarisations $q$ and $u$ were computed by dividing the measured Stokes~$Q$ and $U$ values by a model for Stokes~$I$, rather than by the measured Stokes~$I$ value in each channel, to avoid propagating noise or systematic errors present in $I$ but not in $Q$ and/or $U$. The Stokes~$I$ model was obtained by fitting a simple power-law spectrum of the form $S_\nu = S_0 \nu^{\alpha}$, where $S_\nu$ is the Stokes~$I$ flux density and $\alpha$ is the spectral index. We note that an alternative approach would be to derive the RM and polarisation angle directly from the peak of the Faraday dispersion function. This approach is advantageous when multiple Faraday components are present, for example, due to a significant leakage signal. However, this is not the case for 7C\,1604+5447 (see Fig.~\ref{fig:7C1604+5447}), and we therefore adopt the ``$qu$-fitting'' approach described above.

We adopt $\lambda_{0}^{2} = 4.5$~m$^{2}$, corresponding approximately to the mean $\lambda^{2}$ of our observations. From the fitted parameters, we compute the differences relative to the reference observation L686962 as
\begin{equation}
\Delta \chi_{0} = \chi_{0} - \chi_{0}^{\mathrm{ref}}, \qquad
\Delta \mathrm{RM} = \mathrm{RM} - \mathrm{RM}^{\mathrm{ref}}.
\end{equation}

The required correction to the visibility cross-hand phase as a function of wavelength squared is then given by
\begin{align}
\theta_{\mathrm{corr}}(\lambda^{2}) &= 2\left[\Delta \chi_{0} + \Delta \mathrm{RM}\,(\lambda^{2} - \lambda_{0}^{2})\right] \\
&= 2\left(\Delta \chi_{0} - \Delta \mathrm{RM}\,\lambda_{0}^{2}\right) + 2\,\Delta \mathrm{RM}\,\lambda^{2} \nonumber \\
&\equiv \theta_{0} + 2\,\Delta \mathrm{RM}\,\lambda^{2},
\label{eq:viscorr}
\end{align}
where $\theta_{0} = 2(\Delta \chi_{0} - \Delta \mathrm{RM}\,\lambda_{0}^{2})$ is the intercept of the cross-hand phase correction.

In Table \ref{tab:alignment} we give the RM and $\theta_{0}$ values that we found for each of the observations. Applying the corresponding frequency-dependent corrections on the visibilities allows us to align all observations \citep{2017AJ....154...54H}. An advantage of applying these corrections directly to the visibilities is that it enables deep polarisation imaging with visibility data from all observations together. This contrasts with imaging observations separately and applying corrections in image space, followed by image stacking, as done for example by \cite{ruiz2021,2024A&A...687A.267P}, where the fainter polarised signal only detected in the deeper combined data would not be deconvolved during imaging.

\begin{table}
    \centering
    \caption{RM and $\theta_{0}$ values for different observations. The reference observation ID is highlighted in bold. The three other observations are then corrected (Eq.~\ref{eq:viscorr}) to match the reference observation so that all observations can be jointly imaged.}
    \label{tab:alignment}
    \begin{tabular}{llll}
        \toprule
        Observation ID & RM (rad m$^{-2}$) & $\theta_{0}$ (deg)  \\
        \midrule
        \textbf{L686962}  & $6.30 \pm 0.02$ & $0.9 \pm 1.5$ \\
        L769393  & $5.97 \pm 0.02$ & $-6.9\pm1.6$ \\
        L798074  & $5.69 \pm 0.02$ & $-17.6 \pm 1.6$ \\
        L816272  & $6.11 \pm 0.01$ & $1.6 \pm 1.3$  \\
        \hline
    \end{tabular}
\end{table}

\subsection{RM-synthesis}
\label{sec:RMsynthesis}
To search for linearly polarised emission, we performed RM-synthesis, which enables an efficient search for polarised emission as a function of Faraday depth \citep{2005A&A...441.1217B}. We used the \texttt{RM-Tools} package to perform the RM-synthesis \citep{2020ascl.soft05003P,2026arXiv260120092V}.

The input to the RM-synthesis consists of Stokes~$Q$ and $U$ channel maps. In this work, the number of frequency channels per observation ranges between 115 and 125, reflecting the slightly different frequency coverage of the four observing epochs, although all observations share the same channel width. When combining the full 32~h data set, we therefore produced $Q$ and $U$ channel maps using \texttt{WSClean} with the option \texttt{-pol QU -channels-out 125}, where the 125 channels correspond to the full frequency coverage when the four observations are jointly imaged. For the alignment of the individual observing epochs, described in Sect.~\ref{sec:polanglecalibrator}, the number of output channels was adjusted to match the actual channel count of each observation.

To enable deep cleaning, we used the \texttt{WSClean} options \texttt{-join-polarization}, \texttt{-join-channels}, \texttt{-squared-channel-joining}, and \texttt{-fit-rm}. In addition, we applied a Gaussian $uv$-taper, chosen to correspond to a specific angular scale on the sky, to achieve a similar angular resolution across all channel maps. The images were subsequently restored with identical beams. These $Q$ and $U$ channel maps were then used as input for the two-dimensional RM-synthesis\footnote{Here, ``two-dimensional'' refers to the fact that we perform RM-synthesis on two-dimensional images with right ascension and declination axes.}.

Following \citet{2005A&A...441.1217B}, the maximum observable Faraday depth to which one has more than 50\% sensitivity is given by 
\begin{equation}
\left|\phi_{\rm max}\right| = \frac{\sqrt{3}}{\delta\lambda^2},
\end{equation}
where $\delta\lambda^2$ (0.0254~m$^2$) is the width in $\lambda^2$ corresponding to a single frequency channel. For our data, this gives $\left|\phi_{\rm max}\right| = 68$~rad~m$^{-2}$, computed from the channel frequency resolution (390.625~kHz) at the average observing frequency of 140.2~MHz. The original data have a frequency resolution that is at least four times finer, which would in principle allow a search up to $\pm272$~rad~m$^{-2}$. However, note that well below this value, depolarisation effects already become important. Thus, for studies in which one expects large RMs, or in cases where there are no previous estimates of them, the loss in S/N can be quite significant due to the bandwidth depolarisation effect \citep{fine2023}. Therefore, as a general rule, it is advisable to choose a tailored channel width that minimises S/N losses to maximise polarisation detection. This can be estimated with the \texttt{rmtools\_bwpredict} tool in \texttt{RM-Tools}, based on the adjoint transform method developed in \citet{fine2023} (see fig.~6 in their paper).

The largest scale in Faraday depth to which our data are sensitive is given by
\begin{equation}
\mathrm{max\text{-}scale} = \frac{\pi}{\lambda^2_{\rm min}},
\end{equation}
where $\lambda^2_{\rm min}$ corresponds to the shortest wavelength (1.8145~m) of the observations. This value is 0.95~rad~m$^{-2}$. Since this value is comparable to the resolution in Faraday depth space, the data are only sensitive to Faraday-thin (i.e.\ unresolved) emission.

\begin{figure}
    \centering
    \includegraphics[width=0.5\textwidth]{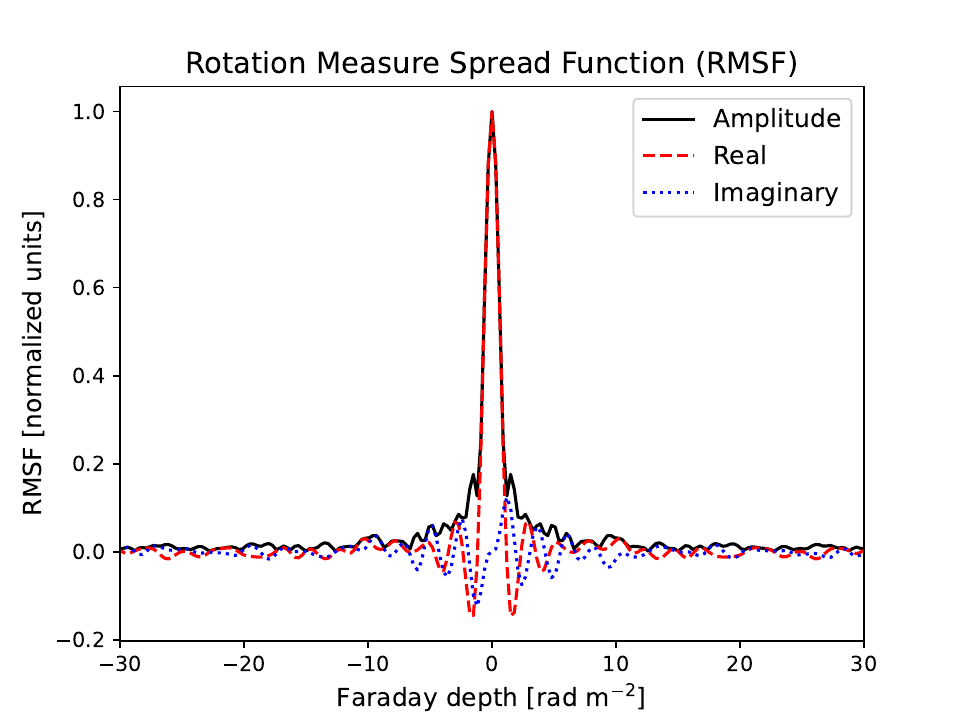}
    \caption{Rotation Measure Spread Function (RMSF) corresponding to the observations used in this work. The RMSF amplitude is shown by the black solid line. The real component is indicated by the red dashed line, and the imaginary component by the blue dotted line.}
    \label{fig:rmsf}
\end{figure}

For the RM-synthesis, we sampled the Faraday dispersion function in steps of 0.3~rad~m$^{-2}$ over the range $-30$ to $+30$~rad~m$^{-2}$. Previous studies \citep{ruiz2021,2024A&A...687A.267P,2025A&A...693A.100P} show that essentially all detected radio sources in the ELAIS-N1 field have RM values of $\lesssim \pm 20$~rad~m$^{-2}$. A larger Faraday depth range could be explored in future work. The input Stokes $Q$ and $U$ images were weighted by the inverse of their variances in order to optimise the sensitivity to faint polarised emission in Faraday depth space.  In Fig.~\ref{fig:rmsf}, we show the Rotation Measure Spread Function (RMSF) of our data.

We applied RM-CLEAN \citep{2009A&A...503..409H} to remove sidelobes introduced by the RMSF. The clean components were restored with a Gaussian of full width at half maximum (FWHM) of 1.09~rad~m$^{-2}$, corresponding to the resolution in Faraday depth space and the adopted ``variance'' parameter in \texttt{RM-Tools}. Within \texttt{RM-Tools}, we used a cleaning threshold of 0.15~mJy, a standard gain of 0.1, and a maximum of 1000 clean iterations.

For polarised source components detected in the Faraday cubes, we performed an additional one-dimensional RM-synthesis step on each detected feature, rather than on a per-pixel basis as described above. This approach allows for summation of the signal and finer sampling along the Faraday depth axis over a larger Faraday depth range, enabling more accurate deconvolution and deeper cleaning. Performing this at full resolution with a two-dimensional RM-synthesis would otherwise result in prohibitively large data cubes.

We summed the Stokes $Q$ and $U$ flux densities over the relevant pixels to obtain integrated spectra, which were then provided to \texttt{RM-Tools} to perform a one-dimensional RM-synthesis using ``variance'' weighting with a sampling of 0.11~rad~m$^{-2}$ over a Faraday depth range of $-60$ to $60$~rad~m$^{-2}$. RM-CLEAN was applied, with the cleaning going down to a threshold of $3\,\sigma_{\rm rms}$.

\section{Results}
\label{sec:results}

In this section, we evaluate the results obtained from the full 32~h data set after applying the polarisation angle alignment procedure described in Sect.~\ref{sec:polanglecalibrator}. We begin with an overview of the achieved noise levels in Stokes~$Q$, $U$, and $V$. We then demonstrate the performance of our polarisation calibration strategy through an analysis of the polarisation angle calibrator 7C\,1604+5447, the brightest radio source in the field, 4C\,+55.32, and an additional polarised source 7C\,1607+5402, previously detected by LOFAR.

For the circular polarisation Stokes~$V$, we investigate the properties of CR~Draconis, the only source detected in circular polarisation in the LoTSS data for this field \citep{callingham2023}. A systematic search for and characterisation of all polarised sources in the ELAIS-N1 field at sub-arcsecond resolution will be presented in future work.

\begin{figure*}
    \centering
    \includegraphics[width=0.48\textwidth]{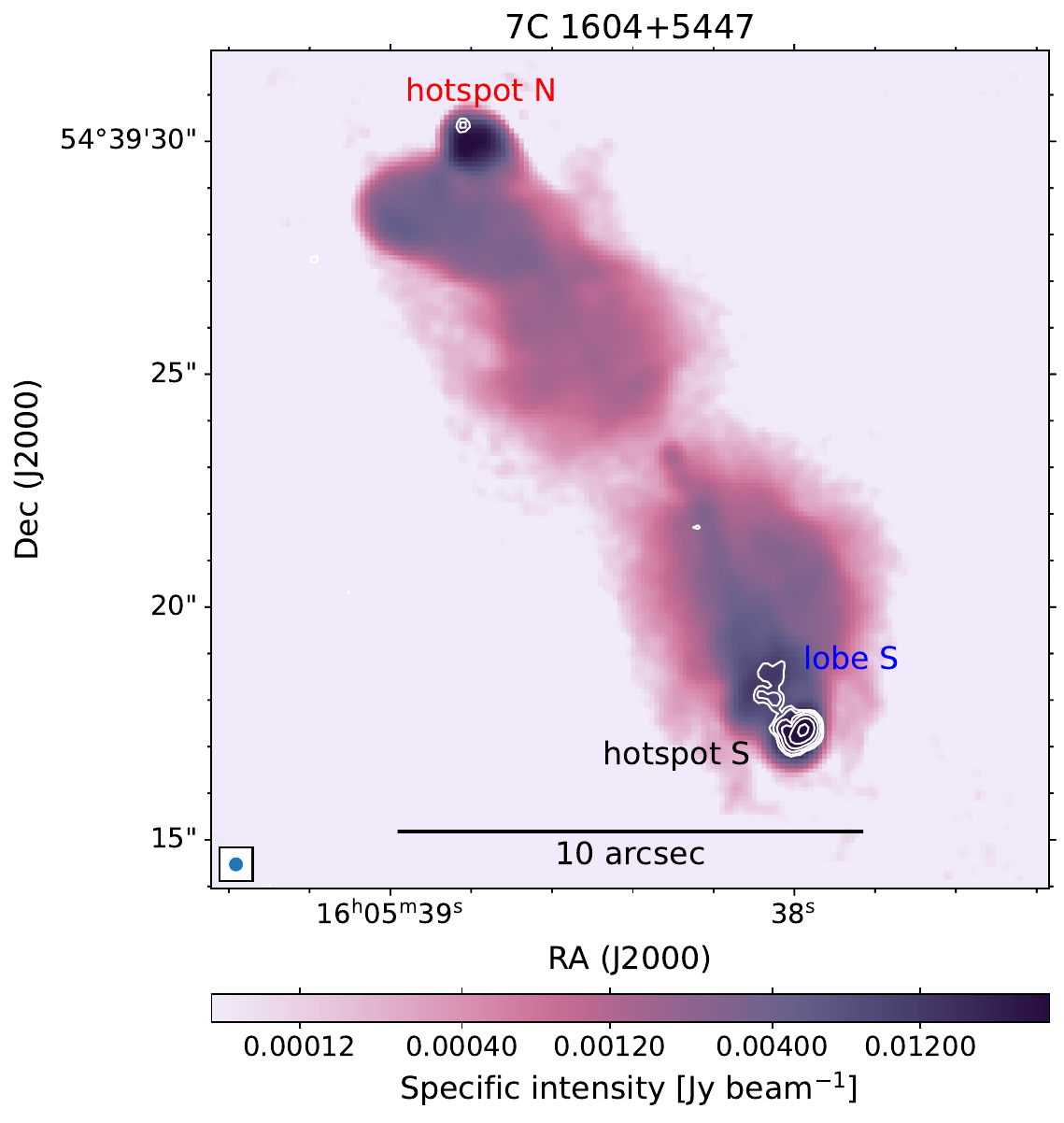}
    \includegraphics[width=0.48\textwidth]{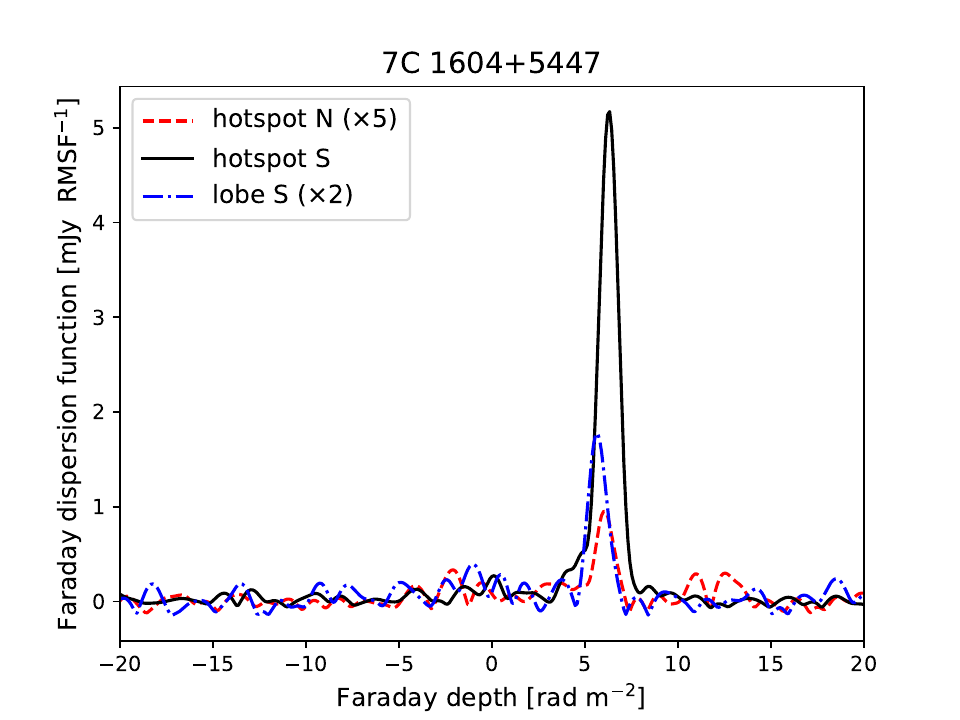}
\caption{Left panel: Stokes~$I$ image of 7C~1604+5447 at a resolution of 0.3\arcsec. Linearly polarised intensity contours are overlaid at levels of $6\times\sigma_{\rm{rms,P}}\left[1,2,4,8,\ldots\right]$, with $\sigma_{\rm{rms,P}}=8$~\textmu Jy~beam$^{-1}$ the noise level in the polarised intensity image. Right panel: Faraday dispersion function for various regions across the source. To improve visibility, the spectra corresponding to the fainter northern hotspot and ``lobe S'' are scaled by factors of 5 and 2, respectively. The uncleaned (dirty) Faraday spectra are shown in Figs.~\ref{fig:APpolcalibratorhotspotS}, \ref{fig:APpolcalibratorlobeS}, and~\ref{fig:APpolcalibratorhotspotN}.}
    \label{fig:7C1604+5447}
\end{figure*}

\begin{figure}
    \centering
    \includegraphics[width=0.48\textwidth]{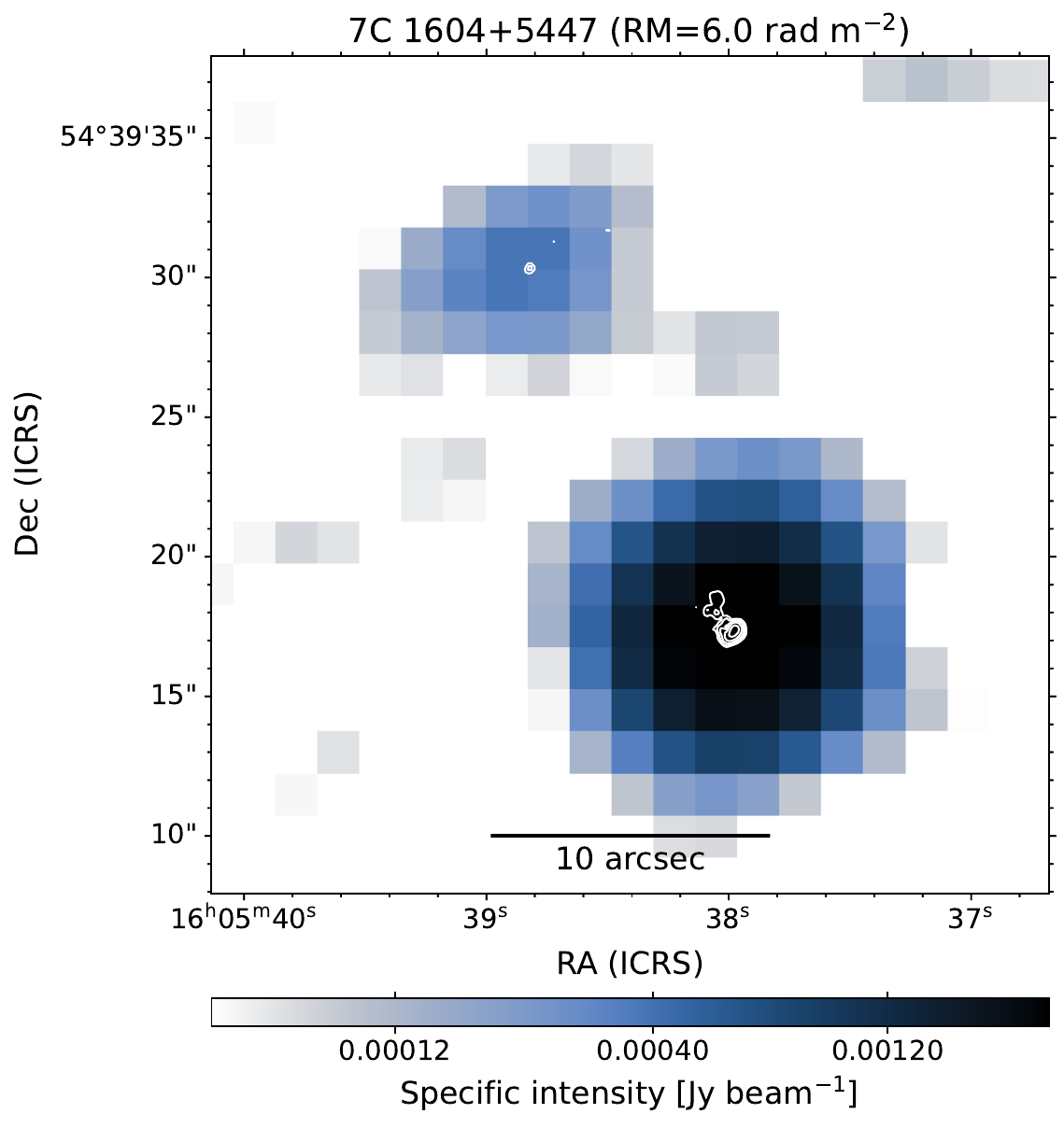}
\caption{Comparison of polarised intensity images extracted from the Faraday cube at a Faraday depth of ${\rm RM}=6~{\rm rad~m^{-2}}$, shown at angular resolutions of $6\arcsec$ (colour scale) and $0.3\arcsec$ (contours). Contours are drawn at levels of $5,\sigma_{\rm rms,P},[1,2,4,8,\ldots]$, where $\sigma_{\rm rms,P}$ denotes the rms noise in polarised intensity. The $6\arcsec$ image was obtained by excluding the international stations prior to imaging.}
    \label{fig:resolutioncomp}
\end{figure}

\subsection{Polarisation image depth}

To directly compare the image depths between the Stokes~$I$ images from \cite{dejong2024} and our Stokes~$Q$, $U$, and $V$ images, we imaged the central region of the field. In Table \ref{tab:depth} we provide the different RMS levels for Stokes~$I$, $Q$, $U$, and $V$ for the three different resolutions from \cite{dejong2024}. The close agreement between the four Stokes parameters indicates that the images are effectively thermal-noise limited (given the visibility weighting scheme), with minimal contamination from polarisation leakage. The same noise levels in Stokes~$Q$ and~$U$ suggest that cross-hand calibration and primary beam polarisation corrections are performing well. At 1.2\arcsec\ resolution, we find the largest difference between the rms noise levels of Stokes~$I$ and Stokes~$V$. This likely arises because the calibration was performed at 0.3\arcsec, and residual calibration artefacts, particularly due to issues with calibrating the Dutch remote stations and possibly due to gaps in the uv-coverage \citep[][]{ye2024}, become more pronounced at the coarser 1.2\arcsec~resolution in Stokes~$I$. This issue has recently been addressed by \cite{dejong2025b}, who implemented an improved calibration strategy for the Dutch LOFAR stations when combined with the international LOFAR stations.

\begin{table}
    \centering
    \caption{RMS noise levels for different Stokes parameters at different resolutions in \textmu Jy~beam$^{-1}$.}
    \label{tab:depth}
    \begin{tabular}{c|cccc}
        \toprule
        Resolution & I & Q & U & V \\
        \midrule
        0.3\arcsec & 14 & 14 & 14 & 14 \\
        0.6\arcsec & 21 & 22 & 22 & 21 \\
        1.2\arcsec & 40 & 39 & 39 & 37 \\
        \hline
    \end{tabular}
\end{table}

In general, the LOFAR point spread function (PSF) deviates strongly from a simple Gaussian shape. Its structure depends on the adopted weighting scheme, any applied \textit{uv}-tapers, and whether stations are phased up, in combination with the highly non-uniform distribution of stations on different spatial scales. To optimise for different science goals, it is therefore beneficial to tailor the weighting scheme to maximise the signal-to-noise ratio of the polarised emission of interest. For example, simply convolving an image with a Gaussian to a lower angular resolution to enhance sensitivity to extended emission may yield different detectability than adopting an alternative weighting scheme that modifies the PSF structure in a more complex way. 

\subsection{7C 1604+5447}
\label{sec:7C1604+5447}

The polarisation angle alignment calibrator 7C~1604+5447 is an FRII type \citep{1974MNRAS.167P..31F} radio galaxy \citep{2024RAA....24c5021L} with a largest angular extent of $15\arcsec$. Hotspots are visible for both the northern and southern lobes in our LOFAR 0.3\arcsec{} resolution Stokes~$I$ image, see Fig.~\ref{fig:7C1604+5447} (left panel).
The source is located at a spectroscopic redshift of $z=0.7683$ \citep{2025arXiv250314745D}. Polarised emission from this source has been previously reported at 1.4~GHz \citep{1998AJ....115.1693C,2007ApJ...666..201T,2009ApJ...702.1230T,2010ApJ...714.1689G,2014ApJS..212...15F}. At 144~MHz at 6\arcsec{} resolution, the source has an integrated flux density of $1.572$~Jy and is catalogued as ILT~J160538.32+543922.6 \citep{shimwell2022}. LOFAR observations at $20\arcsec$ resolution by \cite{ruiz2021} showed polarised emission from the southern lobe, which was imaged in more detail at $6\arcsec$ resolution by \cite{2024A&A...687A.267P,2025A&A...693A.100P}. 

In our 0.3\arcsec{} LOFAR linearly polarised intensity image (left panel Fig.~\ref{fig:7C1604+5447}), the polarised emission predominantly originates from the southern hotspot, with additional faint emission extending northwards. At a resolution of 6\arcsec{}, this emission could not be spatially separated from that of the southern hotspot, see Fig.~\ref{fig:resolutioncomp}. We also detect faint polarised emission from the northern hotspot. For the southern hotspot, we measure an RM of $6.285 \pm 0.008$~rad~m$^{-2}$ (right panel Fig.~\ref{fig:7C1604+5447}). For the faint lobe emission located just north of the southern hotspot, we find an RM of $5.65 \pm 0.05$~rad~m$^{-2}$, indicating a significant difference with respect to the RM of the southern hotspot. This slightly lower RM is also visible in the Faraday depth spectrum shown in Fig.~\ref{fig:7C1604+5447} (right panel), where the blue spectrum peaks at a lower Faraday depth than the black spectrum. For the northern hotspot, we measure an RM of $6.05 \pm 0.08$~rad~m$^{-2}$. A RM map of the polarised emisison from the southern lobe is shown in Fig.~\ref{fig:rmmap} in Appendix~\ref{sec:rmmap}.

\begin{figure*}
    \centering
    \includegraphics[width=0.49\textwidth]{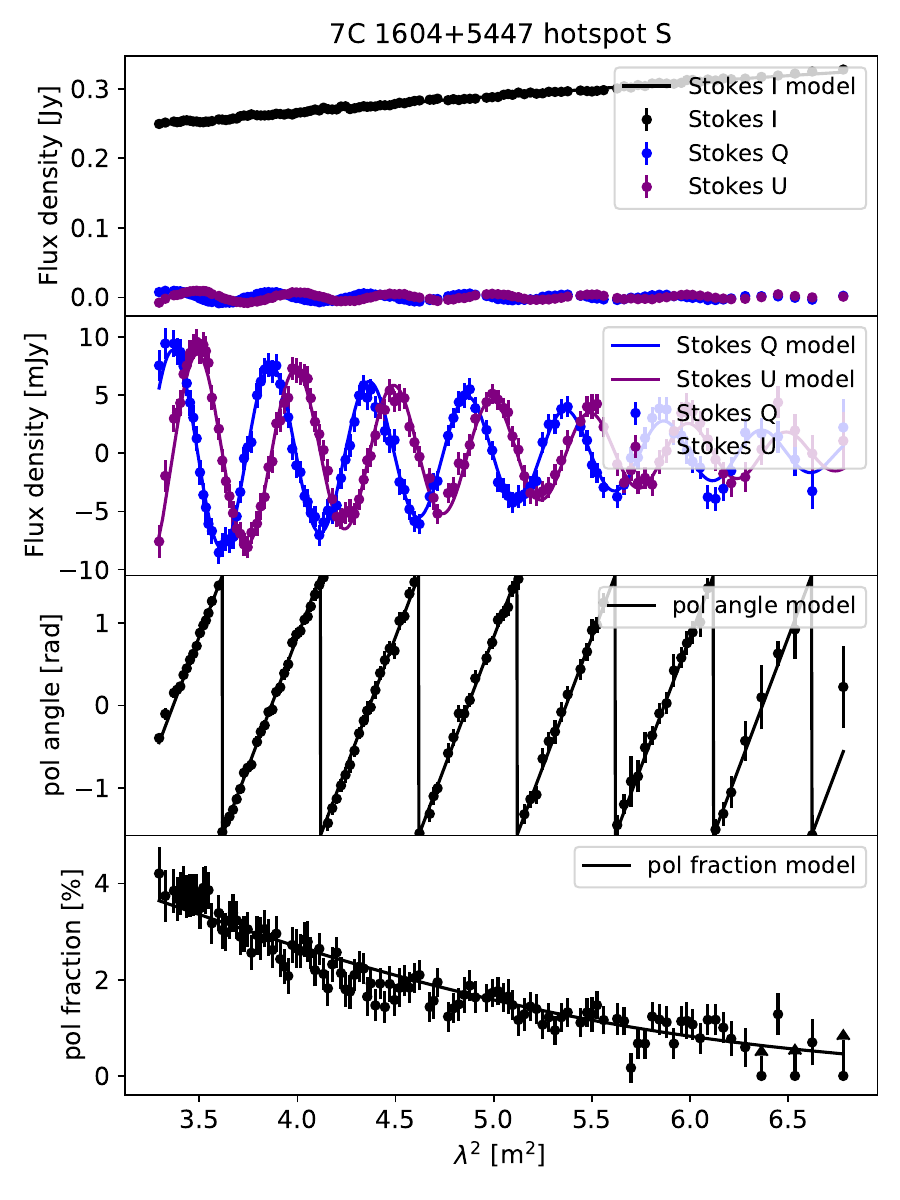}
    \includegraphics[width=0.49\textwidth]{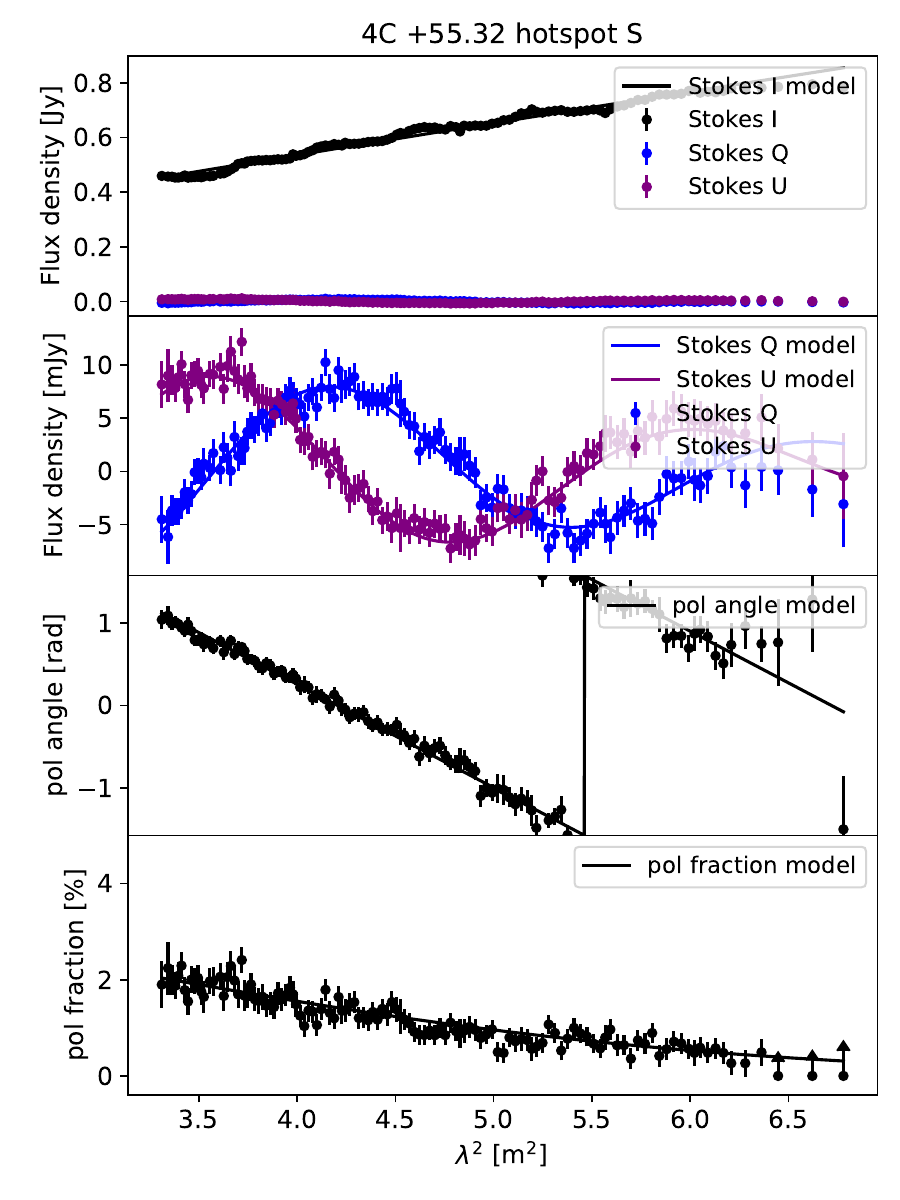}    
    \caption{Left panels:Polarisation properties of 7C\,1604+5447 as a function of $\lambda^{2}$. 
From top to bottom, the panels show the Stokes~$I$, $Q$, and $U$ flux densities; a zoom-in of the $Q$ and $U$ flux densities; the polarisation angle; and the polarisation fraction, defined as ${\sqrt{Q^{2}+U^{2}}}/{I}$. A corresponding model fit (Eq.~\ref{eq:depol}) to these quantities is shown with solid lines. 
Arrows indicate lower limits on the polarisation fraction. The total linearly polarised flux density, ${\sqrt{Q^{2}+U^{2}}}$, was corrected for Ricean bias \citep{1974ApJ...194..249W}.
Right panels: The same quantities, shown for the source 4C\,+55.32. The subtle step-like pattern imprinted on the Stokes~$I$ spectrum of 4C\,+55.32 is a by-product of the Stokes~$I$ amplitude self-calibration, combined with the \texttt{WSClean} wideband imaging setting \texttt{-channels-out 6}, which discretises the self-calibration model. This effect is also present in other sources, but becomes apparent only for the brightest ones with sufficiently high signal-to-noise ratios.}
    \label{fig:polfrac}
\end{figure*}

Given the high signal-to-noise detection of the southern hotspot, we can further characterise its polarisation properties as a function of wavelength, including the polarisation fraction. These results are shown in Fig.~\ref{fig:polfrac} (left panel). \citet{2024A&A...687A.267P} reported a polarisation fraction of 1.75\% for the peak pixel in polarised intensity at $6\arcsec{}$ resolution, while \citet{ruiz2021} reported a lower value of 0.41\% at 20\arcsec{} resolution. In our high-resolution data, we measure a polarisation fraction of $\sim$0.6\% at the longest wavelengths, increasing to $\sim$4\% at the shortest wavelengths, clearly indicating wavelength-dependent depolarisation. We thus fitted these data with an external Faraday rotation depolarisation model of the form
\begin{equation}
\label{eq:depol}
    P(\lambda^2) = p_0 I \exp\left(-2\Sigma_{\text{RM}}^2 \lambda^4\right) \exp\left[2i\left(\chi_{\lambda_{0}^{2}} + \text{RM}\left(\lambda^{2}-\lambda_{0}^{2}\right)\right)\right],
\end{equation}
where $p_0$ is the intrinsic polarisation fraction and $\Sigma_{\text{RM}}$ is the Faraday dispersion within the integration area. This model has the same form as Eqs.~\ref{eq:qu1} and~\ref{eq:qu2}, but includes an exponential depolarisation term. From this fit we obtain a Faraday dispersion of $\Sigma_{\text{RM}} = 0.172 \pm 0.086$~rad~m$^{-2}$. 

A hint of a faint jet-like feature appears to extend southwards from the nucleus (Fig.~\ref{fig:7C1604+5447}), while no jet emission is detected to the north. This asymmetry suggests that the southern jet is relativistically beamed and oriented closer to our line of sight. The band-averaged polarisation fractions of the northern and southern hotspots, $2.12 \pm 0.01$\% and $0.55 \pm 0.06$\%, respectively, may therefore provide evidence for the Laing--Garrington effect \citep{1988Natur.331..149L,1988Natur.331..147G}. In this scenario, the approaching jet is Doppler boosted, placing its associated lobe closer to the observer, such that the radiation traverses a smaller column of the host galaxy's Faraday-rotating plasma and is consequently less depolarised.

\subsection{4C\,+55.32} 
\label{sec:4C+55.32}
4C\,+55.32 is an FR\,II radio source \citep{2024RAA....24c5021L}. The source has an optical counterpart \citep{2023A&A...678A.151H} at a spectroscopic redshift of $z = 0.22795$ \citep{2015ApJS..219...12A}. The integrated flux density at 144~MHz is 3.23~Jy (ILTJ161212.30+552303.7; \citealt{shimwell2022}). Polarised emission from 4C\,+55.32 has previously been detected at 1.4~GHz \citep{1998AJ....115.1693C,2009ApJ...702.1230T,2010ApJ...714.1689G,2014ApJS..212...15F}, but the source was not detected in earlier LOFAR polarisation studies. 

\begin{figure*}
    \centering
    \includegraphics[width=0.47\textwidth]{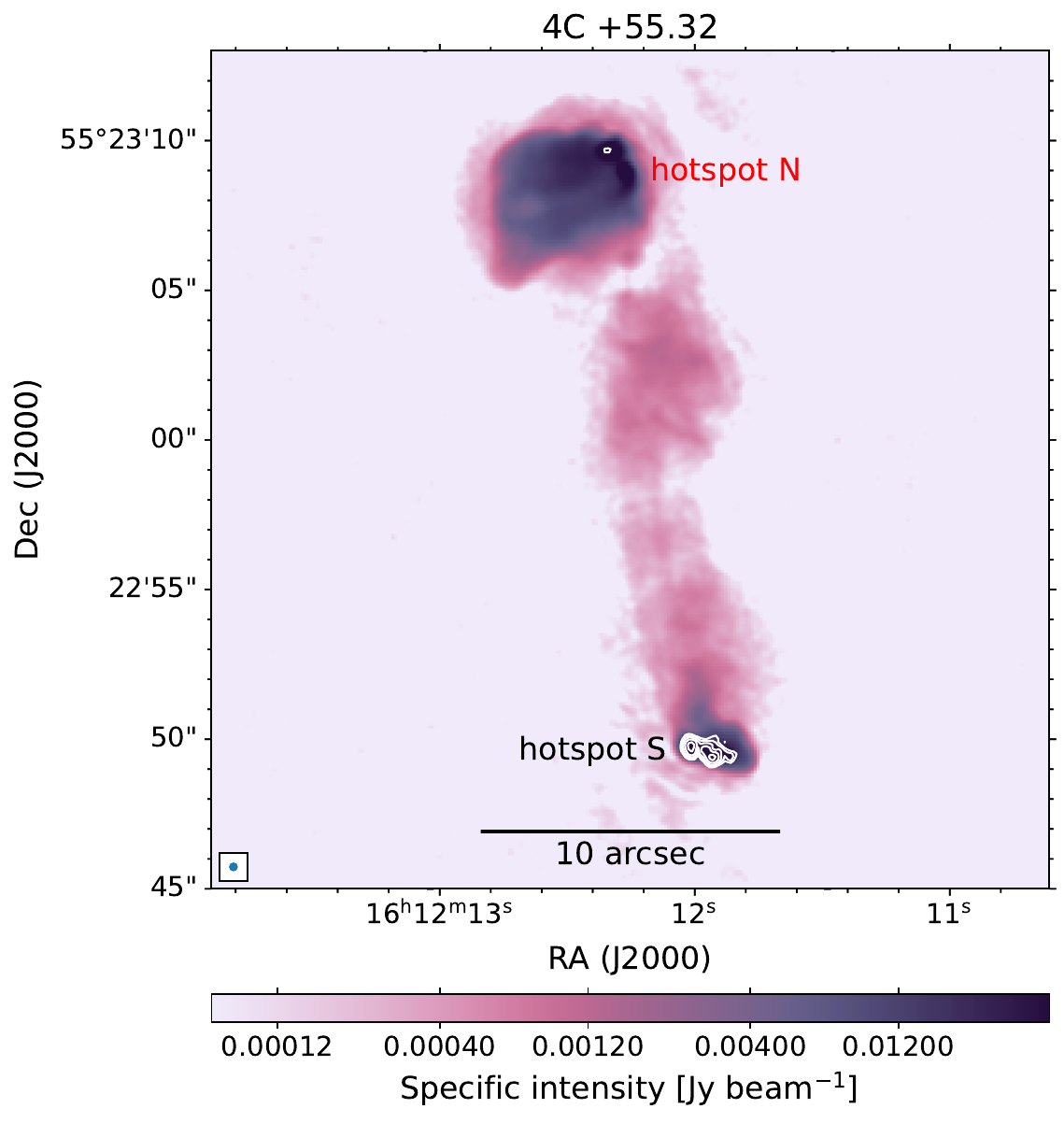}
    \includegraphics[width=0.47\textwidth]{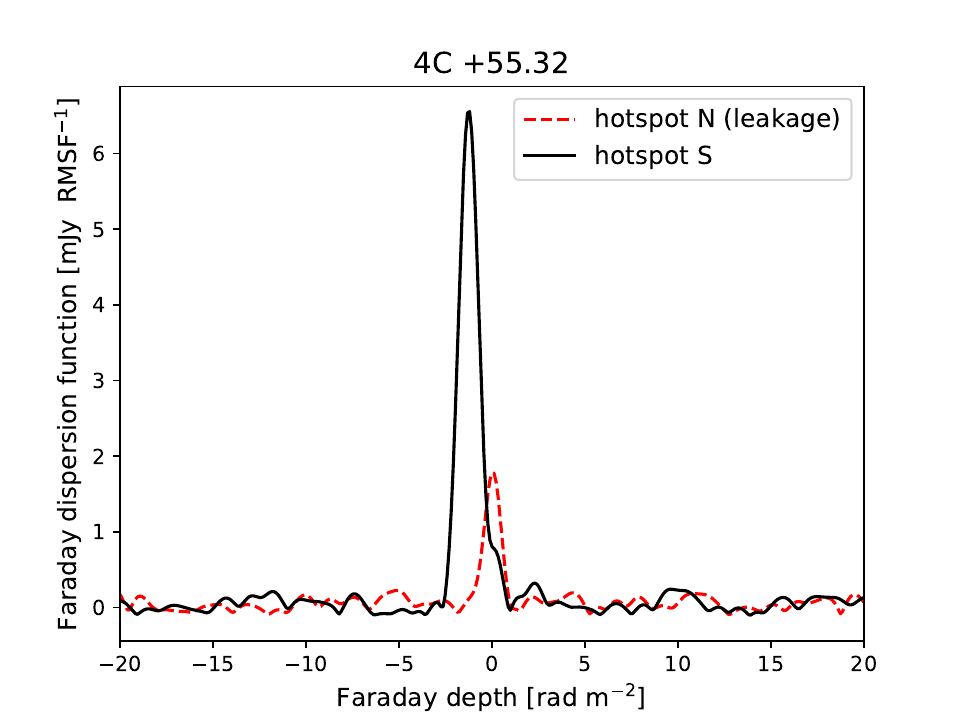}
    \caption{Left panel: Stokes~$I$ image of 4C\,+55.32 at a resolution of 0.3\arcsec. White contours show the linearly polarised intensity at levels of $15 \times \sigma_{\rm rms,P}\,[1,2,4,8,\ldots]$, where $\sigma_{\rm rms,P} = 10$~\textmu Jy~beam$^{-1}$ is the rms noise in the polarised intensity image. Right panel: Faraday dispersion functions extracted from the northern (red) and southern (black) hotspot regions. The spectrum from the northern hotspot peaks close to 0~rad~m$^{-2}$, indicating that the detected signal is likely dominated by instrumental polarisation leakage near zero Faraday depth. This interpretation is supported by the very low measured polarisation fraction of 0.11\%. Despite having a similar Stokes~$I$ intensity, the southern hotspot shows a significantly stronger polarised signal. Although a weaker asymmetric wing consistent with the leakage pattern seen in the northern hotspot is present, the majority of the signal arises from intrinsic polarised emission associated with the source. The uncleaned (dirty) Faraday spectrum is shown in Fig.~\ref{fig:APbrightnewhotspotS}.}
    \label{fig:4C55.32}
\end{figure*}

In Fig.~\ref{fig:4C55.32}, the 0.3\arcsec{} resolution Stokes~$I$ image shows two main lobes, with the northern lobe being more extended than the southern one. In addition, extended emission is visible between the two radio lobes. Both lobes exhibit a double-hotspot structure. In the linearly polarised intensity image, we detect polarised emission from this source at low radio frequencies for the first time. The polarised emission originates from the double-hotspot region at the outer edge of the southern lobe. The Faraday dispersion function peaks at RM=$-1.25\pm 0.01$~rad~m$^{-2}$ for the emission. The detailed polarisation properties of 4C\,+55.32 as a function of wavelength are shown in Fig.~\ref{fig:polfrac} (right panel). We find a polarisation fraction of $\sim$0.4\% at the longest wavelengths, increasing to $\sim$2\% at the shortest wavelengths. Fitting the external depolarisartion model from Eq.~\ref{eq:depol}, we obtain a Faraday dispersion of $\Sigma_{\text{RM}} = 0.164 \pm 0.082$~rad~m$^{-2}$. 


The source was likely missed in previous LOFAR searches for polarised emission because its polarised signal peaks at a Faraday depth close to the region affected by contamination around zero RM (right panel Fig.~\ref{fig:4C55.32}). The leakage signal (i.e., instrumental polarisation) is intrinsically centred at 0~rad~m$^{-2}$. This leakage signal is broadened by the ionospheric RM correction and the RMSF, which has a typical magnitude of $\mathcal{O}(1~\rm rad~m^{-2})$. 
A comparison with the RM spectrum extracted from the northern double hotspot region, which has a similar total intensity, shows that the polarised signal from the southern lobe is not dominated by instrumental leakage. The level of contamination, estimated from the northern hotspot region, is approximately an order of magnitude lower. Moreover, the polarised emission from the southern lobe is clearly offset in Faraday depth from the peak associated with leakage. 

Assuming that all observed polarised emission from the northern double-hotspot region is due to instrumental leakage, as it peaks at zero Faraday depth, we use this signal to estimate the fractional leakage from Stokes~$I$ into linear polarisation in our data. We do so by dividing the maximum detected linearly polarised intensity at the northern hotspot by the corresponding Stokes~$I$ flux density, yielding a leakage level of 0.11\%. Using the same method, we compute the Stokes~$I$ to Stokes~$V$ leakage, finding a value of 0.021\%.

In the LoTSS survey, the typical instrumental polarisation leakage is approximately 0.2\% from Stokes~$I$ into Stokes~$Q/U$, and 0.056\% from Stokes~$I$ into Stokes~$V$ \citep{shimwell2022}. A caveat of our analysis is that this estimate is based on a single measurement; the leakage level is expected to vary with angular separation from the polarisation leakage calibrator due to small deviations in the LOFAR station beam compared to the theoretical beam model.

\subsection{7C\,1607+5402}
\begin{figure}
    \centering
    \includegraphics[width=0.49\textwidth]{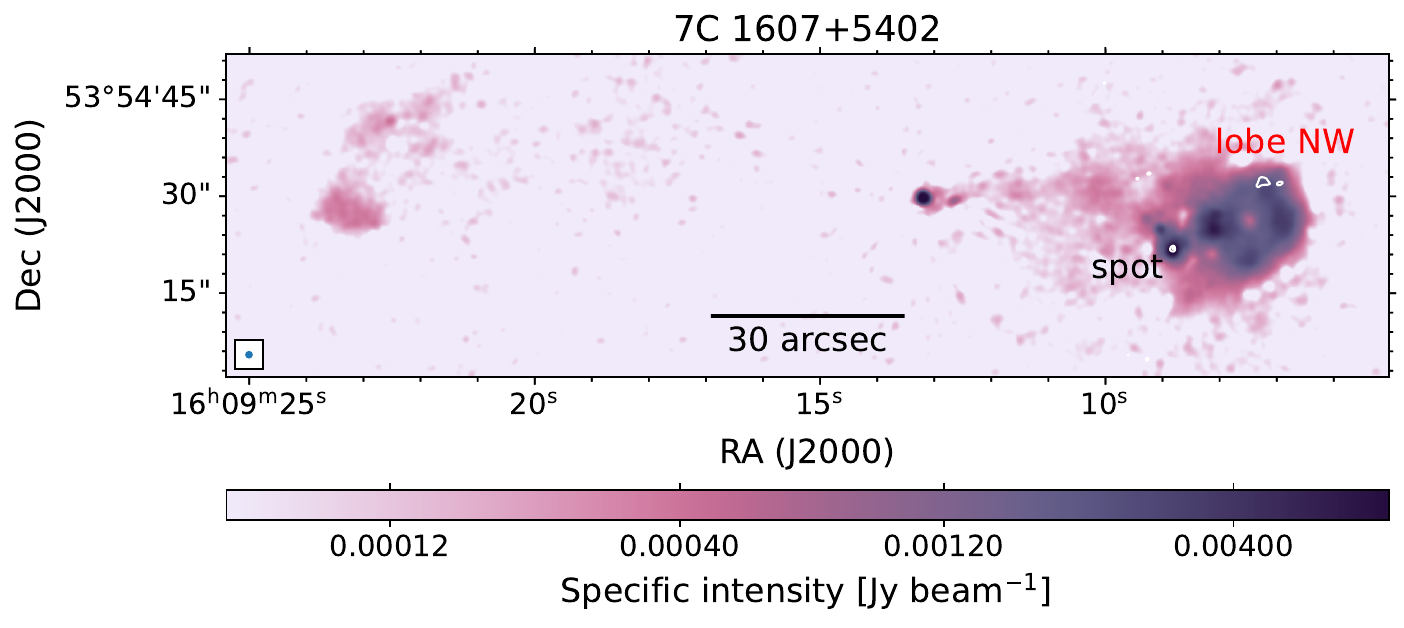}
    \includegraphics[width=0.47\textwidth]{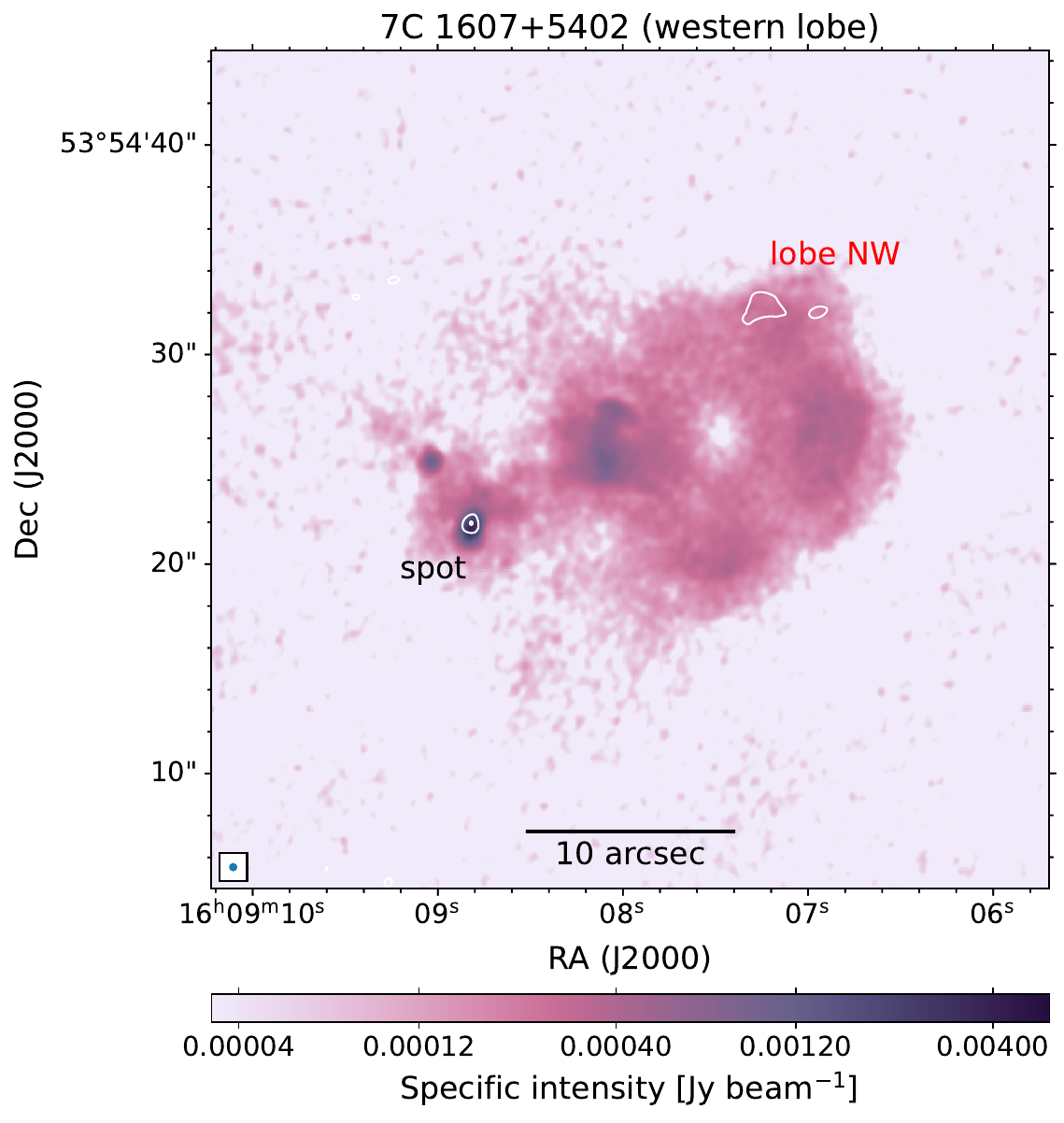}
    \includegraphics[width=0.47\textwidth]{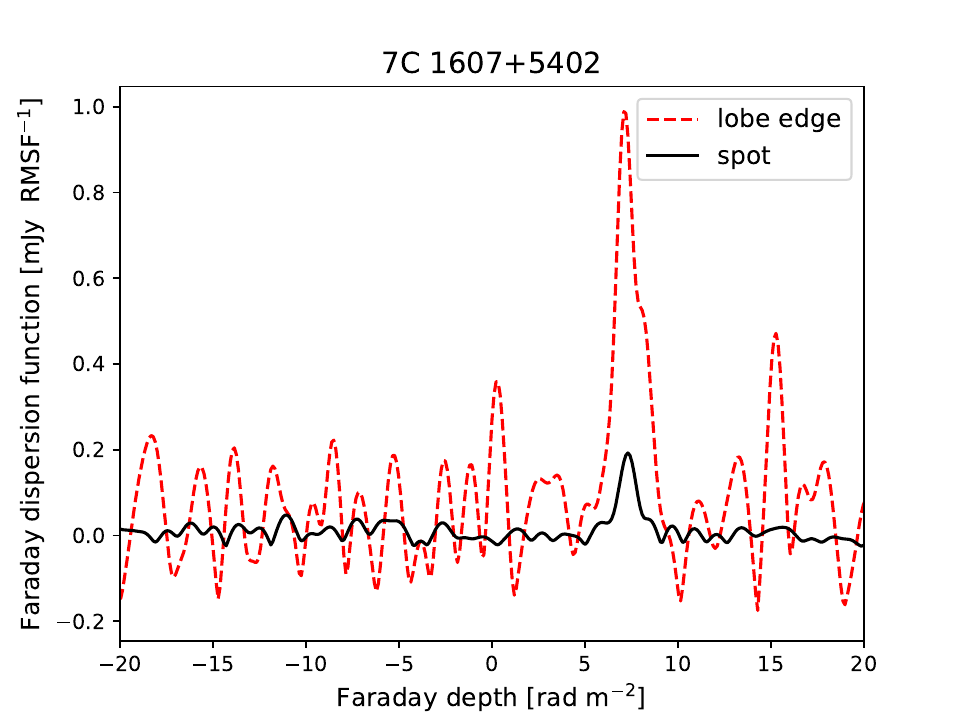}
    \caption{Top panel: Stokes~$I$ image of 7C\,1607+5402 at a resolution of 1.2\arcsec. White contours indicate the linearly polarised intensity at levels of $6 \times \sigma_{\rm rms,P}\,[1,2,4,8,\ldots]$, where $\sigma_{\rm rms,P} = 12$~\textmu Jy~beam$^{-1}$ is the rms noise in the polarised intensity image.  Middle: Stokes~$I$ image of 7C\,1607+5402 at a resolution of 0.4\arcsec. White contours show the linearly polarised intensity from the 1.2\arcsec-resolution polarised intensity image, plotted at the same contour levels as in the top panel.  Bottom panel: Faraday dispersion functions for the ``spot'' (black) and ``NW lobe'' (red) regions, extracted from the polarised intensity cubes at resolutions of 0.4\arcsec\ and 1.2\arcsec, respectively. The uncleaned (dirty) Faraday spectra are shown in Figs.~\ref{fig:APspecialsourcespot} and~\ref{fig:APspecialsourcelobeedge}.}
    \label{fig:7C1607+5402}
\end{figure}

7C\,1607+5402 has been identified as a quasar (QSO) at a redshift of $z = 0.99270$ \citep{2015ApJS..219...12A,2020ApJS..250....8L}. The source is also detected at X-ray wavelengths \citep{2000IAUC.7432....3V,2016A&A...588A.103B,2016ApJS..224...40W}. 7C\,1607+5402 is an example of a giant radio galaxy hosted by a quasar, with two radio lobes spanning a projected linear size of $\sim$1~Mpc \citep{2021ApJS..253...25K,2022A&A...660A..59M,2024A&A...686A..21S,2024A&A...691A.185M}. The integrated flux density of the source at 144~MHz is 532.7~mJy \citep{2023A&A...678A.151H}.

Figure~\ref{fig:7C1607+5402} (top panel) shows a Stokes~$I$ image at 1.2\arcsec\ resolution, revealing an FR\,II morphology with a pronounced asymmetry in lobe brightness. A higher-resolution 0.4\arcsec\ image (middle panel of Fig.~\ref{fig:7C1607+5402}) reveals a compact feature in the western lobe (labelled ``spot''). The nature of this feature is unclear; it may correspond to a jet instability or deflection, or alternatively represent a hotspot. Polarised emission from 7C\,1607+5402 at 1.4~GHz was previously reported by \citet{2010ApJ...714.1689G}. At LOFAR HBA frequencies, polarised emission was detected by \citet{ruiz2021,2024A&A...687A.267P} using images at 20\arcsec\ and 6\arcsec\ resolution, with the polarised signal originating from the western lobe.

In our 0.4\arcsec\ resolution LOFAR image, we detect polarised emission associated with the compact ``spot'' feature. However, the 6\arcsec\ resolution study of \citet{2024A&A...687A.267P} indicates that most of the polarised emission arises from the north-western part of the lobe. This component is not detected at 0.4\arcsec\ resolution, implying that it is resolved out. To recover this emission, we re-imaged the LOFAR data, producing Q and U cubes at a resolution of 1.2\arcsec\ by applying a \textit{uv}-taper and combining the LOFAR core stations into a virtual super-station, thereby reducing contamination from nearby sources. In the resulting 1.2\arcsec\ linearly polarised intensity image, shown as contours in the top and middle panels of Fig.~\ref{fig:7C1607+5402}, we detect the extended polarised emission, while polarised emission from the ``spot'' feature is also recovered.

From the Faraday spectrum (bottom panel Fig.~\ref{fig:7C1607+5402}), we measure an RM of $7.30 \pm 0.08$~rad~m$^{-2}$ for the ``spot'', and $7.26 \pm 0.05$~rad~m$^{-2}$ for the north-western lobe. These values can be compared to those reported by \citet{ruiz2021,2024A&A...687A.267P}, who measured $7.30 \pm 0.02$ and $7.17 \pm 0.02$~rad~m$^{-2}$, respectively.

\subsection{CR Draconis in Stokes $V$}

\citet{callingham2021, callingham2019a} detected at 146~MHz highly circularly polarised Stokes~$V$ radio emission from the binary M-dwarf system CR Draconis (BD+55 1823; GJ 9552) in the ELAIS-N1 deep field. This was the first radio detection of the source, following non-detections in earlier surveys at 325~MHz \citep{sirothia2009}, 610~MHz \citep{garn2008}, and 1.4~GHz \citep{2007ApJ...666..201T}. \cite{callingham2021} argues that this emission originates from coherent radio bursts, powered by the electron-cyclotron maser instability (ECMI; \citealt{wu1979, treumann2006}), which is caused by binary interaction.

The flare star of CR~Draconis exhibits a B-band (\textasciitilde450~nm) flare rate of approximately one every \textasciitilde10~hours \citep{haagen2018}, implying that a flare is likely to occur during most of our 8~hour LOFAR observations. We measure in our Stokes~$I$ images a total intensity that spans from $0.64\pm0.13$~mJy in observation L798074 to $2.07\pm0.41$~mJy in observation L686962. The corresponding Stokes~V images give us total intensities ranging from $0.44\pm0.09$~mJy in observation L816272 to $1.99\pm0.40$~mJy in observation L686962. The inferred circular polarisation fraction reaches a maximum of $96\pm21$\% in observation L686962, while the minimum value of $51\pm21$\% is measured for observation L816272. Although our observations were taken several years after those presented by \citet{callingham2021}, the measured flux density variations and circular polarisation fractions are consistent with their reported values.

\begin{figure*}
 \centering
\includegraphics[width=1\linewidth]{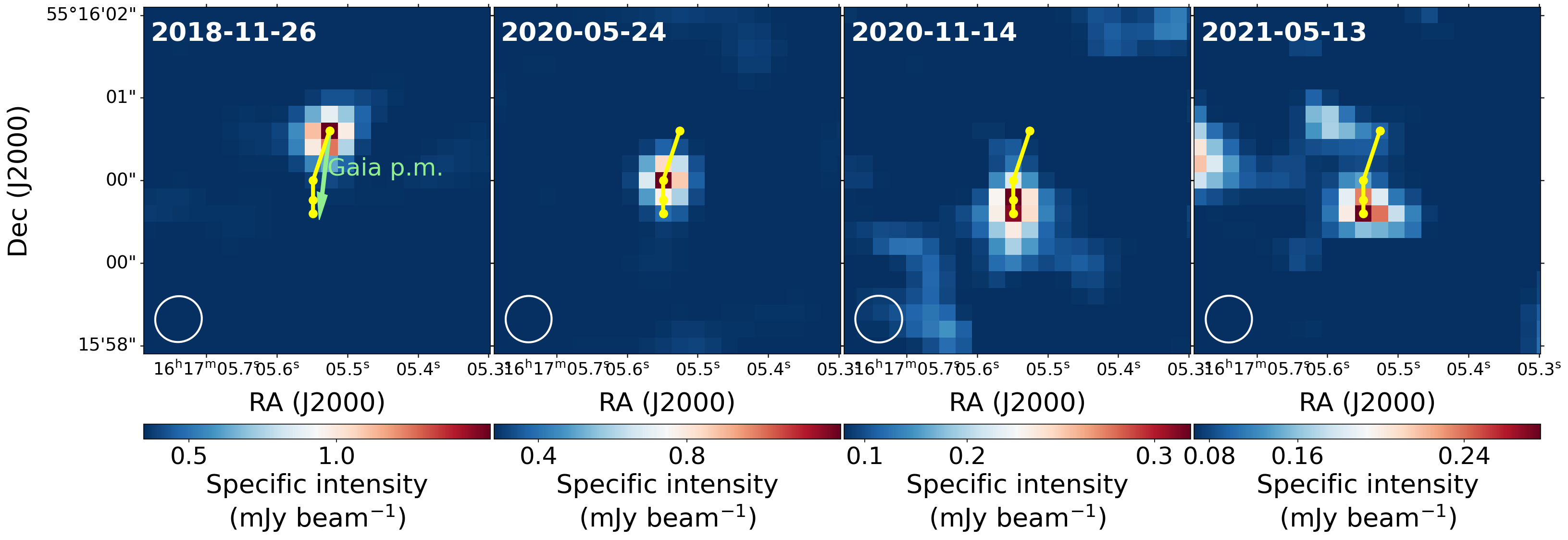}
  \caption{Proper motion of CR Draconis across the four ELAIS-N1 observations at 0.57\arcsec$\times$0.55\arcsec~resolution, imaged in Stokes~$V$. The yellow line indicates the four positions of the sources from the four observations, fitted with \texttt{PyBDSF} \citep{mohan2015}. The light green vector indicates the proper motion over 2.5 years (87.97 mas/yr, -430.963 mas/yr) from Gaia \citep{gaiadr3}. The beam shape is indicated in the bottom left corner.}
\label{fig:drac06}
\end{figure*}

Gaia DR3 measured a proper motion for CR~Draconis of $0.440 \pm 0.001\,\arcsec\,\mathrm{yr}^{-1}$ \citep{gaiadr3}. With our high-resolution observations spanning 2.5~years, we are able to resolve this motion in Stokes~$V$. Since the source lies $\sim$0.9$^{\circ}$ from the pointing centre, time-smearing effects on the longest baselines are significant in two of the observing epochs. To mitigate this, we image all four observations at a resolution of $0.57\arcsec \times 0.55\arcsec$, where time smearing is substantially reduced (Fig.~\ref{fig:drac06}). The proper motion corresponding to Figure \ref{fig:drac06} ranges from 0.44$\pm0.11$\arcsec~$\mathrm{yr}^{-1}$ to 0.33$\pm0.11$\arcsec~$\mathrm{yr}^{-1}$, where the latter corresponds to the proper motion between the two most time-smeared observations. Our proper motion measurements agree with those from Gaia DR3, underscoring the high precision of our astrometry.

\section{Discussion}
\label{sec:discussion}

\subsection{Low-frequency polarised emission at sub-arcsecond resolution}
Our observations provide a first view of polarised emission from three radio galaxies at sub-arcsecond resolution at a central frequency of approximately 140~MHz. At these frequencies, but at much lower angular resolution ($\sim$20\arcsec), FR\,II radio galaxies constitute the largest class of polarised sources, accounting for about 40\% of the polarised population and occurring roughly twice as frequently as FR\,I sources \citep{2023MNRAS.519.5723O}. This contrasts with observations at 1.4~GHz, where FR\,I and FR\,II sources are found in comparable numbers \citep[e.g.,][]{2011ApJ...733...69B,2015ApJ...806...83O}.

Given that radio hotspots in FR\,II sources are intrinsically compact and are high S/N, it is likely that FR\,II sources constitute an even larger fraction of the low-frequency polarised source population when observed with baselines. Indeed, for two of the sources in our sample, the polarised emission detected at 0.3\arcsec{} resolution originates from the hotspots. For the third source, the polarised emission arises from a compact region that may represent a hotspot or another compact feature within the radio jet. In this case, we also detect polarised emission near the edge of the radio lobe; however, this component is extended and only becomes visible at the lower angular resolution of 1.2\arcsec{}. We caution, however, that drawing firm conclusions about the low-frequency polarised source population at sub-arcsecond resolution requires a more thorough analysis and a systematic search for polarised emission across the full dataset.

\subsection{Availability of an unpolarised in-field calibrator}
\label{sec:discussionunpolarised}
In this work, we have shown that the polarisation calibration scheme described can successfully be applied to 32~h of observations with the full international LOFAR array, revealing for the first time low-frequency polarised emission at sub-arcsecond resolution in both linear and circular polarisation. Below, we discuss the applicability of this approach to other LOFAR observations.

As demonstrated by LoTSS studies of linear and circular polarisation \citep{callingham2023,2023MNRAS.519.5723O}, accurate polarisation calibration can be applied to the large majority of LOFAR observations with the Dutch part of the array. The main complication arises in fields that contain very strong, compact polarised emission. In these cases, the underlying DDF-pipeline assumption that the field-averaged polarised signal is approximately zero may break down \citep{tasse2021}. As shown by \citet{2023MNRAS.519.5723O}, this occurs in $\lesssim 10\%$ of LoTSS fields and results in an artificial polarisation signature being imprinted on all sources in the field. Additional steps would be required for such fields to be processed for polarisation with the international baselines, as otherwise the polarisation calibration of the Dutch LOFAR stations would be too heavily compromised \citep[for possible approaches to address this see sect. 2.3.4 in][]{2023MNRAS.519.5723O}.

For international baseline calibration, any source that can serve as a successful in-field calibrator can, in principle, also be used as a leakage calibrator, provided it is unpolarised in the HBA band. The signal-to-noise ratio is generally sufficient because the in-field calibrator is also required to solve for the much more rapidly varying ionospheric total electron content (TEC) effects on timescales of $\lesssim 1$~min. This implies that there should be sufficient S/N to solve for leakage terms on the longer timescales of $\gtrsim 30$~min. The amount of linear polarisation of in-field calibrators can be verified by consulting the LoTSS polarisation catalogues \citep[i.e.,][]{2023MNRAS.519.5723O}. For sources that remain compact, the polarisation properties are expected to remain unchanged. However, for significantly resolved sources, linearly polarised emission may be present on long baselines while remaining undetected at LoTSS resolution due to beam depolarisation. For this reason, it is preferable to select a compact calibrator source. Fortunately, the vast majority of radio sources are unpolarised at metre wavelengths. This assumption is even safer for circular polarisation, as the number of in-field calibrators that are circularly polarised is expected to be extremely small. Based on the Stokes~$V$ catalogue of \citet{callingham2023}, covering 5634~deg$^{2}$, only one detected circularly polarised source has an integrated Stokes~$I$ flux density above 100~mJy (121.7~mJy), whereas in-field calibrators are typically required to have integrated flux densities of $\gtrsim 200$--300~mJy. Consequently, at most one or two potential in-field calibrators are expected to violate the Stokes~$V=0$ assumption for a full northern sky survey.

If the chosen in-field calibrator turns out to be polarised, an alternative bright source within the field can be used instead. Moreover, if an incorrect assumption about the calibrator’s polarisation state is made, this will result in artificial polarisation being introduced into all other sources in the field, providing a clear diagnostic of the issue.

\subsection{Availability of an in-field polarised calibrator}
To align two or more observations in Faraday depth, or to perform such an alignment on shorter time intervals within a single observation if needed, the detection of linearly polarised emission is required. Misalignments in polarisation angles arise primarily due to ionospheric Faraday rotation, which introduces an ionospheric contribution to the rotation measure, RM$_{\rm ion}$. The bulk of this effect (typically a few rad~m$^{-2}$) is removed using satellite-based ionospheric models using \texttt{RMextract} at the start of the data processing, leaving residual offsets of approximately $0.1$--$0.3$~rad~m$^{-2}$ after correction \citep{2013A&A...552A..58S}.

From the four observations listed in Table~\ref{tab:alignment}, we find that our measured values are in approximate agreement with this level of precision, with a standard deviation of 0.2~rad~m$^{-2}$. To obtain better statistics, we also consider the study by \citet{2024A&A...687A.267P}, which provides RM measurements for 19 observing runs (their table~4). From these values, we derive a standard deviation of 0.06~rad~m$^{-2}$, which lies at the lower end of the expected residual range and indicates a high accuracy of the satellite-based ionospheric correction for these observations. We note, however, that the accuracy of these corrections depends on the ionospheric conditions sampled. In Sect.~\ref{sec:intraalignment}, we investigate possible RM variations on shorter timescales within single observing epochs.

At 20\arcsec\ resolution in LoTSS-DR2 (8~hr per pointing), an average surface density of linearly polarised sources of $0.43$~deg$^{-2}$ was reported by \citet{2023MNRAS.519.5723O}. Given the typical field of view of the international LOFAR stations of $\sim$6.5~deg$^{2}$, this implies that, on average, slightly fewer than three polarised sources are expected within a single pointing.  In the ELAIS-N1 field, we aligned the observations in Faraday depth using the source 7C\,1604+5447, which is detected at high S/N in 8~h of observations. The source 4C\,+55.32 would also provide a suitable alternative for this purpose. In contrast, 7C\,1607+5402 represents a more challenging case, as its polarised flux density is approximately a factor of 10--20 lower than that of 7C\,1604+5447. We note that a full search for polarised emission using our LOFAR international baseline data for this field has not yet been performed, and additional suitable polarised sources may be present, as suggested by the larger sample reported by \citet{2024A&A...687A.267P}.

As a practical strategy for efficient pipeline processing of LOFAR data including all international stations \citep{morabito2022}, we suggest first checking for the presence of an in-field polarised calibrator using catalogues of polarised sources detected by LoTSS \citep{2023MNRAS.519.5723O}, or, if none are available, sources known to be polarised at higher frequencies \citep[e.g.,from NVSS;][]{1998AJ....115.1693C,2009ApJ...702.1230T,2014ApJS..212...15F}. This is important because, aside from LoTSS, there are currently no low-frequency polarisation surveys covering most of the northern sky. If no suitable polarised calibrator is found, the next step is to inspect all facet calibrator sources \citep[24 to 30 for this field][]{dejong2024,dejong2025b}, which are already processed by the pipeline and are typically among the brightest sources in the field, thus increasing the likelihood of detectable polarised emission. This approach is computationally inexpensive, as it only requires additional imaging and RM-synthesis steps on small datasets with limited image sizes (since LOFAR core stations are phased-up into a single virtual superstation).

The polarisation angle alignment itself is a global correction, consisting of a single RM and a polarisation angle offset that is common to all antennas. In principle, this correction can also be derived through alternative methods. For example, \cite{2016ApJ...830...38L,2017PASA...34...40L,2023A&A...674A.119S} demonstrated that the polarised Galactic foreground can be used for this purpose. Alternatively, one could exploit the larger field of view of the Dutch LOFAR stations, which covers approximately four times more sky area per pointing \citep[][]{shimwell2022}, thereby increasing the probability of finding a suitable polarised calibrator. Given these considerations, the global polarisation angle alignment could also be performed prior to any international baseline processing, avoiding the need to apply this correction at the international baseline processing stage.

\subsection{Intra-observation polarisation angle alignment}
\label{sec:intraalignment}

\begin{figure*}
 \centering
\includegraphics[width=0.49\textwidth]{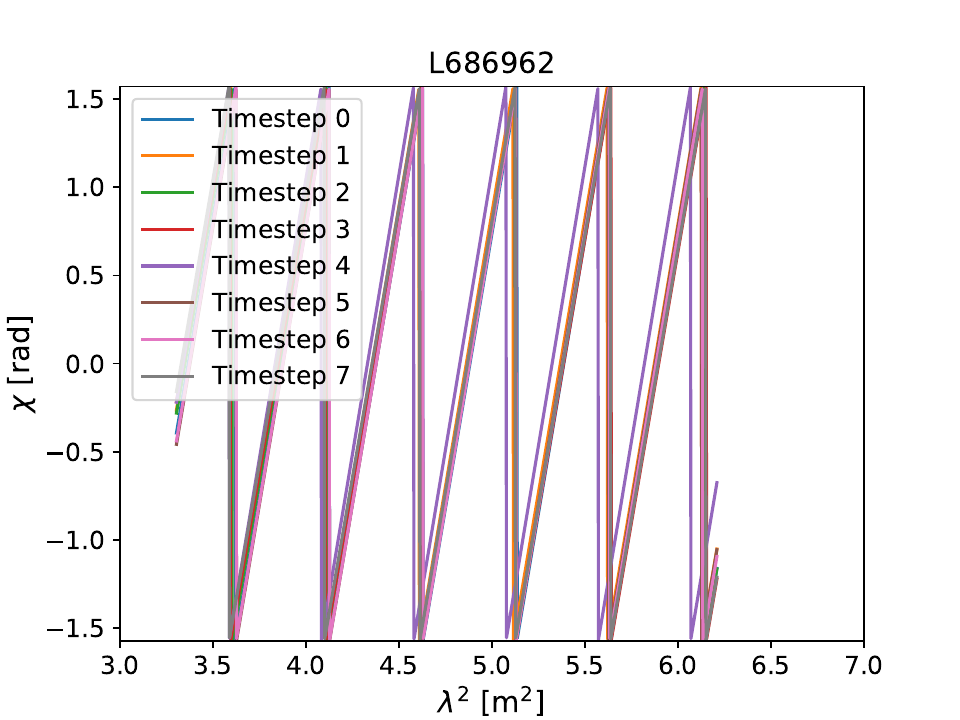}
\includegraphics[width=0.49\textwidth]{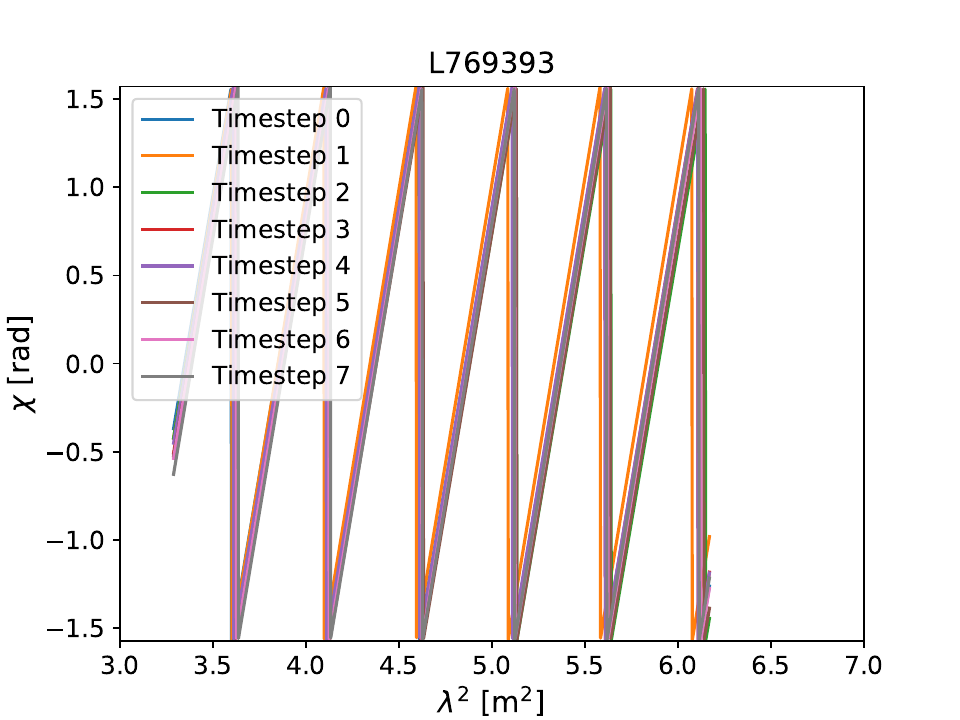}
\includegraphics[width=0.49\textwidth]{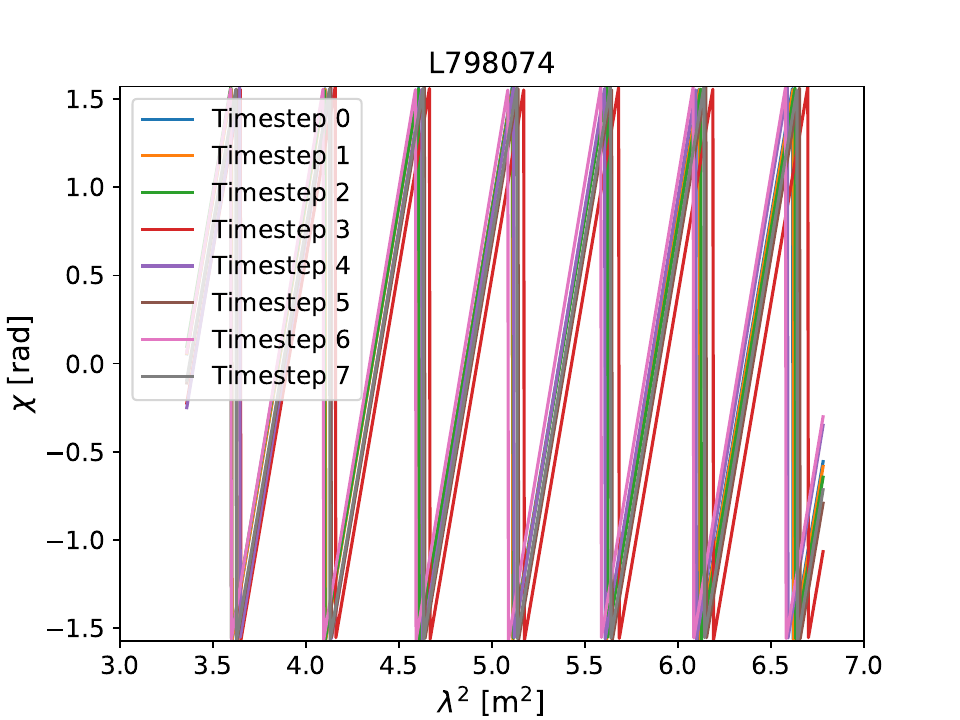}
\includegraphics[width=0.49\textwidth]{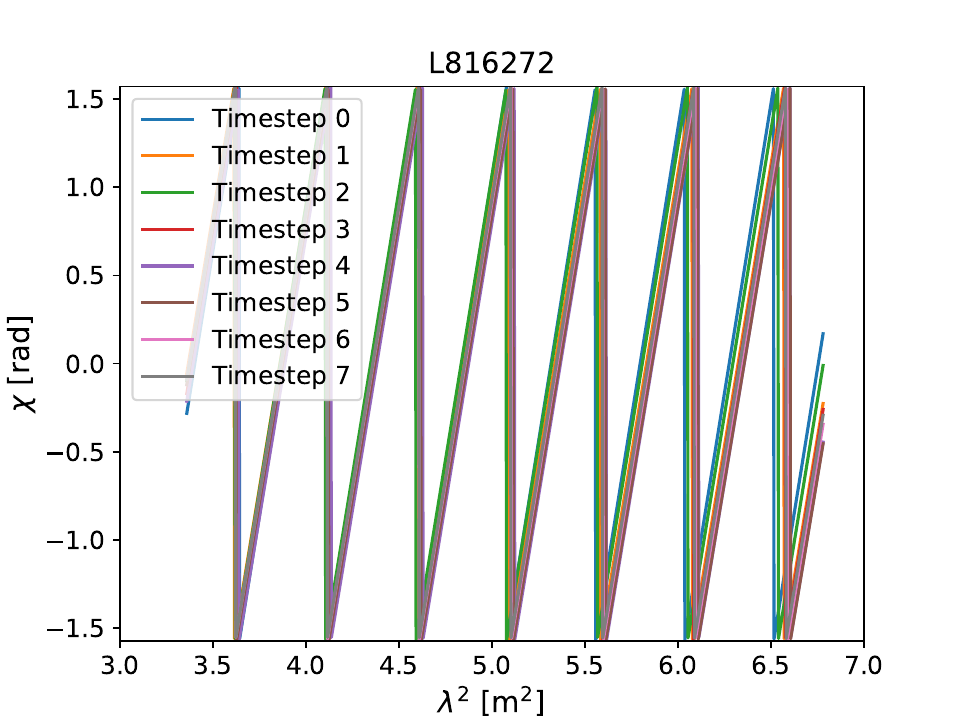}
  \caption{Polarisation angles obtained from fits to the peaks of the RM spectra for eight 1~h subsets of the four observations.}
\label{fig:intrapolangle}    
\end{figure*}

In Sect.~\ref{sec:polanglecalibrator}, we described how the four observing epochs were aligned in Faraday depth and the resultant polarisation angle. Here, we investigate whether an additional correction is required on shorter timescales than the ionospheric corrections applied with \texttt{RMextract} at the start of the data processing. To this end, we constructed polarisation cubes by dividing each observation into 1~h time intervals. For each interval, we fitted the peak of the Faraday spectrum using \texttt{RM-Tools}. The resulting best-fit models, expressed in terms of polarisation angle as a function of $\lambda^{2}$, are shown in Fig.~\ref{fig:intrapolangle}.

The fitted curves are similar across all time intervals, with deviations of less than 1~rad over the full $\lambda^{2}$ range. The largest differences occur at the lowest observing frequencies (highest $\lambda^{2}$), where the deviations reach approximately 0.5~rad for L798074. To assess whether these differences are significant or instead driven by measurement uncertainties, we estimated the uncertainties using a Monte Carlo approach.

\begin{figure}
 \centering
\includegraphics[width=0.49\textwidth]{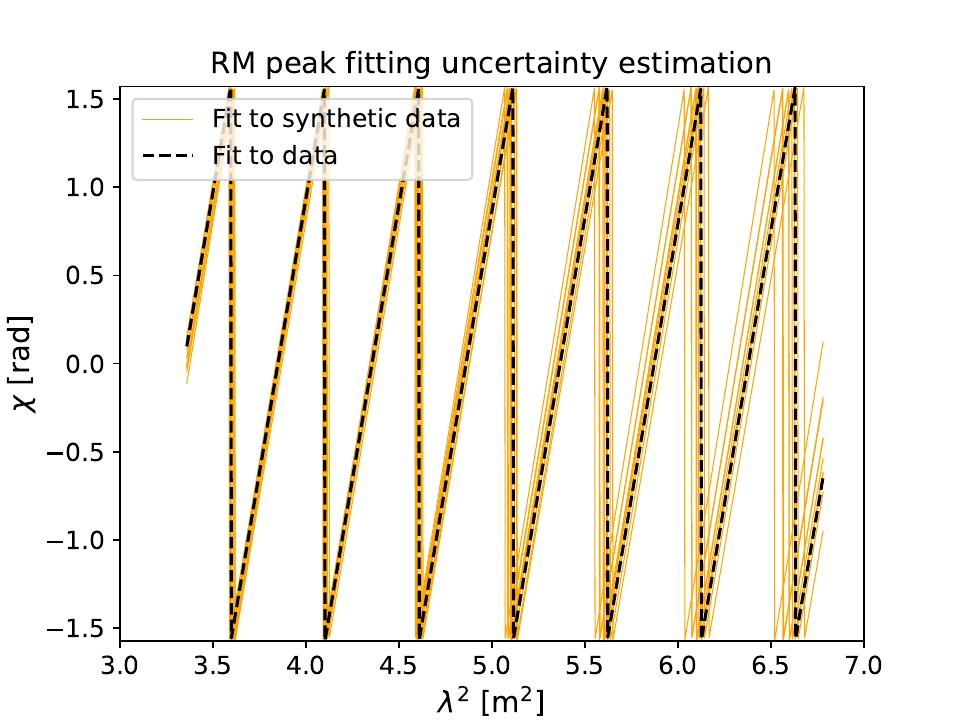}
  \caption{Polarisation angles obtained from a fit to the peak of the RM spectrum for a 1~hr subset of the L798074 data (timestep~2, covering the period 2--3 hr from the start of the observation). The orange lines show fits to models generated from ten synthetic data sets, created by randomly drawing Stokes~$Q$ and $U$ channel flux densities from normal distributions defined by the observed values and their uncertainties. These fits illustrate the uncertainty associated with the RM peak fitting.}
\label{fig:intramcmc}    
\end{figure}

For a representative example (L798074, time interval~2), we generated ten synthetic data sets by randomly drawing Stokes~$Q$ and $U$ flux densities from normal distributions defined by the measured values and their uncertainties. We then refitted the RM peak in Faraday depth space for each realisation. The resulting fits are shown as orange lines in Fig.~\ref{fig:intramcmc}.  Based on this analysis, which was also performed for other time intervals and observing epochs, we find no evidence for significant variations in the polarisation angles within individual observations, as the observed temporal variations are comparable to the uncertainties from the fitting.
 
Further investigations using a larger number of observations, sampling a wider range of ionospheric conditions, and/or employing a polarised calibrator with higher signal-to-noise would be valuable to quantify whether intra-observation polarisation angle corrections are ever required. For our four observations, however, such corrections do not appear to be necessary, suggesting that they are likely not required for the majority of LOFAR observations.

The alignment in Faraday depth and the resulting polarisation angle corrections described in this section and in Sect.~\ref{sec:polanglecalibrator} were performed using Stokes~$Q$ and $U$ polarisation image cubes. An alternative approach would be to obtain these corrections directly in the visibility domain. This could be achieved by predicting a polarised sky model from a reference observation into the \texttt{MODEL\_DATA} column of the measurement set and subsequently solving for the RM and intercept, or by solving for a frequency-dependent rotation angle. In both cases, the derived corrections should be global, that is, identical for all antennas. Future investigations into carrying out such approaches using \texttt{DP3} \citep{dp3,dijkema2023} would be valuable, particularly for data sets with many repeated observations, as this would allow one to bypass the creation of Stokes~$Q$ and $U$ cubes and the associated fitting procedures or RM-synthesis.

\subsection{Outlook}

In this work, we obtained our results using 32~hours of ELAIS-N1 data. Recent work by \citet{shimwell2025} has imaged ELAIS-N1 using 505~hours of LOFAR observations with only the Dutch LOFAR stations. From this dataset, de Jong et al. (in prep.) have selected 200~hours of data that were taken with the international LOFAR stations and processed them with the recently improved calibration strategy \citep{dejong2025b}, combined with the polarisation calibration discussed in this paper. With a central Stokes~$I$ RMS sensitivity of 5.6~\textmu Jy~beam$^{-1}$, these data are 2.5 deeper compared to this work and therefore enable a search for previously undetected polarised sources in ELAIS-N1.

Combining this with the recent work by \cite{gustafsson2025}, which implements a three-dimensional (3D) transformation from spatial and frequency domains into sky–Faraday depth space, including a 3D deconvolution scheme, could further improve deconvolution depth and the reconstruction of the polarised signal. Furthermore, the use of Sidereal Visibility Averaging \citep[SVA;][]{dejong2025a}\footnote{\url{https://github.com/jurjen93/sidereal_visibility_avg}} makes deep imaging with more than 100 frequency channels and hundreds of hours of LOFAR observations computationally feasible, whereas such processing would otherwise be prohibitively expensive.

To further standardise and automate the calibration process described in this paper, we are implementing automated polarisation calibration within Pipeline for the International LOFAR Telescope (PILOT; van der Wild et al., subm.)\footnote{\url{https://github.com/LOFAR-VLBI/pilot}}. This data processing pipeline will be used for future surveys, such as the planned International LoTSS (ILoTSS) survey, which aims to image most of the northern sky at sub-arcsecond resolution. Additional extensions to this pipeline and \texttt{facetselfcal}, as part of the OSCARS project\footnote{\url{https://oscars-project.eu/}} \citep{oscars, dejong_copli}, will further improve the findability, accessibility, interoperability, and reproducibility of the data products and software, bringing them in line with FAIR principles \citep[e.g.,][]{wilkinson2016, otoole2022}.

\section{Conclusions}
\label{sec:conclusions}

In this work, we have presented a strategy to calibrate the polarisation of International LOFAR Telescope (ILT) observations. This strategy exploits an unpolarised in-field calibrator to correct for slowly varying, time- and frequency-dependent instrumental leakage terms. In addition, by using a linearly polarised in-field reference source, multiple observations from different epochs can be aligned in Faraday depth, and the resulting corrections can be applied directly to the visibility data. This enables deeper imaging and more effective deconvolution than previous approaches for LOFAR, which typically relied on stacking images from different epochs.

We applied this calibration strategy to four 8~hr LOFAR observations of the ELAIS-N1 field previously presented in \cite{dejong2024}, including the international stations with baselines up to 1980~km, resulting in a total of 32~hr of data at 115--166~MHz. Using these observations, we successfully imaged and detected three radio sources in linear polarisation at sub-arcsecond resolution. Two of these sources were previously detected with LOFAR in linear polarisation, while one represents a new polarised detection. Our sub-arcsecond-resolution images allow us to precisely localise the polarised emission, which in all cases arises from compact hotspot regions or compact features in the jets of FR\,II radio galaxies. In addition, we detect circularly polarised (Stokes~$V$) emission from the binary M-dwarf system CR~Draconis and, thanks to the high angular resolution, measure its proper motion across the observing epochs.

Finally, we estimate the instrumental leakage from Stokes~$I$ into linear polarisation and Stokes~$V$ to be 0.11\% and 0.021\%, respectively, approximately a factor of two lower than the typical leakage levels reported for LoTSS \citep{shimwell2022}.

Given the availability of suitable in-field calibrators and the prospect of implementing this approach in a pipeline, this work opens up new opportunities to study the metre-wavelength sky in full polarisation at sub-arcsecond resolution over large areas. This can be achieved by exploiting the international baseline data already recorded as part of LoTSS. In addition, this strategy enables ultra-deep polarimetric imaging of well-studied deep fields.

\section*{Acknowledgements}
We thank the anonymous reviewer for helpful comments. 
The authors acknowledge the OSCARS project, which has received funding from the European Commission’s Horizon Europe Research and Innovation programme under grant agreement No.~101129751. 
JMGHJdJ acknowledges support from the project CORTEX (NWA.1160.18.316) of the research programme NWA-ORC which is (partly) financed by the Dutch Research Council (NWO).
EDR acknowledges support by the Deutsche Forschungsgemeinschaft (DFG).
SPO acknowledges support from the Comunidad de Madrid Atracción de Talento program via grant 2022-T1/TIC-23797, and grant PID2023-146372OB-I00 funded by MICIU/AEI/10.13039/ 501100011033 and by ERDF, EU.  FS appreciates the support of STFC [ST/Y004159/1].  LKM is grateful for support from a UKRI FLF [MR/Y020405/1] and LOFAR-UK via STFC [ST/V002406/1]. AB acknowledges support from the ERC CoG $\vec{B}$ELOVED, GA N.101169773. MvdW is grateful for the support of the Science and Technology Facilities Council [ST/V002406/1] and [ST/T000244/1]. DAL also acknowledges support from the Universidad Complutense de Madrid and Banco Santander through the predoctoral grant CT25/24.
LOFAR is the Low Frequency Array designed and constructed by ASTRON \citep{haarlem2013}. It has observing, data processing, and data storage facilities in several countries, which are owned by various parties (each with their own funding sources), and which are collectively operated by the LOFAR ERIC under a joint scientific policy. The LOFAR resources have benefited from the following recent major funding sources: CNRS-INSU, Observatoire de Paris and Université d'Orléans, France; BMBF, MIWF-NRW, MPG, Germany; Science Foundation Ireland (SFI), Department of Business, Enterprise and Innovation (DBEI), Ireland; NWO, The Netherlands; The Science and Technology Facilities Council, UK; Ministry of Science and Higher Education, Poland; The Istituto Nazionale di Astrofisica (INAF), Italy.

This research made use of the Dutch national e-infrastructure with support of the SURF Cooperative (e-infra 180169) and the LOFAR e-infra group. The J\"ulich LOFAR Long Term Archive and the German LOFAR network are both coordinated and operated by the Jülich Supercomputing Centre (JSC), and computing resources on the supercomputer JUWELS at JSC were provided by the Gauss Centre for Supercomputing e.V. (grant CHTB00) through the John von Neumann Institute for Computing (NIC).

This research made use of the University of Hertfordshire high-performance computing facility and the LOFAR-UK computing facility located at the University of Hertfordshire and supported by STFC [ST/P000096/1], and of the Italian LOFAR IT computing infrastructure supported and operated by INAF, and by the Physics Department of Turin University (under an agreement with Consorzio Interuniversitario per la Fisica Spaziale) at the C3S Supercomputing Centre, Italy.

This publication is part of the project Deep high-resolution LOFAR imaging with file number 2023.040 of the research programme Computing Time on National Computing Facilities which is (partly) financed by the Dutch Research Council (NWO).
This research is part of the project LOFAR Data Valorization (LDV) [project numbers 2020.031, 2022.033, and 2024.047] of the research programme Computing Time on National Computer Facilities using SPIDER that is (co-)funded by the Dutch Research Council (NWO), hosted by SURF through the call for proposals of Computing Time on National Computer Facilities. 

The Scientific colour maps \citep{2023zndo...8409685C} are used in
this study to prevent visual distortion of the data and exclusion of
readers with colour-vision deficiencies \citep{2020NatCo..11.5444C}.

Artificial Intelligence tool ChatGPT 5.3 was utilised for copy editing and linguistic refinement to ensure clarity. The final manuscript was reviewed and approved by the human authors.
\section*{Data Availability}
The raw data used in this work are publicly available from the LOFAR Long Term Archive (\url{https://lta.lofar.eu}). The radio images in FITS format are available from the corresponding author upon reasonable request.




\bibliographystyle{mnras}
\bibliography{bib} 

@ARTICLE{sweijen2022,
       author = {{Sweijen}, F. and {van Weeren}, R.~J. and {R{\"o}ttgering}, H.~J.~A. and {Morabito}, L.~K. and {Jackson}, N. and {Offringa}, A.~R. and {van der Tol}, S. and {Veenboer}, B. and {Oonk}, J.~B.~R. and {Best}, P.~N. and {Bondi}, M. and {Shimwell}, T.~W. and {Tasse}, C. and {Thomson}, A.~P.},
        title = "{Deep sub-arcsecond wide-field imaging of the Lockman Hole field at 144 MHz}",
      journal = {Nature Astronomy},
         year = 2022,
        month = jan,
       volume = {6},
        pages = {350-356},
          doi = {10.1038/s41550-021-01573-z},
archivePrefix = {arXiv},
       eprint = {2202.01608},
 primaryClass = {astro-ph.IM},
       adsurl = {https://ui.adsabs.harvard.edu/abs/2022NatAs...6..350S}
}

@ARTICLE{2020NatCo..11.5444C,
       author = {{Crameri}, Fabio and {Shephard}, Grace E. and {Heron}, Philip J.},
        title = "{The misuse of colour in science communication}",
      journal = {Nature Communications},
         year = 2020,
        month = oct,
       volume = {11},
          eid = {5444},
        pages = {5444},
          doi = {10.1038/s41467-020-19160-7},
       adsurl = {https://ui.adsabs.harvard.edu/abs/2020NatCo..11.5444C}
}

@software{2023zndo...8409685C,
       author = {{Crameri}, Fabio},
        title = "{Scientific colour maps}",
         year = 2023,
        month = oct,
          eid = {10.5281/zenodo.8409685},
          doi = {10.5281/zenodo.8409685},
      version = {8.0.1},
    publisher = {Zenodo},
       adsurl = {https://ui.adsabs.harvard.edu/abs/2023zndo...8409685C}
}

@ARTICLE{2026arXiv260120092V,
       author = {{Van Eck}, Cameron L. and {Purcell}, Cormac R. and {Baidoo}, Lerato and {Thomson}, Alec J.~M. and {Ma}, Yik Ki and {Oberhelman}, Lindsey and {Osinga}, Erik and {Vanderwoude}, Shannon and {West}, Jennifer L. and {Ideguchi}, Shinsuke and {Par{\'e}}, Dylan M. and {Kaczmarek}, Jane F. and {Willis}, Tony and {Akahori}, Takuya and {Anderson}, Craig S. and {Gaensler}, B.~M. and {O'Sullivan}, Shane and {Sun}, Xiaohui and {Amaral}, Ariel D. and {Riseley}, C.~J. and {Stil}, Jeroen and {Zhang}, Xiang},
        title = "{RM-Tools: Software for Analyzing Polarized Radio Spectra}",
      journal = {arXiv e-prints},
         year = 2026,
        month = jan,
          eid = {arXiv:2601.20092},
        pages = {arXiv:2601.20092},
          doi = {10.48550/arXiv.2601.20092},
archivePrefix = {arXiv},
       eprint = {2601.20092},
 primaryClass = {astro-ph.IM},
       adsurl = {https://ui.adsabs.harvard.edu/abs/2026arXiv260120092V}
}

@ARTICLE{dejong2025b,
       author = {{de Jong}, J.~M.~G.~H.~J. and {Veefkind}, L. and {van Weeren}, R.~J. and {Oonk}, J.~B.~R. and {Schlimbach}, R.~J. and {Kampert}, D.~N.~G. and {van der Wild}, M. and {Morabito}, L.~K. and {Sweijen}, F. and {Offringa}, A.~R. and {R{\"o}ttgering}, H.~J.~A.},
        title = "{Scalable and robust wide-field facet calibration with LOFAR's longest baselines}",
      journal = {\mnras},
         year = 2025,
        month = oct,
       volume = {542},
       number = {4},
        pages = {3253-3276},
          doi = {10.1093/mnras/staf1373},
archivePrefix = {arXiv},
       eprint = {2508.12115},
 primaryClass = {astro-ph.IM},
       adsurl = {https://ui.adsabs.harvard.edu/abs/2025MNRAS.542.3253D}
}

@misc{dejong_copli,
  author       = {de Jong, Jurjen},
  title        = {COPLI poster - OSCARS 2nd AGM 2026},
  month        = feb,
  year         = 2026,
  publisher    = {Zenodo},
  doi          = {10.5281/zenodo.18761240},
  url          = {https://doi.org/10.5281/zenodo.18761240},
}

@ARTICLE{dejong2025a,
       author = {{de Jong}, J.~M.~G.~H.~J. and {van Weeren}, R.~J. and {Dijkema}, T.~J. and {Oonk}, J.~B.~R. and {R{\"o}ttgering}, H.~J.~A. and {Sweijen}, F.},
        title = "{Unlocking ultra-deep wide-field imaging with sidereal visibility averaging}",
      journal = {\aap},
         year = 2025,
        month = feb,
       volume = {694},
          eid = {A98},
        pages = {A98},
          doi = {10.1051/0004-6361/202452492},
archivePrefix = {arXiv},
       eprint = {2501.07374},
 primaryClass = {astro-ph.IM},
       adsurl = {https://ui.adsabs.harvard.edu/abs/2025A&A...694A..98D}
}

@misc{otoole2022,
      title={FAIR standards for astronomical data}, 
      author={Simon O'Toole and James Tocknell},
      year={2022},
      eprint={2203.10710},
      archivePrefix={arXiv},
      primaryClass={astro-ph.IM},
      url={https://arxiv.org/abs/2203.10710}, 
}

@article{wilkinson2016,
  title={The FAIR Guiding Principles for scientific data management and stewardship},
  author={Wilkinson, Mark D and Dumontier, Michel and Aalbersberg, IJsbrand Jan and others},
  journal={Scientific data},
  volume={3},
  pages={160018},
  year={2016},
  publisher={Nature Publishing Group},
  doi={10.1038/sdata.2016.18},
  url={https://www.nature.com/articles/sdata201618}
}

@ARTICLE{gustafsson2025,
       author = {{Gustafsson}, V. and {Br{\"u}ggen}, M. and {Tasse}, C. and {En{\ss}lin}, T. and {O'Sullivan}, S.~P. and {de Gasperin}, F.},
        title = "{Faraday synthesis in direction-dependent imaging}",
      journal = {\aap},
         year = 2025,
        month = aug,
       volume = {700},
          eid = {A221},
        pages = {A221},
          doi = {10.1051/0004-6361/202554903},
archivePrefix = {arXiv},
       eprint = {2504.00141},
 primaryClass = {astro-ph.IM},
       adsurl = {https://ui.adsabs.harvard.edu/abs/2025A&A...700A.221G}
}

@misc{oscars,
  title        = {Building Capacity for FAIR and Open Science: Insights from the World Café sessions at the OSCARS 1st AGM},
  author       = {David, Romain and Schmidt-Tremmel, Friederike and Carboni, Nicoletta and Aguilar Gómez, Fernando and Atienza Lopez, Pablo and Arvanitidis, Christos and Bederke, Paul and Bo, Carles and Bolmsten, Fredrik and Bodera Sempere, Jordi and Boichuk, Volodymyra and Chambers, Sally and Conesa, Ana and Ewbank, Jonathan and Feudo, Fabio and Fišer, Darja and Fuhrmann, Patrick and García-Díez, Markel and Goble, Carole and Grilo, Clara and Guerrieri, Giovanni and Harb, Robert and Heger, Tina and Hienola, Anca and Hinrichsen, Bernd and Idmhand, Fatiha and Iess, Alberto and de Jong, Jurjen and Kaftan, Lea and Knodel, Oliver and Kotsiubynskyi, Volodymyr and Lamanna, Giovanni and Le Dévédec, Sylvia E. and Legger, Federica and van der Lek, Iulianna and Liaskos, Nektarios and Liborio, Leandro and Makula, Lorenz and Millar, Paul and Monteiro, Pedro T. and Morgan, Sarah and Müller, Roman-Ulrich and Nentwich, Melanie and Novelli, Massimiliano and Panov, Panče and Papadimitriou, Eleftheria and Revilla, Reynier I. and dos Santos Oliveira, Jorge and Saunders, Gary and Serjeant, Stephen and Stansby, David and Tedds, Jonathan and Tessier, Laurent and Thompson, Warren and Wiśniewska, Aneta and Zhao, Zhiming},
  year         = {2025},
  month        = {dec},
  doi          = {10.5281/zenodo.17854379},
  url          = {https://doi.org/10.5281/zenodo.17854379},
  publisher    = {Zenodo},
  note         = {Version v1, published Dec 8, 2025}
}

@ARTICLE{jackson2022,
       author = {{Jackson}, N. and {Badole}, S. and {Morgan}, J. and {Chhetri}, R. and {Pr{\={u}}sis}, K. and {Nikolajevs}, A. and {Morabito}, L. and {Brentjens}, M. and {Sweijen}, F. and {Iacobelli}, M. and {Orr{\`u}}, E. and {Sluman}, J. and {Blaauw}, R. and {Mulder}, H. and {van Dijk}, P. and {Mooney}, S. and {Deller}, A. and {Moldon}, J. and {Callingham}, J.~R. and {Harwood}, J. and {Hardcastle}, M. and {Heald}, G. and {Drabent}, A. and {McKean}, J.~P. and {Asgekar}, A. and {Avruch}, I.~M. and {Bentum}, M.~J. and {Bonafede}, A. and {Brouw}, W.~N. and {Br{\"u}ggen}, M. and {Butcher}, H.~R. and {Ciardi}, B. and {Coolen}, A. and {Corstanje}, A. and {Damstra}, S. and {Duscha}, S. and {Eisl{\"o}ffel}, J. and {Falcke}, H. and {Garrett}, M. and {de Gasperin}, F. and {Griessmeier}, J. -M. and {Gunst}, A.~W. and {van Haarlem}, M.~P. and {Hoeft}, M. and {van der Horst}, A.~J. and {J{\"u}tte}, E. and {Koopmans}, L.~V.~E. and {Krankowski}, A. and {Maat}, P. and {Mann}, G. and {Miley}, G.~K. and {Nelles}, A. and {Norden}, M. and {Paas}, M. and {Pandey}, V.~N. and {Pandey-Pommier}, M. and {Pizzo}, R.~F. and {Reich}, W. and {Rothkaehl}, H. and {Rowlinson}, A. and {Ruiter}, M. and {Shulevski}, A. and {Schwarz}, D.~J. and {Smirnov}, O. and {Tagger}, M. and {Vocks}, C. and {van Weeren}, R.~J. and {Wijers}, R. and {Wucknitz}, O. and {Zarka}, P. and {Zensus}, J.~A. and {Zucca}, P.},
        title = "{Sub-arcsecond imaging with the International LOFAR Telescope. II. Completion of the LOFAR Long-Baseline Calibrator Survey}",
      journal = {\aap},
         year = 2022,
        month = feb,
       volume = {658},
          eid = {A2},
        pages = {A2},
          doi = {10.1051/0004-6361/202140756},
archivePrefix = {arXiv},
       eprint = {2108.07284},
 primaryClass = {astro-ph.GA},
       adsurl = {https://ui.adsabs.harvard.edu/abs/2022A&A...658A...2J}
}

@ARTICLE{shimwell2019,
       author = {{Shimwell}, T.~W. and {Tasse}, C. and {Hardcastle}, M.~J. and {Mechev}, A.~P. and {Williams}, W.~L. and {Best}, P.~N. and {R{\"o}ttgering}, H.~J.~A. and {Callingham}, J.~R. and {Dijkema}, T.~J. and {de Gasperin}, F. and {Hoang}, D.~N. and {Hugo}, B. and {Mirmont}, M. and {Oonk}, J.~B.~R. and {Prandoni}, I. and {Rafferty}, D. and {Sabater}, J. and {Smirnov}, O. and {van Weeren}, R.~J. and {White}, G.~J. and {Atemkeng}, M. and {Bester}, L. and {Bonnassieux}, E. and {Br{\"u}ggen}, M. and {Brunetti}, G. and {Chy{\.z}y}, K.~T. and {Cochrane}, R. and {Conway}, J.~E. and {Croston}, J.~H. and {Danezi}, A. and {Duncan}, K. and {Haverkorn}, M. and {Heald}, G.~H. and {Iacobelli}, M. and {Intema}, H.~T. and {Jackson}, N. and {Jamrozy}, M. and {Jarvis}, M.~J. and {Lakhoo}, R. and {Mevius}, M. and {Miley}, G.~K. and {Morabito}, L. and {Morganti}, R. and {Nisbet}, D. and {Orr{\'u}}, E. and {Perkins}, S. and {Pizzo}, R.~F. and {Schrijvers}, C. and {Smith}, D.~J.~B. and {Vermeulen}, R. and {Wise}, M.~W. and {Alegre}, L. and {Bacon}, D.~J. and {van Bemmel}, I.~M. and {Beswick}, R.~J. and {Bonafede}, A. and {Botteon}, A. and {Bourke}, S. and {Brienza}, M. and {Calistro Rivera}, G. and {Cassano}, R. and {Clarke}, A.~O. and {Conselice}, C.~J. and {Dettmar}, R.~J. and {Drabent}, A. and {Dumba}, C. and {Emig}, K.~L. and {En{\ss}lin}, T.~A. and {Ferrari}, C. and {Garrett}, M.~A. and {G{\'e}nova-Santos}, R.~T. and {Goyal}, A. and {G{\"u}rkan}, G. and {Hale}, C. and {Harwood}, J.~J. and {Heesen}, V. and {Hoeft}, M. and {Horellou}, C. and {Jackson}, C. and {Kokotanekov}, G. and {Kondapally}, R. and {Kunert-Bajraszewska}, M. and {Mahatma}, V. and {Mahony}, E.~K. and {Mandal}, S. and {McKean}, J.~P. and {Merloni}, A. and {Mingo}, B. and {Miskolczi}, A. and {Mooney}, S. and {Nikiel-Wroczy{\'n}ski}, B. and {O'Sullivan}, S.~P. and {Quinn}, J. and {Reich}, W. and {Roskowi{\'n}ski}, C. and {Rowlinson}, A. and {Savini}, F. and {Saxena}, A. and {Schwarz}, D.~J. and {Shulevski}, A. and {Sridhar}, S.~S. and {Stacey}, H.~R. and {Urquhart}, S. and {van der Wiel}, M.~H.~D. and {Varenius}, E. and {Webster}, B. and {Wilber}, A.},
        title = "{The LOFAR Two-metre Sky Survey. II. First data release}",
      journal = {\aap},
         year = 2019,
        month = feb,
       volume = {622},
          eid = {A1},
        pages = {A1},
          doi = {10.1051/0004-6361/201833559},
archivePrefix = {arXiv},
       eprint = {1811.07926},
 primaryClass = {astro-ph.GA},
       adsurl = {https://ui.adsabs.harvard.edu/abs/2019A&A...622A...1S}
}

@ARTICLE{2007P&SS...55..598Z,
       author = {{Zarka}, Philippe},
        title = "{Plasma interactions of exoplanets with their parent star and associated radio emissions}",
      journal = {\planss},
         year = 2007,
        month = apr,
       volume = {55},
       number = {5},
        pages = {598-617},
          doi = {10.1016/j.pss.2006.05.045},
       adsurl = {https://ui.adsabs.harvard.edu/abs/2007P&SS...55..598Z}
}

@ARTICLE{2011A&A...531A..29H,
       author = {{Hess}, S.~L.~G. and {Zarka}, P.},
        title = "{Modeling the radio signature of the orbital parameters, rotation, and magnetic field of exoplanets}",
      journal = {\aap},
         year = 2011,
        month = jul,
       volume = {531},
          eid = {A29},
        pages = {A29},
          doi = {10.1051/0004-6361/201116510},
       adsurl = {https://ui.adsabs.harvard.edu/abs/2011A&A...531A..29H}
}

@ARTICLE{morabito2022,
       author = {{Morabito}, L.~K. and {Jackson}, N.~J. and {Mooney}, S. and {Sweijen}, F. and {Badole}, S. and {Kukreti}, P. and {Venkattu}, D. and {Groeneveld}, C. and {Kappes}, A. and {Bonnassieux}, E. and {Drabent}, A. and {Iacobelli}, M. and {Croston}, J.~H. and {Best}, P.~N. and {Bondi}, M. and {Callingham}, J.~R. and {Conway}, J.~E. and {Deller}, A.~T. and {Hardcastle}, M.~J. and {McKean}, J.~P. and {Miley}, G.~K. and {Moldon}, J. and {R{\"o}ttgering}, H.~J.~A. and {Tasse}, C. and {Shimwell}, T.~W. and {van Weeren}, R.~J. and {Anderson}, J.~M. and {Asgekar}, A. and {Avruch}, I.~M. and {van Bemmel}, I.~M. and {Bentum}, M.~J. and {Bonafede}, A. and {Brouw}, W.~N. and {Butcher}, H.~R. and {Ciardi}, B. and {Corstanje}, A. and {Coolen}, A. and {Damstra}, S. and {de Gasperin}, F. and {Duscha}, S. and {Eisl{\"o}ffel}, J. and {Engels}, D. and {Falcke}, H. and {Garrett}, M.~A. and {Griessmeier}, J. and {Gunst}, A.~W. and {van Haarlem}, M.~P. and {Hoeft}, M. and {van der Horst}, A.~J. and {J{\"u}tte}, E. and {Kadler}, M. and {Koopmans}, L.~V.~E. and {Krankowski}, A. and {Mann}, G. and {Nelles}, A. and {Oonk}, J.~B.~R. and {Orru}, E. and {Paas}, H. and {Pandey}, V.~N. and {Pizzo}, R.~F. and {Pandey-Pommier}, M. and {Reich}, W. and {Rothkaehl}, H. and {Ruiter}, M. and {Schwarz}, D.~J. and {Shulevski}, A. and {Soida}, M. and {Tagger}, M. and {Vocks}, C. and {Wijers}, R.~A.~M.~J. and {Wijnholds}, S.~J. and {Wucknitz}, O. and {Zarka}, P. and {Zucca}, P.},
        title = "{Sub-arcsecond imaging with the International LOFAR Telescope. I. Foundational calibration strategy and pipeline}",
      journal = {\aap},
         year = 2022,
        month = feb,
       volume = {658},
          eid = {A1},
        pages = {A1},
          doi = {10.1051/0004-6361/202140649},
archivePrefix = {arXiv},
       eprint = {2108.07283},
 primaryClass = {astro-ph.IM},
       adsurl = {https://ui.adsabs.harvard.edu/abs/2022A&A...658A...1M}
}

@ARTICLE{sabater2021,
       author = {{Sabater}, J. and {Best}, P.~N. and {Tasse}, C. and {Hardcastle}, M.~J. and {Shimwell}, T.~W. and {Nisbet}, D. and {Jelic}, V. and {Callingham}, J.~R. and {R{\"o}ttgering}, H.~J.~A. and {Bonato}, M. and {Bondi}, M. and {Ciardi}, B. and {Cochrane}, R.~K. and {Jarvis}, M.~J. and {Kondapally}, R. and {Koopmans}, L.~V.~E. and {O'Sullivan}, S.~P. and {Prandoni}, I. and {Schwarz}, D.~J. and {Smith}, D.~J.~B. and {Wang}, L. and {Williams}, W.~L. and {Zaroubi}, S.},
        title = "{The LOFAR Two-meter Sky Survey: Deep Fields Data Release 1. II. The ELAIS-N1 LOFAR deep field}",
      journal = {\aap},
         year = 2021,
        month = apr,
       volume = {648},
          eid = {A2},
        pages = {A2},
          doi = {10.1051/0004-6361/202038828},
archivePrefix = {arXiv},
       eprint = {2011.08211},
 primaryClass = {astro-ph.GA},
       adsurl = {https://ui.adsabs.harvard.edu/abs/2021A&A...648A...2S}
}

@ARTICLE{tasse2021,
       author = {{Tasse}, C. and {Shimwell}, T. and {Hardcastle}, M.~J. and {O'Sullivan}, S.~P. and {van Weeren}, R. and {Best}, P.~N. and {Bester}, L. and {Hugo}, B. and {Smirnov}, O. and {Sabater}, J. and {Calistro-Rivera}, G. and {de Gasperin}, F. and {Morabito}, L.~K. and {R{\"o}ttgering}, H. and {Williams}, W.~L. and {Bonato}, M. and {Bondi}, M. and {Botteon}, A. and {Br{\"u}ggen}, M. and {Brunetti}, G. and {Chy{\.z}y}, K.~T. and {Garrett}, M.~A. and {G{\"u}rkan}, G. and {Jarvis}, M.~J. and {Kondapally}, R. and {Mandal}, S. and {Prandoni}, I. and {Repetti}, A. and {Retana-Montenegro}, E. and {Schwarz}, D.~J. and {Shulevski}, A. and {Wiaux}, Y.},
        title = "{The LOFAR Two-meter Sky Survey: Deep Fields Data Release 1. I. Direction-dependent calibration and imaging}",
      journal = {\aap},
         year = 2021,
        month = apr,
       volume = {648},
          eid = {A1},
        pages = {A1},
          doi = {10.1051/0004-6361/202038804},
archivePrefix = {arXiv},
       eprint = {2011.08328},
 primaryClass = {astro-ph.IM},
       adsurl = {https://ui.adsabs.harvard.edu/abs/2021A&A...648A...1T}
}

@ARTICLE{shimwell2017,
       author = {{Shimwell}, T.~W. and {R{\"o}ttgering}, H.~J.~A. and {Best}, P.~N. and {Williams}, W.~L. and {Dijkema}, T.~J. and {de Gasperin}, F. and {Hardcastle}, M.~J. and {Heald}, G.~H. and {Hoang}, D.~N. and {Horneffer}, A. and {Intema}, H. and {Mahony}, E.~K. and {Mandal}, S. and {Mechev}, A.~P. and {Morabito}, L. and {Oonk}, J.~B.~R. and {Rafferty}, D. and {Retana-Montenegro}, E. and {Sabater}, J. and {Tasse}, C. and {van Weeren}, R.~J. and {Br{\"u}ggen}, M. and {Brunetti}, G. and {Chy{\.z}y}, K.~T. and {Conway}, J.~E. and {Haverkorn}, M. and {Jackson}, N. and {Jarvis}, M.~J. and {McKean}, J.~P. and {Miley}, G.~K. and {Morganti}, R. and {White}, G.~J. and {Wise}, M.~W. and {van Bemmel}, I.~M. and {Beck}, R. and {Brienza}, M. and {Bonafede}, A. and {Calistro Rivera}, G. and {Cassano}, R. and {Clarke}, A.~O. and {Cseh}, D. and {Deller}, A. and {Drabent}, A. and {van Driel}, W. and {Engels}, D. and {Falcke}, H. and {Ferrari}, C. and {Fr{\"o}hlich}, S. and {Garrett}, M.~A. and {Harwood}, J.~J. and {Heesen}, V. and {Hoeft}, M. and {Horellou}, C. and {Israel}, F.~P. and {Kapi{\'n}ska}, A.~D. and {Kunert-Bajraszewska}, M. and {McKay}, D.~J. and {Mohan}, N.~R. and {Orr{\'u}}, E. and {Pizzo}, R.~F. and {Prandoni}, I. and {Schwarz}, D.~J. and {Shulevski}, A. and {Sipior}, M. and {Smith}, D.~J.~B. and {Sridhar}, S.~S. and {Steinmetz}, M. and {Stroe}, A. and {Varenius}, E. and {van der Werf}, P.~P. and {Zensus}, J.~A. and {Zwart}, J.~T.~L.},
        title = "{The LOFAR Two-metre Sky Survey. I. Survey description and preliminary data release}",
      journal = {\aap},
         year = 2017,
        month = feb,
       volume = {598},
          eid = {A104},
        pages = {A104},
          doi = {10.1051/0004-6361/201629313},
archivePrefix = {arXiv},
       eprint = {1611.02700},
 primaryClass = {astro-ph.IM},
       adsurl = {https://ui.adsabs.harvard.edu/abs/2017A&A...598A.104S}
}

@ARTICLE{2015ApJS..219...12A,
       author = {{Alam}, Shadab and {Albareti}, Franco D. and {Allende Prieto}, Carlos and {Anders}, F. and {Anderson}, Scott F. and {Anderton}, Timothy and {Andrews}, Brett H. and {Armengaud}, Eric and {Aubourg}, {\'E}ric and {Bailey}, Stephen and {Basu}, Sarbani and {Bautista}, Julian E. and {Beaton}, Rachael L. and {Beers}, Timothy C. and {Bender}, Chad F. and {Berlind}, Andreas A. and {Beutler}, Florian and {Bhardwaj}, Vaishali and {Bird}, Jonathan C. and {Bizyaev}, Dmitry and {Blake}, Cullen H. and {Blanton}, Michael R. and {Blomqvist}, Michael and {Bochanski}, John J. and {Bolton}, Adam S. and {Bovy}, Jo and {Shelden Bradley}, A. and {Brandt}, W.~N. and {Brauer}, D.~E. and {Brinkmann}, J. and {Brown}, Peter J. and {Brownstein}, Joel R. and {Burden}, Angela and {Burtin}, Etienne and {Busca}, Nicol{\'a}s G. and {Cai}, Zheng and {Capozzi}, Diego and {Carnero Rosell}, Aurelio and {Carr}, Michael A. and {Carrera}, Ricardo and {Chambers}, K.~C. and {Chaplin}, William James and {Chen}, Yen-Chi and {Chiappini}, Cristina and {Chojnowski}, S. Drew and {Chuang}, Chia-Hsun and {Clerc}, Nicolas and {Comparat}, Johan and {Covey}, Kevin and {Croft}, Rupert A.~C. and {Cuesta}, Antonio J. and {Cunha}, Katia and {da Costa}, Luiz N. and {Da Rio}, Nicola and {Davenport}, James R.~A. and {Dawson}, Kyle S. and {De Lee}, Nathan and {Delubac}, Timoth{\'e}e and {Deshpande}, Rohit and {Dhital}, Saurav and {Dutra-Ferreira}, Let{\'\i}cia and {Dwelly}, Tom and {Ealet}, Anne and {Ebelke}, Garrett L. and {Edmondson}, Edward M. and {Eisenstein}, Daniel J. and {Ellsworth}, Tristan and {Elsworth}, Yvonne and {Epstein}, Courtney R. and {Eracleous}, Michael and {Escoffier}, Stephanie and {Esposito}, Massimiliano and {Evans}, Michael L. and {Fan}, Xiaohui and {Fern{\'a}ndez-Alvar}, Emma and {Feuillet}, Diane and {Filiz Ak}, Nurten and {Finley}, Hayley and {Finoguenov}, Alexis and {Flaherty}, Kevin and {Fleming}, Scott W. and {Font-Ribera}, Andreu and {Foster}, Jonathan and {Frinchaboy}, Peter M. and {Galbraith-Frew}, J.~G. and {Garc{\'\i}a}, Rafael A. and {Garc{\'\i}a-Hern{\'a}ndez}, D.~A. and {Garc{\'\i}a P{\'e}rez}, Ana E. and {Gaulme}, Patrick and {Ge}, Jian and {G{\'e}nova-Santos}, R. and {Georgakakis}, A. and {Ghezzi}, Luan and {Gillespie}, Bruce A. and {Girardi}, L{\'e}o and {Goddard}, Daniel and {Gontcho}, Satya Gontcho A. and {Gonz{\'a}lez Hern{\'a}ndez}, Jonay I. and {Grebel}, Eva K. and {Green}, Paul J. and {Grieb}, Jan Niklas and {Grieves}, Nolan and {Gunn}, James E. and {Guo}, Hong and {Harding}, Paul and {Hasselquist}, Sten and {Hawley}, Suzanne L. and {Hayden}, Michael and {Hearty}, Fred R. and {Hekker}, Saskia and {Ho}, Shirley and {Hogg}, David W. and {Holley-Bockelmann}, Kelly and {Holtzman}, Jon A. and {Honscheid}, Klaus and {Huber}, Daniel and {Huehnerhoff}, Joseph and {Ivans}, Inese I. and {Jiang}, Linhua and {Johnson}, Jennifer A. and {Kinemuchi}, Karen and {Kirkby}, David and {Kitaura}, Francisco and {Klaene}, Mark A. and {Knapp}, Gillian R. and {Kneib}, Jean-Paul and {Koenig}, Xavier P. and {Lam}, Charles R. and {Lan}, Ting-Wen and {Lang}, Dustin and {Laurent}, Pierre and {Le Goff}, Jean-Marc and {Leauthaud}, Alexie and {Lee}, Khee-Gan and {Lee}, Young Sun and {Licquia}, Timothy C. and {Liu}, Jian and {Long}, Daniel C. and {L{\'o}pez-Corredoira}, Mart{\'\i}n and {Lorenzo-Oliveira}, Diego and {Lucatello}, Sara and {Lundgren}, Britt and {Lupton}, Robert H. and {Mack}, III, Claude E. and {Mahadevan}, Suvrath and {Maia}, Marcio A.~G. and {Majewski}, Steven R. and {Malanushenko}, Elena and {Malanushenko}, Viktor and {Manchado}, A. and {Manera}, Marc and {Mao}, Qingqing and {Maraston}, Claudia and {Marchwinski}, Robert C. and {Margala}, Daniel and {Martell}, Sarah L. and {Martig}, Marie and {Masters}, Karen L. and {Mathur}, Savita and {McBride}, Cameron K. and {McGehee}, Peregrine M. and {McGreer}, Ian D. and {McMahon}, Richard G. and {M{\'e}nard}, Brice and {Menzel}, Marie-Luise and {Merloni}, Andrea and {M{\'e}sz{\'a}ros}, Szabolcs and {Miller}, Adam A. and {Miralda-Escud{\'e}}, Jordi and {Miyatake}, Hironao and {Montero-Dorta}, Antonio D. and {More}, Surhud and {Morganson}, Eric and {Morice-Atkinson}, Xan and {Morrison}, Heather L. and {Mosser}, Ben{\^o}it and {Muna}, Demitri and {Myers}, Adam D. and {Nandra}, Kirpal and {Newman}, Jeffrey A. and {Neyrinck}, Mark and {Nguyen}, Duy Cuong and {Nichol}, Robert C. and {Nidever}, David L. and {Noterdaeme}, Pasquier and {Nuza}, Sebasti{\'a}n E. and {O'Connell}, Julia E. and {O'Connell}, Robert W. and {O'Connell}, Ross and {Ogando}, Ricardo L.~C. and {Olmstead}, Matthew D. and {Oravetz}, Audrey E. and {Oravetz}, Daniel J. and {Osumi}, Keisuke and {Owen}, Russell and {Padgett}, Deborah L. and {Padmanabhan}, Nikhil and {Paegert}, Martin and {Palanque-Delabrouille}, Nathalie and {Pan}, Kaike},
        title = "{The Eleventh and Twelfth Data Releases of the Sloan Digital Sky Survey: Final Data from SDSS-III}",
      journal = {\apjs},
         year = 2015,
        month = jul,
       volume = {219},
       number = {1},
          eid = {12},
        pages = {12},
          doi = {10.1088/0067-0049/219/1/12},
archivePrefix = {arXiv},
       eprint = {1501.00963},
 primaryClass = {astro-ph.IM},
       adsurl = {https://ui.adsabs.harvard.edu/abs/2015ApJS..219...12A}
}

@ARTICLE{2009A&A...503..409H,
       author = {{Heald}, G. and {Braun}, R. and {Edmonds}, R.},
        title = "{The Westerbork SINGS survey. II Polarization, Faraday rotation, and magnetic fields}",
      journal = {\aap},
         year = 2009,
        month = aug,
       volume = {503},
       number = {2},
        pages = {409-435},
          doi = {10.1051/0004-6361/200912240},
archivePrefix = {arXiv},
       eprint = {0905.3995},
 primaryClass = {astro-ph.GA},
       adsurl = {https://ui.adsabs.harvard.edu/abs/2009A&A...503..409H}
}

@ARTICLE{shimwell2025,
       author = {{Shimwell}, T.~W. and {Hale}, C.~L. and {Best}, P.~N. and {Botteon}, A. and {Drabent}, A. and {Hardcastle}, M.~J. and {Jeli{\'c}}, V. and {de Jong}, J.~M.~G.~H.~J. and {Kondapally}, R. and {R{\"o}ttgering}, H.~J.~A. and {Tasse}, C. and {van Weeren}, R.~J. and {Williams}, W.~L. and {Bonafede}, A. and {Bondi}, M. and {Br{\"u}ggen}, M. and {Brunetti}, G. and {Callingham}, J.~R. and {De Gasperin}, F. and {Duncan}, K.~J. and {Horellou}, C. and {Iyer}, S. and {de Ruiter}, I. and {Ma{\l}ek}, K. and {Nair}, D.~G. and {Morabito}, L.~K. and {Prandoni}, I. and {Rowlinson}, A. and {Sabater}, J. and {Shulevski}, A. and {Smith}, D.~J.~B. and {Sweijen}, F.},
        title = "{The LOFAR Two-metre Sky Survey: Deep Fields Data Release 2: I. The ELAIS-N1 field}",
      journal = {\aap},
         year = 2025,
        month = mar,
       volume = {695},
          eid = {A80},
        pages = {A80},
          doi = {10.1051/0004-6361/202452930},
archivePrefix = {arXiv},
       eprint = {2501.04093},
 primaryClass = {astro-ph.CO},
       adsurl = {https://ui.adsabs.harvard.edu/abs/2025A&A...695A..80S}
}

@ARTICLE{shimwell2022,
       author = {{Shimwell}, T.~W. and {Hardcastle}, M.~J. and {Tasse}, C. and {Best}, P.~N. and {R{\"o}ttgering}, H.~J.~A. and {Williams}, W.~L. and {Botteon}, A. and {Drabent}, A. and {Mechev}, A. and {Shulevski}, A. and {van Weeren}, R.~J. and {Bester}, L. and {Br{\"u}ggen}, M. and {Brunetti}, G. and {Callingham}, J.~R. and {Chy{\.z}y}, K.~T. and {Conway}, J.~E. and {Dijkema}, T.~J. and {Duncan}, K. and {de Gasperin}, F. and {Hale}, C.~L. and {Haverkorn}, M. and {Hugo}, B. and {Jackson}, N. and {Mevius}, M. and {Miley}, G.~K. and {Morabito}, L.~K. and {Morganti}, R. and {Offringa}, A. and {Oonk}, J.~B.~R. and {Rafferty}, D. and {Sabater}, J. and {Smith}, D.~J.~B. and {Schwarz}, D.~J. and {Smirnov}, O. and {O'Sullivan}, S.~P. and {Vedantham}, H. and {White}, G.~J. and {Albert}, J.~G. and {Alegre}, L. and {Asabere}, B. and {Bacon}, D.~J. and {Bonafede}, A. and {Bonnassieux}, E. and {Brienza}, M. and {Bilicki}, M. and {Bonato}, M. and {Calistro Rivera}, G. and {Cassano}, R. and {Cochrane}, R. and {Croston}, J.~H. and {Cuciti}, V. and {Dallacasa}, D. and {Danezi}, A. and {Dettmar}, R.~J. and {Di Gennaro}, G. and {Edler}, H.~W. and {En{\ss}lin}, T.~A. and {Emig}, K.~L. and {Franzen}, T.~M.~O. and {Garc{\'\i}a-Vergara}, C. and {Grange}, Y.~G. and {G{\"u}rkan}, G. and {Hajduk}, M. and {Heald}, G. and {Heesen}, V. and {Hoang}, D.~N. and {Hoeft}, M. and {Horellou}, C. and {Iacobelli}, M. and {Jamrozy}, M. and {Jeli{\'c}}, V. and {Kondapally}, R. and {Kukreti}, P. and {Kunert-Bajraszewska}, M. and {Magliocchetti}, M. and {Mahatma}, V. and {Ma{\l}ek}, K. and {Mandal}, S. and {Massaro}, F. and {Meyer-Zhao}, Z. and {Mingo}, B. and {Mostert}, R.~I.~J. and {Nair}, D.~G. and {Nakoneczny}, S.~J. and {Nikiel-Wroczy{\'n}ski}, B. and {Orr{\'u}}, E. and {Pajdosz-{\'S}mierciak}, U. and {Pasini}, T. and {Prandoni}, I. and {van Piggelen}, H.~E. and {Rajpurohit}, K. and {Retana-Montenegro}, E. and {Riseley}, C.~J. and {Rowlinson}, A. and {Saxena}, A. and {Schrijvers}, C. and {Sweijen}, F. and {Siewert}, T.~M. and {Timmerman}, R. and {Vaccari}, M. and {Vink}, J. and {West}, J.~L. and {Wo{\l}owska}, A. and {Zhang}, X. and {Zheng}, J.},
        title = "{The LOFAR Two-metre Sky Survey. V. Second data release}",
      journal = {\aap},
         year = 2022,
        month = mar,
       volume = {659},
          eid = {A1},
        pages = {A1},
          doi = {10.1051/0004-6361/202142484},
archivePrefix = {arXiv},
       eprint = {2202.11733},
 primaryClass = {astro-ph.GA},
       adsurl = {https://ui.adsabs.harvard.edu/abs/2022A&A...659A...1S}
}

@ARTICLE{ye2022,
       author = {{Ye}, Haoyang and {Gull}, Stephen F. and {Tan}, Sze M. and {Nikolic}, Bojan},
        title = "{High accuracy wide-field imaging method in radio interferometry}",
      journal = {\mnras},
         year = 2022,
        month = mar,
       volume = {510},
       number = {3},
        pages = {4110-4125},
          doi = {10.1093/mnras/stab3548},
archivePrefix = {arXiv},
       eprint = {2101.11172},
 primaryClass = {astro-ph.IM},
       adsurl = {https://ui.adsabs.harvard.edu/abs/2022MNRAS.510.4110Y}
}

@ARTICLE{arras2021,
       author = {{Arras}, Philipp and {Reinecke}, Martin and {Westermann}, R{\"u}diger and {En{\ss}lin}, Torsten A.},
        title = "{Efficient wide-field radio interferometry response}",
      journal = {\aap},
         year = 2021,
        month = feb,
       volume = {646},
          eid = {A58},
        pages = {A58},
          doi = {10.1051/0004-6361/202039723},
archivePrefix = {arXiv},
       eprint = {2010.10122},
 primaryClass = {astro-ph.IM},
       adsurl = {https://ui.adsabs.harvard.edu/abs/2021A&A...646A..58A}
}

@ARTICLE{tasse2014,
       author = {{Tasse}, Cyril},
        title = "{Applying Wirtinger derivatives to the radio interferometry calibration problem}",
      journal = {arXiv e-prints},
         year = 2014,
        month = oct,
          eid = {arXiv:1410.8706},
        pages = {arXiv:1410.8706},
          doi = {10.48550/arXiv.1410.8706},
archivePrefix = {arXiv},
       eprint = {1410.8706},
 primaryClass = {astro-ph.IM},
       adsurl = {https://ui.adsabs.harvard.edu/abs/2014arXiv1410.8706T}
}

@ARTICLE{vanweeren2021,
       author = {{van Weeren}, R.~J. and {Shimwell}, T.~W. and {Botteon}, A. and {Brunetti}, G. and {Br{\"u}ggen}, M. and {Boxelaar}, J.~M. and {Cassano}, R. and {Di Gennaro}, G. and {Andrade-Santos}, F. and {Bonnassieux}, E. and {Bonafede}, A. and {Cuciti}, V. and {Dallacasa}, D. and {de Gasperin}, F. and {Gastaldello}, F. and {Hardcastle}, M.~J. and {Hoeft}, M. and {Kraft}, R.~P. and {Mandal}, S. and {Rossetti}, M. and {R{\"o}ttgering}, H.~J.~A. and {Tasse}, C. and {Wilber}, A.~G.},
        title = "{LOFAR observations of galaxy clusters in HETDEX. Extraction and self-calibration of individual LOFAR targets}",
      journal = {\aap},
         year = 2021,
        month = jul,
       volume = {651},
          eid = {A115},
        pages = {A115},
          doi = {10.1051/0004-6361/202039826},
archivePrefix = {arXiv},
       eprint = {2011.02387},
 primaryClass = {astro-ph.CO},
       adsurl = {https://ui.adsabs.harvard.edu/abs/2021A&A...651A.115V}
}

@ARTICLE{dejong2024,
       author = {{de Jong}, J.~M.~G.~H.~J. and {van Weeren}, R.~J. and {Sweijen}, F. and {Oonk}, J.~B.~R. and {Shimwell}, T.~W. and {Offringa}, A.~R. and {Morabito}, L.~K. and {R{\"o}ttgering}, H.~J.~A. and {Kondapally}, R. and {Escott}, E.~L. and {Best}, P.~N. and {Bondi}, M. and {Ye}, H. and {Petley}, J.~W.},
        title = "{Into the depths: Unveiling ELAIS-N1 with LOFAR's deepest sub-arcsecond wide-field images}",
      journal = {\aap},
         year = 2024,
        month = sep,
       volume = {689},
          eid = {A80},
        pages = {A80},
          doi = {10.1051/0004-6361/202450595},
archivePrefix = {arXiv},
       eprint = {2407.13247},
 primaryClass = {astro-ph.IM},
       adsurl = {https://ui.adsabs.harvard.edu/abs/2024A&A...689A..80D}
}

@ARTICLE{wsclean,
       author = {{Offringa}, A.~R. and {McKinley}, B. and {Hurley-Walker}, N. and {Briggs}, F.~H. and {Wayth}, R.~B. and {Kaplan}, D.~L. and {Bell}, M.~E. and {Feng}, L. and {Neben}, A.~R. and {Hughes}, J.~D. and {Rhee}, J. and {Murphy}, T. and {Bhat}, N.~D.~R. and {Bernardi}, G. and {Bowman}, J.~D. and {Cappallo}, R.~J. and {Corey}, B.~E. and {Deshpande}, A.~A. and {Emrich}, D. and {Ewall-Wice}, A. and {Gaensler}, B.~M. and {Goeke}, R. and {Greenhill}, L.~J. and {Hazelton}, B.~J. and {Hindson}, L. and {Johnston-Hollitt}, M. and {Jacobs}, D.~C. and {Kasper}, J.~C. and {Kratzenberg}, E. and {Lenc}, E. and {Lonsdale}, C.~J. and {Lynch}, M.~J. and {McWhirter}, S.~R. and {Mitchell}, D.~A. and {Morales}, M.~F. and {Morgan}, E. and {Kudryavtseva}, N. and {Oberoi}, D. and {Ord}, S.~M. and {Pindor}, B. and {Procopio}, P. and {Prabu}, T. and {Riding}, J. and {Roshi}, D.~A. and {Shankar}, N. Udaya and {Srivani}, K.~S. and {Subrahmanyan}, R. and {Tingay}, S.~J. and {Waterson}, M. and {Webster}, R.~L. and {Whitney}, A.~R. and {Williams}, A. and {Williams}, C.~L.},
        title = "{WSCLEAN: an implementation of a fast, generic wide-field imager for radio astronomy}",
      journal = {\mnras},
         year = 2014,
        month = oct,
       volume = {444},
       number = {1},
        pages = {606-619},
          doi = {10.1093/mnras/stu1368},
archivePrefix = {arXiv},
       eprint = {1407.1943},
 primaryClass = {astro-ph.IM},
       adsurl = {https://ui.adsabs.harvard.edu/abs/2014MNRAS.444..606O}
}

@software{dijkema2023,
       author = {{Dijkema}, T.~J. and {Nijhuis}, M. and {van Diepen}, G. and {Offringa}, A. and {Krombeen}, L. and {de Wever}, M. and {Maljaars}, J. and {Loose}, M.},
        title = "{DP3: Streaming processing pipeline for radio interferometric data}",
 howpublished = {Astrophysics Source Code Library, record ascl:2305.014},
         year = 2023,
        month = may,
          eid = {ascl:2305.014},
       adsurl = {https://ui.adsabs.harvard.edu/abs/2023ascl.soft05014D}
}

@MISC{dp3,
       author = {{van Diepen}, Ger and {Dijkema}, Tammo Jan and {Offringa}, Andr{\'e}},
        title = "{DPPP: Default Pre-Processing Pipeline}",
 howpublished = {Astrophysics Source Code Library, record ascl:1804.003},
         year = 2018,
        month = apr,
          eid = {ascl:1804.003},
        pages = {ascl:1804.003},
archivePrefix = {ascl},
       eprint = {1804.003},
       adsurl = {https://ui.adsabs.harvard.edu/abs/2018ascl.soft04003V}
}

@ARTICLE{garn2008,
       author = {{Garn}, Timothy and {Green}, David A. and {Riley}, Julia M. and {Alexander}, Paul},
        title = "{A 610-MHz survey of the ELAIS-N1 field with the Giant Metrewave Radio Telescope - observations, data analysis and source catalogue}",
      journal = {\mnras},
         year = 2008,
        month = jan,
       volume = {383},
       number = {1},
        pages = {75-85},
          doi = {10.1111/j.1365-2966.2007.12562.x},
archivePrefix = {arXiv},
       eprint = {0710.1500},
 primaryClass = {astro-ph},
       adsurl = {https://ui.adsabs.harvard.edu/abs/2008MNRAS.383...75G}
}

@ARTICLE{smirnov2015,
       author = {{Smirnov}, O.~M. and {Tasse}, C.},
        title = "{Radio interferometric gain calibration as a complex optimization problem}",
      journal = {\mnras},
         year = 2015,
        month = may,
       volume = {449},
       number = {3},
        pages = {2668-2684},
          doi = {10.1093/mnras/stv418},
archivePrefix = {arXiv},
       eprint = {1502.06974},
 primaryClass = {astro-ph.IM},
       adsurl = {https://ui.adsabs.harvard.edu/abs/2015MNRAS.449.2668S}
}

@ARTICLE{tasse2014b,
       author = {{Tasse}, C.},
        title = "{Nonlinear Kalman filters for calibration in radio interferometry}",
      journal = {\aap},
         year = 2014,
        month = jun,
       volume = {566},
          eid = {A127},
        pages = {A127},
          doi = {10.1051/0004-6361/201423503},
archivePrefix = {arXiv},
       eprint = {1403.6308},
 primaryClass = {astro-ph.IM},
       adsurl = {https://ui.adsabs.harvard.edu/abs/2014A&A...566A.127T}
}

@ARTICLE{2015ApJ...806...83O,
       author = {{O'Sullivan}, S.~P. and {Gaensler}, B.~M. and {Lara-L{\'o}pez}, M.~A. and {van Velzen}, S. and {Banfield}, J.~K. and {Farnes}, J.~S.},
        title = "{The Magnetic Field and Polarization Properties of Radio Galaxies in Different Accretion States}",
      journal = {\apj},
         year = 2015,
        month = jun,
       volume = {806},
       number = {1},
          eid = {83},
        pages = {83},
          doi = {10.1088/0004-637X/806/1/83},
archivePrefix = {arXiv},
       eprint = {1504.06679},
 primaryClass = {astro-ph.GA},
       adsurl = {https://ui.adsabs.harvard.edu/abs/2015ApJ...806...83O}
}

@ARTICLE{2011ApJ...733...69B,
       author = {{Banfield}, Julie K. and {George}, Samuel J. and {Taylor}, A. Russ and {Stil}, Jeroen M. and {Kothes}, Roland and {Scott}, Douglas},
        title = "{Polarized Radio Sources: A Study of Luminosity, Redshift, and Infrared Colors}",
      journal = {\apj},
         year = 2011,
        month = may,
       volume = {733},
       number = {1},
          eid = {69},
        pages = {69},
          doi = {10.1088/0004-637X/733/1/69},
archivePrefix = {arXiv},
       eprint = {1103.4228},
 primaryClass = {astro-ph.CO},
       adsurl = {https://ui.adsabs.harvard.edu/abs/2011ApJ...733...69B}
}

@ARTICLE{1974MNRAS.167P..31F,
       author = {{Fanaroff}, B.~L. and {Riley}, J.~M.},
        title = "{The morphology of extragalactic radio sources of high and low luminosity}",
      journal = {\mnras},
         year = 1974,
        month = may,
       volume = {167},
        pages = {31P-36P},
          doi = {10.1093/mnras/167.1.31P},
       adsurl = {https://ui.adsabs.harvard.edu/abs/1974MNRAS.167P..31F}
}

@MISC{mevius2018,
       author = {{Mevius}, Maaijke},
        title = "{RMextract: Ionospheric Faraday Rotation calculator}",
 howpublished = {Astrophysics Source Code Library, record ascl:1806.024},
         year = 2018,
        month = jun,
          eid = {ascl:1806.024},
        pages = {ascl:1806.024},
archivePrefix = {ascl},
       eprint = {1806.024},
       adsurl = {https://ui.adsabs.harvard.edu/abs/2018ascl.soft06024M}
}

@MISC{mohan2015,
       author = {{Mohan}, Niruj and {Rafferty}, David},
        title = "{PyBDSF: Python Blob Detection and Source Finder}",
 howpublished = {Astrophysics Source Code Library, record ascl:1502.007},
         year = 2015,
        month = feb,
          eid = {ascl:1502.007},
        pages = {ascl:1502.007},
archivePrefix = {ascl},
       eprint = {1502.007},
       adsurl = {https://ui.adsabs.harvard.edu/abs/2015ascl.soft02007M}
}

@ARTICLE{sexton2022,
       author = {{Sexton}, Remington O. and {Secrest}, Nathan J. and {Johnson}, Megan C. and {Dorland}, Bryan N.},
        title = "{The US Naval Observatory VLBI Spectroscopic Catalog}",
      journal = {\apjs},
         year = 2022,
        month = jun,
       volume = {260},
       number = {2},
          eid = {33},
        pages = {33},
          doi = {10.3847/1538-4365/ac609f},
archivePrefix = {arXiv},
       eprint = {2203.13089},
 primaryClass = {astro-ph.GA},
       adsurl = {https://ui.adsabs.harvard.edu/abs/2022ApJS..260...33S}
}

@ARTICLE{charlot2020,
       author = {{Charlot}, P. and {Jacobs}, C.~S. and {Gordon}, D. and {Lambert}, S. and {de Witt}, A. and {B{\"o}hm}, J. and {Fey}, A.~L. and {Heinkelmann}, R. and {Skurikhina}, E. and {Titov}, O. and {Arias}, E.~F. and {Bolotin}, S. and {Bourda}, G. and {Ma}, C. and {Malkin}, Z. and {Nothnagel}, A. and {Mayer}, D. and {MacMillan}, D.~S. and {Nilsson}, T. and {Gaume}, R.},
        title = "{The third realization of the International Celestial Reference Frame by very long baseline interferometry}",
      journal = {\aap},
         year = 2020,
        month = dec,
       volume = {644},
          eid = {A159},
        pages = {A159},
          doi = {10.1051/0004-6361/202038368},
archivePrefix = {arXiv},
       eprint = {2010.13625},
 primaryClass = {astro-ph.GA},
       adsurl = {https://ui.adsabs.harvard.edu/abs/2020A&A...644A.159C}
}

@ARTICLE{tremblay2016,
       author = {{Tremblay}, S.~E. and {Taylor}, G.~B. and {Ortiz}, A.~A. and {Tremblay}, C.~D. and {Helmboldt}, J.~F. and {Romani}, R.~W.},
        title = "{Compact symmetric objects and supermassive binary black holes in the VLBA Imaging and Polarimetry Survey}",
      journal = {\mnras},
         year = 2016,
        month = jun,
       volume = {459},
       number = {1},
        pages = {820-840},
          doi = {10.1093/mnras/stw592},
archivePrefix = {arXiv},
       eprint = {1603.03094},
 primaryClass = {astro-ph.GA},
       adsurl = {https://ui.adsabs.harvard.edu/abs/2016MNRAS.459..820T}
}

@ARTICLE{2007ApJ...666..201T,
       author = {{Taylor}, A.~R. and {Stil}, J.~M. and {Grant}, J.~K. and {Landecker}, T.~L. and {Kothes}, R. and {Reid}, R.~I. and {Gray}, A.~D. and {Scott}, Douglas and {Martin}, P.~G. and {Boothroyd}, A.~I. and {Joncas}, G. and {Lockman}, Felix J. and {English}, J. and {Sajina}, A. and {Bond}, J.~R.},
        title = "{Radio Polarimetry of the ELAIS N1 Field: Polarized Compact Sources}",
      journal = {\apj},
         year = 2007,
        month = sep,
       volume = {666},
       number = {1},
        pages = {201-211},
          doi = {10.1086/519786},
archivePrefix = {arXiv},
       eprint = {0705.2736},
 primaryClass = {astro-ph},
       adsurl = {https://ui.adsabs.harvard.edu/abs/2007ApJ...666..201T}
}

@ARTICLE{2010ApJ...714.1689G,
       author = {{Grant}, J.~K. and {Taylor}, A.~R. and {Stil}, J.~M. and {Landecker}, T.~L. and {Kothes}, R. and {Ransom}, R.~R. and {Scott}, Douglas},
        title = "{The DRAO Planck Deep Fields: The Polarization Properties of Radio Galaxies at 1.4 GHz}",
      journal = {\apj},
         year = 2010,
        month = may,
       volume = {714},
       number = {2},
        pages = {1689-1701},
          doi = {10.1088/0004-637X/714/2/1689},
archivePrefix = {arXiv},
       eprint = {1003.4460},
 primaryClass = {astro-ph.CO},
       adsurl = {https://ui.adsabs.harvard.edu/abs/2010ApJ...714.1689G}
}

@ARTICLE{2009ApJ...702.1230T,
       author = {{Taylor}, A.~R. and {Stil}, J.~M. and {Sunstrum}, C.},
        title = "{A Rotation Measure Image of the Sky}",
      journal = {\apj},
         year = 2009,
        month = sep,
       volume = {702},
       number = {2},
        pages = {1230-1236},
          doi = {10.1088/0004-637X/702/2/1230},
       adsurl = {https://ui.adsabs.harvard.edu/abs/2009ApJ...702.1230T}
}

@ARTICLE{2024RAA....24c5021L,
       author = {{Lao}, Bao-Qiang and {Yang}, Xiao-Long and {Jaiswal}, Sumit and {Mohan}, Prashanth and {Sun}, Xiao-Hui and {Qin}, Sheng-Li and {Zhao}, Ru-Shuang},
        title = "{A Machine Learning Made Catalog of FR-II Radio Galaxies from the FIRST Survey}",
      journal = {Research in Astronomy and Astrophysics},
         year = 2024,
        month = mar,
       volume = {24},
       number = {3},
          eid = {035021},
        pages = {035021},
          doi = {10.1088/1674-4527/ad204f},
archivePrefix = {arXiv},
       eprint = {2401.08048},
 primaryClass = {astro-ph.GA},
       adsurl = {https://ui.adsabs.harvard.edu/abs/2024RAA....24c5021L}
}

@ARTICLE{1998AJ....115.1693C,
       author = {{Condon}, J.~J. and {Cotton}, W.~D. and {Greisen}, E.~W. and {Yin}, Q.~F. and {Perley}, R.~A. and {Taylor}, G.~B. and {Broderick}, J.~J.},
        title = "{The NRAO VLA Sky Survey}",
      journal = {\aj},
         year = 1998,
        month = may,
       volume = {115},
       number = {5},
        pages = {1693-1716},
          doi = {10.1086/300337},
       adsurl = {https://ui.adsabs.harvard.edu/abs/1998AJ....115.1693C}
}

@ARTICLE{2014ApJS..212...15F,
       author = {{Farnes}, J.~S. and {Gaensler}, B.~M. and {Carretti}, E.},
        title = "{A Broadband Polarization Catalog of Extragalactic Radio Sources}",
      journal = {\apjs},
         year = 2014,
        month = may,
       volume = {212},
       number = {1},
          eid = {15},
        pages = {15},
          doi = {10.1088/0067-0049/212/1/15},
archivePrefix = {arXiv},
       eprint = {1403.2391},
 primaryClass = {astro-ph.GA},
       adsurl = {https://ui.adsabs.harvard.edu/abs/2014ApJS..212...15F}
}

@ARTICLE{ruiz2021,
       author = {{Herrera Ruiz}, N. and {O'Sullivan}, S.~P. and {Vacca}, V. and {Jeli{\'c}}, V. and {Nikiel-Wroczy{\'n}ski}, B. and {Bourke}, S. and {Sabater}, J. and {Dettmar}, R.-J. and {Heald}, G. and {Horellou}, C. and {Piras}, S. and {Sobey}, C. and {Shimwell}, T.~W. and {Tasse}, C. and {Hardcastle}, M.~J. and {Kondapally}, R. and {Chy{\.z}y}, K.~T. and {Iacobelli}, M. and {Best}, P.~N. and {Br{\"u}ggen}, M. and {Carretti}, E. and {Prandoni}, I.},
        title = "{LOFAR Deep Fields: probing a broader population of polarized radio galaxies in ELAIS-N1}",
      journal = {\aap},
         year = 2021,
        month = apr,
       volume = {648},
          eid = {A12},
        pages = {A12},
          doi = {10.1051/0004-6361/202038896},
archivePrefix = {arXiv},
       eprint = {2011.08292},
 primaryClass = {astro-ph.GA},
       adsurl = {https://ui.adsabs.harvard.edu/abs/2021A&A...648A..12H}
}

@INPROCEEDINGS{sirothia2009,
       author = {{Sirothia}, S.~K. and {Dennefeld}, M. and {Saikia}, D.~J. and {Dole}, H. and {Ricquebourg}, F. and {Roland}, J.},
        title = "{325-MHz Observations of the ELAIS-N1 Field}",
    booktitle = {The Low-Frequency Radio Universe},
         year = 2009,
       editor = {{Saikia}, D.~J. and {Green}, D.~A. and {Gupta}, Y. and {Venturi}, T.},
       series = {Astronomical Society of the Pacific Conference Series},
       volume = {407},
        month = sep,
        pages = {27},
       adsurl = {https://ui.adsabs.harvard.edu/abs/2009ASPC..407...27S}
}

@ARTICLE{ye2024,
       author = {{Ye}, Haoyang and {Sweijen}, Frits and {van Weeren}, Reinout J. and {Williams}, Wendy and {de Jong}, Jurjen and {Morabito}, Leah K. and {Rottgering}, Huub and {Shimwell}, Timothy W. and {Best}, P.~N. and {Bondi}, Marco and {Br{\"u}ggen}, Marcus and {de Gasperin}, Francesco and {Tasse}, Cyril},
        title = "{1-arcsecond imaging of the ELAIS-N1 field at 144MHz using the LoTSS survey with the international LOFAR telescope}",
      journal = {\aap},
         year = 2024,
        month = nov,
       volume = {691},
          eid = {A347},
        pages = {A347},
          doi = {10.1051/0004-6361/202348103},
archivePrefix = {arXiv},
       eprint = {2309.16560},
 primaryClass = {astro-ph.IM},
       adsurl = {https://ui.adsabs.harvard.edu/abs/2024A&A...691A.347Y}
}

@ARTICLE{2018MNRAS.475.4263O,
       author = {{O'Sullivan}, S.~P. and {Lenc}, E. and {Anderson}, C.~S. and {Gaensler}, B.~M. and {Murphy}, T.},
        title = "{Faraday rotation at low frequencies: magnetoionic material of the large FRII radio galaxy PKS J0636-2036}",
      journal = {\mnras},
         year = 2018,
        month = apr,
       volume = {475},
       number = {3},
        pages = {4263-4277},
          doi = {10.1093/mnras/sty171},
archivePrefix = {arXiv},
       eprint = {1801.02452},
 primaryClass = {astro-ph.GA},
       adsurl = {https://ui.adsabs.harvard.edu/abs/2018MNRAS.475.4263O}
}

@ARTICLE{2017PASA...34...40L,
       author = {{Lenc}, E. and {Anderson}, C.~S. and {Barry}, N. and {Bowman}, J.~D. and {Cairns}, I.~H. and {Farnes}, J.~S. and {Gaensler}, B.~M. and {Heald}, G. and {Johnston-Hollitt}, M. and {Kaplan}, D.~L. and {Lynch}, C.~R. and {McCauley}, P.~I. and {Mitchell}, D.~A. and {Morgan}, J. and {Morales}, M.~F. and {Murphy}, Tara and {Offringa}, A.~R. and {Ord}, S.~M. and {Pindor}, B. and {Riseley}, C. and {Sadler}, E.~M. and {Sobey}, C. and {Sokolowski}, M. and {Sullivan}, I.~S. and {O'Sullivan}, S.~P. and {Sun}, X.~H. and {Tremblay}, S.~E. and {Trott}, C.~M. and {Wayth}, R.~B.},
        title = "{The Challenges of Low-Frequency Radio Polarimetry: Lessons from the Murchison Widefield Array}",
      journal = {\pasa},
         year = 2017,
        month = sep,
       volume = {34},
          eid = {e040},
        pages = {e040},
          doi = {10.1017/pasa.2017.36},
archivePrefix = {arXiv},
       eprint = {1708.05799},
 primaryClass = {astro-ph.IM},
       adsurl = {https://ui.adsabs.harvard.edu/abs/2017PASA...34...40L}
}

@ARTICLE{2016ApJ...830...38L,
       author = {{Lenc}, E. and {Gaensler}, B.~M. and {Sun}, X.~H. and {Sadler}, E.~M. and {Willis}, A.~G. and {Barry}, N. and {Beardsley}, A.~P. and {Bell}, M.~E. and {Bernardi}, G. and {Bowman}, J.~D. and {Briggs}, F. and {Callingham}, J.~R. and {Cappallo}, R.~J. and {Carroll}, P. and {Corey}, B.~E. and {de Oliveira-Costa}, A. and {Deshpande}, A.~A. and {Dillon}, J.~S. and {Dwarkanath}, K.~S. and {Emrich}, D. and {Ewall-Wice}, A. and {Feng}, L. and {For}, B.-Q. and {Goeke}, R. and {Greenhill}, L.~J. and {Hancock}, P. and {Hazelton}, B.~J. and {Hewitt}, J.~N. and {Hindson}, L. and {Hurley-Walker}, N. and {Johnston-Hollitt}, M. and {Jacobs}, D.~C. and {Kapi{\'n}ska}, A.~D. and {Kaplan}, D.~L. and {Kasper}, J.~C. and {Kim}, H.-S. and {Kratzenberg}, E. and {Line}, J. and {Loeb}, A. and {Lonsdale}, C.~J. and {Lynch}, M.~J. and {McKinley}, B. and {McWhirter}, S.~R. and {Mitchell}, D.~A. and {Morales}, M.~F. and {Morgan}, E. and {Morgan}, J. and {Murphy}, T. and {Neben}, A.~R. and {Oberoi}, D. and {Offringa}, A.~R. and {Ord}, S.~M. and {Paul}, S. and {Pindor}, B. and {Pober}, J.~C. and {Prabu}, T. and {Procopio}, P. and {Riding}, J. and {Rogers}, A.~E.~E. and {Roshi}, A. and {Udaya Shankar}, N. and {Sethi}, S.~K. and {Srivani}, K.~S. and {Staveley-Smith}, L. and {Subrahmanyan}, R. and {Sullivan}, I.~S. and {Tegmark}, M. and {Thyagarajan}, Nithyanandan and {Tingay}, S.~J. and {Trott}, C. and {Waterson}, M. and {Wayth}, R.~B. and {Webster}, R.~L. and {Whitney}, A.~R. and {Williams}, A. and {Williams}, C.~L. and {Wu}, C. and {Wyithe}, J.~S.~B. and {Zheng}, Q.},
        title = "{Low-frequency Observations of Linearly Polarized Structures in the Interstellar Medium near the South Galactic Pole}",
      journal = {\apj},
         year = 2016,
        month = oct,
       volume = {830},
       number = {1},
          eid = {38},
        pages = {38},
          doi = {10.3847/0004-637X/830/1/38},
archivePrefix = {arXiv},
       eprint = {1607.05779},
 primaryClass = {astro-ph.GA},
       adsurl = {https://ui.adsabs.harvard.edu/abs/2016ApJ...830...38L}
}

@ARTICLE{callingham2023,
       author = {{Callingham}, J.~R. and {Shimwell}, T.~W. and {Vedantham}, H.~K. and {Bassa}, C.~G. and {O'Sullivan}, S.~P. and {Yiu}, T.~W.~H. and {Bloot}, S. and {Best}, P.~N. and {Hardcastle}, M.~J. and {Haverkorn}, M. and {Kavanagh}, R.~D. and {Lamy}, L. and {Pope}, B.~J.~S. and {R{\"o}ttgering}, H.~J.~A. and {Schwarz}, D.~J. and {Tasse}, C. and {van Weeren}, R.~J. and {White}, G.~J. and {Zarka}, P. and {Bomans}, D.~J. and {Bonafede}, A. and {Bonato}, M. and {Botteon}, A. and {Bruggen}, M. and {Chy{\.z}y}, K.~T. and {Drabent}, A. and {Emig}, K.~L. and {Gloudemans}, A.~J. and {G{\"u}rkan}, G. and {Hajduk}, M. and {Hoang}, D.~N. and {Hoeft}, M. and {Iacobelli}, M. and {Kadler}, M. and {Kunert-Bajraszewska}, M. and {Mingo}, B. and {Morabito}, L.~K. and {Nair}, D.~G. and {P{\'e}rez-Torres}, M. and {Ray}, T.~P. and {Riseley}, C.~J. and {Rowlinson}, A. and {Shulevski}, A. and {Sweijen}, F. and {Timmerman}, R. and {Vaccari}, M. and {Zheng}, J.},
        title = "{V-LoTSS: The circularly polarised LOFAR Two-metre Sky Survey}",
      journal = {\aap},
         year = 2023,
        month = feb,
       volume = {670},
          eid = {A124},
        pages = {A124},
          doi = {10.1051/0004-6361/202245567},
archivePrefix = {arXiv},
       eprint = {2212.09815},
 primaryClass = {astro-ph.GA},
       adsurl = {https://ui.adsabs.harvard.edu/abs/2023A&A...670A.124C}
}

@ARTICLE{2025A&A...694A..44O,
       author = {{Osinga}, E. and {van Weeren}, R.~J. and {Rudnick}, L. and {Andrade-Santos}, F. and {Bonafede}, A. and {Clarke}, T. and {Duncan}, K. and {Giacintucci}, S. and {R{\"o}ttgering}, H.~J.~A.},
        title = "{Probing cluster magnetism with embedded and background radio sources in Planck clusters}",
      journal = {\aap},
         year = 2025,
        month = feb,
       volume = {694},
          eid = {A44},
        pages = {A44},
          doi = {10.1051/0004-6361/202451885},
archivePrefix = {arXiv},
       eprint = {2408.07178},
 primaryClass = {astro-ph.CO},
       adsurl = {https://ui.adsabs.harvard.edu/abs/2025A&A...694A..44O}
}

@ARTICLE{2020NatAs...4..577V,
       author = {{Vedantham}, H.~K. and {Callingham}, J.~R. and {Shimwell}, T.~W. and {Tasse}, C. and {Pope}, B.~J.~S. and {Bedell}, M. and {Snellen}, I. and {Best}, P. and {Hardcastle}, M.~J. and {Haverkorn}, M. and {Mechev}, A. and {O'Sullivan}, S.~P. and {R{\"o}ttgering}, H.~J.~A. and {White}, G.~J.},
        title = "{Coherent radio emission from a quiescent red dwarf indicative of star-planet interaction}",
      journal = {Nature Astronomy},
         year = 2020,
        month = feb,
       volume = {4},
        pages = {577-583},
          doi = {10.1038/s41550-020-1011-9},
archivePrefix = {arXiv},
       eprint = {2002.08727},
 primaryClass = {astro-ph.EP},
       adsurl = {https://ui.adsabs.harvard.edu/abs/2020NatAs...4..577V}
}

@ARTICLE{2022A&A...663A...7E,
       author = {{Erceg}, Ana and {Jeli{\'c}}, Vibor and {Haverkorn}, Marijke and {Bracco}, Andrea and {Shimwell}, Timothy W. and {Tasse}, Cyril and {Dickey}, John M. and {Ceraj}, Lana and {Drabent}, Alexander and {Hardcastle}, Martin J. and {Turi{\'c}}, Luka},
        title = "{Faraday tomography of LoTSS-DR2 data. I. Faraday moments in the high-latitude outer Galaxy and revealing Loop III in polarisation}",
      journal = {\aap},
         year = 2022,
        month = jul,
       volume = {663},
          eid = {A7},
        pages = {A7},
          doi = {10.1051/0004-6361/202142244},
archivePrefix = {arXiv},
       eprint = {2203.01351},
 primaryClass = {astro-ph.GA},
       adsurl = {https://ui.adsabs.harvard.edu/abs/2022A&A...663A...7E}
}

@ARTICLE{2025MNRAS.540..416S,
       author = {{Sweijen}, F. and {Pierce}, J.~C.~S. and {Hardcastle}, M.~J. and {Croston}, J.~H. and {Morabito}, L.~K. and {Bondi}, M. and {Callingham}, J.~R. and {Jurlin}, N. and {Prandoni}, I. and {R{\"o}ttgering}, H.~J.~A. and {van Weeren}, R.~J.},
        title = "{The low-frequency size distribution of radio sources in the Lockman Hole}",
      journal = {\mnras},
         year = 2025,
        month = jun,
       volume = {540},
       number = {1},
        pages = {416-432},
          doi = {10.1093/mnras/staf630},
       adsurl = {https://ui.adsabs.harvard.edu/abs/2025MNRAS.540..416S}
}

@ARTICLE{2025Ap&SS.370...19M,
       author = {{Morabito}, Leah K. and {Jackson}, Neal and {de Jong}, Jurjen and {Escott}, Emmy and {Groeneveld}, Christian and {Mahatma}, Vijay and {Petley}, James and {Sweijen}, Frits and {Timmerman}, Roland and {van Weeren}, Reinout J.},
        title = "{A decade of sub-arcsecond imaging with the International LOFAR Telescope}",
      journal = {\apss},
         year = 2025,
        month = feb,
       volume = {370},
       number = {2},
          eid = {19},
        pages = {19},
          doi = {10.1007/s10509-025-04406-x},
archivePrefix = {arXiv},
       eprint = {2502.06946},
 primaryClass = {astro-ph.IM},
       adsurl = {https://ui.adsabs.harvard.edu/abs/2025Ap&SS.370...19M}
}

@ARTICLE{2013PASA...30....7T,
       author = {{Tingay}, S.~J. and {Goeke}, R. and {Bowman}, J.~D. and {Emrich}, D. and {Ord}, S.~M. and {Mitchell}, D.~A. and {Morales}, M.~F. and {Booler}, T. and {Crosse}, B. and {Wayth}, R.~B. and {Lonsdale}, C.~J. and {Tremblay}, S. and {Pallot}, D. and {Colegate}, T. and {Wicenec}, A. and {Kudryavtseva}, N. and {Arcus}, W. and {Barnes}, D. and {Bernardi}, G. and {Briggs}, F. and {Burns}, S. and {Bunton}, J.~D. and {Cappallo}, R.~J. and {Corey}, B.~E. and {Deshpande}, A. and {Desouza}, L. and {Gaensler}, B.~M. and {Greenhill}, L.~J. and {Hall}, P.~J. and {Hazelton}, B.~J. and {Herne}, D. and {Hewitt}, J.~N. and {Johnston-Hollitt}, M. and {Kaplan}, D.~L. and {Kasper}, J.~C. and {Kincaid}, B.~B. and {Koenig}, R. and {Kratzenberg}, E. and {Lynch}, M.~J. and {Mckinley}, B. and {Mcwhirter}, S.~R. and {Morgan}, E. and {Oberoi}, D. and {Pathikulangara}, J. and {Prabu}, T. and {Remillard}, R.~A. and {Rogers}, A.~E.~E. and {Roshi}, A. and {Salah}, J.~E. and {Sault}, R.~J. and {Udaya-Shankar}, N. and {Schlagenhaufer}, F. and {Srivani}, K.~S. and {Stevens}, J. and {Subrahmanyan}, R. and {Waterson}, M. and {Webster}, R.~L. and {Whitney}, A.~R. and {Williams}, A. and {Williams}, C.~L. and {Wyithe}, J.~S.~B.},
        title = "{The Murchison Widefield Array: The Square Kilometre Array Precursor at Low Radio Frequencies}",
      journal = {\pasa},
         year = 2013,
        month = jan,
       volume = {30},
          eid = {e007},
        pages = {e007},
          doi = {10.1017/pasa.2012.007},
archivePrefix = {arXiv},
       eprint = {1206.6945},
 primaryClass = {astro-ph.IM},
       adsurl = {https://ui.adsabs.harvard.edu/abs/2013PASA...30....7T}
}

@ARTICLE{2020A&A...638A..48S,
       author = {{Stuardi}, C. and {O'Sullivan}, S.~P. and {Bonafede}, A. and {Br{\"u}ggen}, M. and {Dabhade}, P. and {Horellou}, C. and {Morganti}, R. and {Carretti}, E. and {Heald}, G. and {Iacobelli}, M. and {Vacca}, V.},
        title = "{The LOFAR view of intergalactic magnetic fields with giant radio galaxies}",
      journal = {\aap},
         year = 2020,
        month = jun,
       volume = {638},
          eid = {A48},
        pages = {A48},
          doi = {10.1051/0004-6361/202037635},
archivePrefix = {arXiv},
       eprint = {2004.05169},
 primaryClass = {astro-ph.GA},
       adsurl = {https://ui.adsabs.harvard.edu/abs/2020A&A...638A..48S}
}

@ARTICLE{2023A&A...670L..23H,
       author = {{Heesen}, V. and {O'Sullivan}, S.~P. and {Br{\"u}ggen}, M. and {Basu}, A. and {Beck}, R. and {Seta}, A. and {Carretti}, E. and {Krause}, M.~G.~H. and {Haverkorn}, M. and {Hutschenreuter}, S. and {Bracco}, A. and {Stein}, M. and {Bomans}, D.~J. and {Dettmar}, R. -J. and {Chy{\.z}y}, K.~T. and {Heald}, G.~H. and {Paladino}, R. and {Horellou}, C.},
        title = "{Detection of magnetic fields in the circumgalactic medium of nearby galaxies using Faraday rotation}",
      journal = {\aap},
         year = 2023,
        month = feb,
       volume = {670},
          eid = {L23},
        pages = {L23},
          doi = {10.1051/0004-6361/202346008},
archivePrefix = {arXiv},
       eprint = {2302.06617},
 primaryClass = {astro-ph.GA},
       adsurl = {https://ui.adsabs.harvard.edu/abs/2023A&A...670L..23H}
}

@ARTICLE{2023A&A...678A.151H,
       author = {{Hardcastle}, M.~J. and {Horton}, M.~A. and {Williams}, W.~L. and {Duncan}, K.~J. and {Alegre}, L. and {Barkus}, B. and {Croston}, J.~H. and {Dickinson}, H. and {Osinga}, E. and {R{\"o}ttgering}, H.~J.~A. and {Sabater}, J. and {Shimwell}, T.~W. and {Smith}, D.~J.~B. and {Best}, P.~N. and {Botteon}, A. and {Br{\"u}ggen}, M. and {Drabent}, A. and {de Gasperin}, F. and {G{\"u}rkan}, G. and {Hajduk}, M. and {Hale}, C.~L. and {Hoeft}, M. and {Jamrozy}, M. and {Kunert-Bajraszewska}, M. and {Kondapally}, R. and {Magliocchetti}, M. and {Mahatma}, V.~H. and {Mostert}, R.~I.~J. and {O'Sullivan}, S.~P. and {Pajdosz-{\'S}mierciak}, U. and {Petley}, J. and {Pierce}, J.~C.~S. and {Prandoni}, I. and {Schwarz}, D.~J. and {Shulewski}, A. and {Siewert}, T.~M. and {Stott}, J.~P. and {Tang}, H. and {Vaccari}, M. and {Zheng}, X. and {Bailey}, T. and {Desbled}, S. and {Goyal}, A. and {Gonano}, V. and {Hanset}, M. and {Kurtz}, W. and {Lim}, S.~M. and {Mielle}, L. and {Molloy}, C.~S. and {Roth}, R. and {Terentev}, I.~A. and {Torres}, M.},
        title = "{The LOFAR Two-Metre Sky Survey. VI. Optical identifications for the second data release}",
      journal = {\aap},
         year = 2023,
        month = oct,
       volume = {678},
          eid = {A151},
        pages = {A151},
          doi = {10.1051/0004-6361/202347333},
archivePrefix = {arXiv},
       eprint = {2309.00102},
 primaryClass = {astro-ph.GA},
       adsurl = {https://ui.adsabs.harvard.edu/abs/2023A&A...678A.151H}
}

@ARTICLE{2000IAUC.7432....3V,
       author = {{Voges}, W. and {Aschenbach}, B. and {Boller}, T. and {Brauninger}, H. and {Briel}, U. and {Burkert}, W. and {Dennerl}, K. and {Englhauser}, J. and {Gruber}, R. and {Haberl}, F. and {Hartner}, G. and {Hasinger}, G. and {Pfeffermann}, E. and {Pietsch}, W. and {Predehl}, P. and {Schmitt}, J. and {Trumper}, J. and {Zimmermann}, U.},
        title = "{Rosat All-Sky Survey Faint Source Catalogue}",
      journal = {\iaucirc},
         year = 2000,
        month = may,
       volume = {7432},
        pages = {3},
       adsurl = {https://ui.adsabs.harvard.edu/abs/2000IAUC.7432....3V}
}

@ARTICLE{2016ApJS..224...40W,
       author = {{Wang}, Song and {Liu}, Jifeng and {Qiu}, Yanli and {Bai}, Yu and {Yang}, Huiqin and {Guo}, Jincheng and {Zhang}, Peng},
        title = "{CHANDRA ACIS Survey of X-Ray Point Sources: The Source Catalog}",
      journal = {\apjs},
         year = 2016,
        month = jun,
       volume = {224},
       number = {2},
          eid = {40},
        pages = {40},
          doi = {10.3847/0067-0049/224/2/40},
archivePrefix = {arXiv},
       eprint = {1603.08353},
 primaryClass = {astro-ph.SR},
       adsurl = {https://ui.adsabs.harvard.edu/abs/2016ApJS..224...40W}
}

@ARTICLE{2022A&A...660A..59M,
       author = {{Mahato}, Mousumi and {Dabhade}, Pratik and {Saikia}, D.~J. and {Combes}, Fran{\c{c}}oise and {Bagchi}, Joydeep and {Ho}, L.~C. and {Raychaudhury}, Somak},
        title = "{Search and analysis of giant radio galaxies with associated nuclei (SAGAN). III. New insights into giant radio quasars}",
      journal = {\aap},
         year = 2022,
        month = apr,
       volume = {660},
          eid = {A59},
        pages = {A59},
          doi = {10.1051/0004-6361/202141928},
archivePrefix = {arXiv},
       eprint = {2111.11905},
 primaryClass = {astro-ph.GA},
       adsurl = {https://ui.adsabs.harvard.edu/abs/2022A&A...660A..59M}
}

@ARTICLE{2021ApJS..253...25K,
       author = {{Ku{\'z}micz}, Agnieszka and {Jamrozy}, Marek},
        title = "{Giant Radio Quasars: Sample and Basic Properties}",
      journal = {\apjs},
         year = 2021,
        month = mar,
       volume = {253},
       number = {1},
          eid = {25},
        pages = {25},
          doi = {10.3847/1538-4365/abd483},
archivePrefix = {arXiv},
       eprint = {2012.08857},
 primaryClass = {astro-ph.GA},
       adsurl = {https://ui.adsabs.harvard.edu/abs/2021ApJS..253...25K}
}

@ARTICLE{2024A&A...686A..21S,
       author = {{Simonte}, M. and {Andernach}, H. and {Br{\"u}ggen}, M. and {Miley}, G.~K. and {Barthel}, P.},
        title = "{Giant radio galaxies in the LOFAR deep fields}",
      journal = {\aap},
         year = 2024,
        month = jun,
       volume = {686},
          eid = {A21},
        pages = {A21},
          doi = {10.1051/0004-6361/202348904},
archivePrefix = {arXiv},
       eprint = {2403.08037},
 primaryClass = {astro-ph.GA},
       adsurl = {https://ui.adsabs.harvard.edu/abs/2024A&A...686A..21S}
}

@ARTICLE{2020ApJS..250....8L,
       author = {{Lyke}, Brad W. and {Higley}, Alexandra N. and {McLane}, J.~N. and {Schurhammer}, Danielle P. and {Myers}, Adam D. and {Ross}, Ashley J. and {Dawson}, Kyle and {Chabanier}, Sol{\`e}ne and {Martini}, Paul and {Busca}, Nicol{\'a}s G. and {Mas des Bourboux}, H{\'e}lion du and {Salvato}, Mara and {Streblyanska}, Alina and {Zarrouk}, Pauline and {Burtin}, Etienne and {Anderson}, Scott F. and {Bautista}, Julian and {Bizyaev}, Dmitry and {Brandt}, W.~N. and {Brinkmann}, Jonathan and {Brownstein}, Joel R. and {Comparat}, Johan and {Green}, Paul and {de la Macorra}, Axel and {Mu{\~n}oz Guti{\'e}rrez}, Andrea and {Hou}, Jiamin and {Newman}, Jeffrey A. and {Palanque-Delabrouille}, Nathalie and {P{\^a}ris}, Isabelle and {Percival}, Will J. and {Petitjean}, Patrick and {Rich}, James and {Rossi}, Graziano and {Schneider}, Donald P. and {Smith}, Alexander and {Vivek}, M. and {Weaver}, Benjamin Alan},
        title = "{The Sloan Digital Sky Survey Quasar Catalog: Sixteenth Data Release}",
      journal = {\apjs},
         year = 2020,
        month = sep,
       volume = {250},
       number = {1},
          eid = {8},
        pages = {8},
          doi = {10.3847/1538-4365/aba623},
archivePrefix = {arXiv},
       eprint = {2007.09001},
 primaryClass = {astro-ph.GA},
       adsurl = {https://ui.adsabs.harvard.edu/abs/2020ApJS..250....8L}
}

@ARTICLE{1988Natur.331..147G,
       author = {{Garrington}, S.~T. and {Leahy}, J.~P. and {Conway}, R.~G. and {Laing}, R.~A.},
        title = "{A systematic asymmetry in the polarization properties of double radio sources with one jet}",
      journal = {\nat},
         year = 1988,
        month = jan,
       volume = {331},
       number = {6152},
        pages = {147-149},
          doi = {10.1038/331147a0},
       adsurl = {https://ui.adsabs.harvard.edu/abs/1988Natur.331..147G}
}

@ARTICLE{1988Natur.331..149L,
       author = {{Laing}, R.~A.},
        title = "{The sidedness of jets and depolarization in powerful extragalactic radio sources}",
      journal = {\nat},
         year = 1988,
        month = jan,
       volume = {331},
       number = {6152},
        pages = {149-151},
          doi = {10.1038/331149a0},
       adsurl = {https://ui.adsabs.harvard.edu/abs/1988Natur.331..149L}
}

@ARTICLE{2024A&A...691A..23D,
       author = {{De Rubeis}, E. and {Stuardi}, C. and {Bonafede}, A. and {Vazza}, F. and {van Weeren}, R.~J. and {de Gasperin}, F. and {Br{\"u}ggen}, M.},
        title = "{Magnetic fields in the outskirts of PSZ2 G096.88+24.18 from a depolarization analysis of radio relics}",
      journal = {\aap},
         year = 2024,
        month = nov,
       volume = {691},
          eid = {A23},
        pages = {A23},
          doi = {10.1051/0004-6361/202450892},
archivePrefix = {arXiv},
       eprint = {2408.08603},
 primaryClass = {astro-ph.CO},
       adsurl = {https://ui.adsabs.harvard.edu/abs/2024A&A...691A..23D}
}

@ARTICLE{1974ApJ...194..249W,
       author = {{Wardle}, J.~F.~C. and {Kronberg}, P.~P.},
        title = "{The linear polarization of quasi-stellar radio sources at 3.71 and 11.1 centimeters.}",
      journal = {\apj},
         year = 1974,
        month = jan,
       volume = {194},
          eid = {249},
        pages = {249},
          doi = {10.1086/153240},
       adsurl = {https://ui.adsabs.harvard.edu/abs/1974ApJ...194..249W}
}

@ARTICLE{2016A&A...588A.103B,
       author = {{Boller}, Th. and {Freyberg}, M.~J. and {Tr{\"u}mper}, J. and {Haberl}, F. and {Voges}, W. and {Nandra}, K.},
        title = "{Second ROSAT all-sky survey (2RXS) source catalogue}",
      journal = {\aap},
         year = 2016,
        month = apr,
       volume = {588},
          eid = {A103},
        pages = {A103},
          doi = {10.1051/0004-6361/201525648},
archivePrefix = {arXiv},
       eprint = {1609.09244},
 primaryClass = {astro-ph.HE},
       adsurl = {https://ui.adsabs.harvard.edu/abs/2016A&A...588A.103B}
}

@ARTICLE{2020MNRAS.495.2607O,
       author = {{O'Sullivan}, S.~P. and {Br{\"u}ggen}, M. and {Vazza}, F. and {Carretti}, E. and {Locatelli}, N.~T. and {Stuardi}, C. and {Vacca}, V. and {Vernstrom}, T. and {Heald}, G. and {Horellou}, C. and {Shimwell}, T.~W. and {Hardcastle}, M.~J. and {Tasse}, C. and {R{\"o}ttgering}, H.},
        title = "{New constraints on the magnetization of the cosmic web using LOFAR Faraday rotation observations}",
      journal = {\mnras},
         year = 2020,
        month = jul,
       volume = {495},
       number = {3},
        pages = {2607-2619},
          doi = {10.1093/mnras/staa1395},
archivePrefix = {arXiv},
       eprint = {2002.06924},
 primaryClass = {astro-ph.CO},
       adsurl = {https://ui.adsabs.harvard.edu/abs/2020MNRAS.495.2607O}
}

@ARTICLE{2018PASA...35...43R,
       author = {{Riseley}, C.~J. and {Lenc}, E. and {Van Eck}, C.~L. and {Heald}, G. and {Gaensler}, B.~M. and {Anderson}, C.~S. and {Hancock}, P.~J. and {Hurley-Walker}, N. and {Sridhar}, S.~S. and {White}, S.~V.},
        title = "{The POlarised GLEAM Survey (POGS) I: First results from a low-frequency radio linear polarisation survey of the southern sky}",
      journal = {\pasa},
         year = 2018,
        month = dec,
       volume = {35},
          eid = {e043},
        pages = {e043},
          doi = {10.1017/pasa.2018.39},
archivePrefix = {arXiv},
       eprint = {1809.09327},
 primaryClass = {astro-ph.GA},
       adsurl = {https://ui.adsabs.harvard.edu/abs/2018PASA...35...43R}
}

@ARTICLE{2020PASA...37...29R,
       author = {{Riseley}, C.~J. and {Galvin}, T.~J. and {Sobey}, C. and {Vernstrom}, T. and {White}, S.~V. and {Zhang}, X. and {Gaensler}, B.~M. and {Heald}, G. and {Anderson}, C.~S. and {Franzen}, T.~M.~O. and {Hancock}, P.~J. and {Hurley-Walker}, N. and {Lenc}, E. and {Van Eck}, C.~L.},
        title = "{The POlarised GLEAM Survey (POGS) II: Results from an all-sky rotation measure synthesis survey at long wavelengths}",
      journal = {\pasa},
         year = 2020,
        month = jul,
       volume = {37},
          eid = {e029},
        pages = {e029},
          doi = {10.1017/pasa.2020.20},
archivePrefix = {arXiv},
       eprint = {2005.09266},
 primaryClass = {astro-ph.GA},
       adsurl = {https://ui.adsabs.harvard.edu/abs/2020PASA...37...29R}
}

@ARTICLE{2021MNRAS.502..273M,
       author = {{Mahatma}, V.~H. and {Hardcastle}, M.~J. and {Harwood}, J. and {O'Sullivan}, S.~P. and {Heald}, G. and {Horellou}, C. and {Smith}, D.~J.~B.},
        title = "{A low-frequency study of linear polarization in radio galaxies}",
      journal = {\mnras},
         year = 2021,
        month = mar,
       volume = {502},
       number = {1},
        pages = {273-292},
          doi = {10.1093/mnras/staa3980},
archivePrefix = {arXiv},
       eprint = {2012.11990},
 primaryClass = {astro-ph.GA},
       adsurl = {https://ui.adsabs.harvard.edu/abs/2021MNRAS.502..273M}
}

@ARTICLE{2019A&A...623A..71V,
       author = {{Van Eck}, C.~L. and {Haverkorn}, M. and {Alves}, M.~I.~R. and {Beck}, R. and {Best}, P. and {Carretti}, E. and {Chy{\.z}y}, K.~T. and {En{\ss}lin}, T. and {Farnes}, J.~S. and {Ferri{\`e}re}, K. and {Heald}, G. and {Iacobelli}, M. and {Jeli{\'c}}, V. and {Reich}, W. and {R{\"o}ttgering}, H.~J.~A. and {Schnitzeler}, D.~H.~F.~M.},
        title = "{Diffuse polarized emission in the LOFAR Two-meter Sky Survey}",
      journal = {\aap},
         year = 2019,
        month = mar,
       volume = {623},
          eid = {A71},
        pages = {A71},
          doi = {10.1051/0004-6361/201834777},
archivePrefix = {arXiv},
       eprint = {1902.00531},
 primaryClass = {astro-ph.GA},
       adsurl = {https://ui.adsabs.harvard.edu/abs/2019A&A...623A..71V}
}

@ARTICLE{2017A&A...597A..98V,
       author = {{Van Eck}, C.~L. and {Haverkorn}, M. and {Alves}, M.~I.~R. and {Beck}, R. and {de Bruyn}, A.~G. and {En{\ss}lin}, T. and {Farnes}, J.~S. and {Ferri{\`e}re}, K. and {Heald}, G. and {Horellou}, C. and {Horneffer}, A. and {Iacobelli}, M. and {Jeli{\'c}}, V. and {Mart{\'\i}-Vidal}, I. and {Mulcahy}, D.~D. and {Reich}, W. and {R{\"o}ttgering}, H.~J.~A. and {Scaife}, A.~M.~M. and {Schnitzeler}, D.~H.~F.~M. and {Sobey}, C. and {Sridhar}, S.~S.},
        title = "{Faraday tomography of the local interstellar medium with LOFAR: Galactic foregrounds towards IC 342}",
      journal = {\aap},
         year = 2017,
        month = jan,
       volume = {597},
          eid = {A98},
        pages = {A98},
          doi = {10.1051/0004-6361/201629707},
archivePrefix = {arXiv},
       eprint = {1612.00710},
 primaryClass = {astro-ph.GA},
       adsurl = {https://ui.adsabs.harvard.edu/abs/2017A&A...597A..98V}
}

@ARTICLE{2018A&A...615L...3J,
       author = {{Jeli{\'c}}, Vibor and {Prelogovi{\'c}}, David and {Haverkorn}, Marijke and {Remeijn}, Jur and {Klind{\v{z}}i{\'c}}, Dora},
        title = "{Magnetically aligned straight depolarization canals and the rolling Hough transform}",
      journal = {\aap},
         year = 2018,
        month = jul,
       volume = {615},
          eid = {L3},
        pages = {L3},
          doi = {10.1051/0004-6361/201833291},
archivePrefix = {arXiv},
       eprint = {1806.06634},
 primaryClass = {astro-ph.GA},
       adsurl = {https://ui.adsabs.harvard.edu/abs/2018A&A...615L...3J}
}

@ARTICLE{2015A&A...583A.137J,
       author = {{Jeli{\'c}}, V. and {de Bruyn}, A.~G. and {Pandey}, V.~N. and {Mevius}, M. and {Haverkorn}, M. and {Brentjens}, M.~A. and {Koopmans}, L.~V.~E. and {Zaroubi}, S. and {Abdalla}, F.~B. and {Asad}, K.~M.~B. and {Bus}, S. and {Chapman}, E. and {Ciardi}, B. and {Fernandez}, E.~R. and {Ghosh}, A. and {Harker}, G. and {Iliev}, I.~T. and {Jensen}, H. and {Kazemi}, S. and {Mellema}, G. and {Offringa}, A.~R. and {Patil}, A.~H. and {Vedantham}, H.~K. and {Yatawatta}, S.},
        title = "{Linear polarization structures in LOFAR observations of the interstellar medium in the 3C 196 field}",
      journal = {\aap},
         year = 2015,
        month = nov,
       volume = {583},
          eid = {A137},
        pages = {A137},
          doi = {10.1051/0004-6361/201526638},
archivePrefix = {arXiv},
       eprint = {1508.06650},
 primaryClass = {astro-ph.GA},
       adsurl = {https://ui.adsabs.harvard.edu/abs/2015A&A...583A.137J}
}

@ARTICLE{2013A&A...558A..72I,
       author = {{Iacobelli}, M. and {Haverkorn}, M. and {Orr{\'u}}, E. and {Pizzo}, R.~F. and {Anderson}, J. and {Beck}, R. and {Bell}, M.~R. and {Bonafede}, A. and {Chyzy}, K. and {Dettmar}, R. -J. and {En{\ss}lin}, T.~A. and {Heald}, G. and {Horellou}, C. and {Horneffer}, A. and {Jurusik}, W. and {Junklewitz}, H. and {Kuniyoshi}, M. and {Mulcahy}, D.~D. and {Paladino}, R. and {Reich}, W. and {Scaife}, A. and {Sobey}, C. and {Sotomayor-Beltran}, C. and {Alexov}, A. and {Asgekar}, A. and {Avruch}, I.~M. and {Bell}, M.~E. and {van Bemmel}, I. and {Bentum}, M.~J. and {Bernardi}, G. and {Best}, P. and {B{\i}rzan}, L. and {Breitling}, F. and {Broderick}, J. and {Brouw}, W.~N. and {Br{\"u}ggen}, M. and {Butcher}, H.~R. and {Ciardi}, B. and {Conway}, J.~E. and {de Gasperin}, F. and {de Geus}, E. and {Duscha}, S. and {Eisl{\"o}ffel}, J. and {Engels}, D. and {Falcke}, H. and {Fallows}, R.~A. and {Ferrari}, C. and {Frieswijk}, W. and {Garrett}, M.~A. and {Grie{\ss}meier}, J. and {Gunst}, A.~W. and {Hamaker}, J.~P. and {Hassall}, T.~E. and {Hessels}, J.~W.~T. and {Hoeft}, M. and {H{\"o}randel}, J. and {Jelic}, V. and {Karastergiou}, A. and {Kondratiev}, V.~I. and {Koopmans}, L.~V.~E. and {Kramer}, M. and {Kuper}, G. and {van Leeuwen}, J. and {Macario}, G. and {Mann}, G. and {McKean}, J.~P. and {Munk}, H. and {Pandey-Pommier}, M. and {Polatidis}, A.~G. and {R{\"o}ttgering}, H. and {Schwarz}, D. and {Sluman}, J. and {Smirnov}, O. and {Stappers}, B.~W. and {Steinmetz}, M. and {Tagger}, M. and {Tang}, Y. and {Tasse}, C. and {Toribio}, C. and {Vermeulen}, R. and {Vocks}, C. and {Vogt}, C. and {van Weeren}, R.~J. and {Wise}, M.~W. and {Wucknitz}, O. and {Yatawatta}, S. and {Zarka}, P. and {Zensus}, A.},
        title = "{Studying Galactic interstellar turbulence through fluctuations in synchrotron emission. First LOFAR Galactic foreground detection}",
      journal = {\aap},
         year = 2013,
        month = oct,
       volume = {558},
          eid = {A72},
        pages = {A72},
          doi = {10.1051/0004-6361/201322013},
archivePrefix = {arXiv},
       eprint = {1308.2804},
 primaryClass = {astro-ph.GA},
       adsurl = {https://ui.adsabs.harvard.edu/abs/2013A&A...558A..72I}
}

@ARTICLE{2024A&A...688A.200E,
       author = {{Erceg}, Ana and {Jeli{\'c}}, Vibor and {Haverkorn}, Marijke and {Gajovi{\'c}}, Lovorka and {Hardcastle}, Martin and {Shimwell}, Timothy W. and {Tasse}, Cyril},
        title = "{Faraday tomography of LoTSS-DR2 data. III. Revealing the Local Bubble and the complex of local interstellar clouds in the high-latitude inner Galaxy}",
      journal = {\aap},
         year = 2024,
        month = aug,
       volume = {688},
          eid = {A200},
        pages = {A200},
          doi = {10.1051/0004-6361/202450082},
archivePrefix = {arXiv},
       eprint = {2406.14679},
 primaryClass = {astro-ph.GA},
       adsurl = {https://ui.adsabs.harvard.edu/abs/2024A&A...688A.200E}
}

@ARTICLE{2024AJ....167..226V,
       author = {{Vanderwoude}, S. and {West}, J.~L. and {Gaensler}, B.~M. and {Rudnick}, L. and {Van Eck}, C.~L. and {Thomson}, A.~J.~M. and {Andernach}, H. and {Anderson}, C.~S. and {Carretti}, E. and {Heald}, G.~H. and {Leahy}, J.~P. and {McClure-Griffiths}, N.~M. and {O'Sullivan}, S.~P. and {Tahani}, M. and {Willis}, A.~G.},
        title = "{Prototype Faraday Rotation Measure Catalogs from the Polarisation Sky Survey of the Universe's Magnetism (POSSUM) Pilot Observations}",
      journal = {\aj},
         year = 2024,
        month = may,
       volume = {167},
       number = {5},
          eid = {226},
        pages = {226},
          doi = {10.3847/1538-3881/ad2fc8},
archivePrefix = {arXiv},
       eprint = {2403.15668},
 primaryClass = {astro-ph.GA},
       adsurl = {https://ui.adsabs.harvard.edu/abs/2024AJ....167..226V}
}

@ARTICLE{2013A&A...552A..58S,
       author = {{Sotomayor-Beltran}, C. and {Sobey}, C. and {Hessels}, J.~W.~T. and {de Bruyn}, G. and {Noutsos}, A. and {Alexov}, A. and {Anderson}, J. and {Asgekar}, A. and {Avruch}, I.~M. and {Beck}, R. and {Bell}, M.~E. and {Bell}, M.~R. and {Bentum}, M.~J. and {Bernardi}, G. and {Best}, P. and {Birzan}, L. and {Bonafede}, A. and {Breitling}, F. and {Broderick}, J. and {Brouw}, W.~N. and {Br{\"u}ggen}, M. and {Ciardi}, B. and {de Gasperin}, F. and {Dettmar}, R.-J. and {van Duin}, A. and {Duscha}, S. and {Eisl{\"o}ffel}, J. and {Falcke}, H. and {Fallows}, R.~A. and {Fender}, R. and {Ferrari}, C. and {Frieswijk}, W. and {Garrett}, M.~A. and {Grie{\ss}meier}, J. and {Grit}, T. and {Gunst}, A.~W. and {Hassall}, T.~E. and {Heald}, G. and {Hoeft}, M. and {Horneffer}, A. and {Iacobelli}, M. and {Juette}, E. and {Karastergiou}, A. and {Keane}, E. and {Kohler}, J. and {Kramer}, M. and {Kondratiev}, V.~I. and {Koopmans}, L.~V.~E. and {Kuniyoshi}, M. and {Kuper}, G. and {van Leeuwen}, J. and {Maat}, P. and {Macario}, G. and {Markoff}, S. and {McKean}, J.~P. and {Mulcahy}, D.~D. and {Munk}, H. and {Orru}, E. and {Paas}, H. and {Pandey-Pommier}, M. and {Pilia}, M. and {Pizzo}, R. and {Polatidis}, A.~G. and {Reich}, W. and {R{\"o}ttgering}, H. and {Serylak}, M. and {Sluman}, J. and {Stappers}, B.~W. and {Tagger}, M. and {Tang}, Y. and {Tasse}, C. and {ter Veen}, S. and {Vermeulen}, R. and {van Weeren}, R.~J. and {Wijers}, R.~A.~M.~J. and {Wijnholds}, S.~J. and {Wise}, M.~W. and {Wucknitz}, O. and {Yatawatta}, S. and {Zarka}, P.},
        title = "{Calibrating high-precision Faraday rotation measurements for LOFAR and the next generation of low-frequency radio telescopes}",
      journal = {\aap},
         year = 2013,
        month = apr,
       volume = {552},
          eid = {A58},
        pages = {A58},
          doi = {10.1051/0004-6361/201220728},
archivePrefix = {arXiv},
       eprint = {1303.6230},
 primaryClass = {astro-ph.IM},
       adsurl = {https://ui.adsabs.harvard.edu/abs/2013A&A...552A..58S}
}

@ARTICLE{2023MNRAS.519.5723O,
       author = {{O'Sullivan}, S.~P. and {Shimwell}, T.~W. and {Hardcastle}, M.~J. and {Tasse}, C. and {Heald}, G. and {Carretti}, E. and {Br{\"u}ggen}, M. and {Vacca}, V. and {Sobey}, C. and {Van Eck}, C.~L. and {Horellou}, C. and {Beck}, R. and {Bilicki}, M. and {Bourke}, S. and {Botteon}, A. and {Croston}, J.~H. and {Drabent}, A. and {Duncan}, K. and {Heesen}, V. and {Ideguchi}, S. and {Kirwan}, M. and {Lawlor}, L. and {Mingo}, B. and {Nikiel-Wroczy{\'n}ski}, B. and {Piotrowska}, J. and {Scaife}, A.~M.~M. and {van Weeren}, R.~J.},
        title = "{The Faraday Rotation Measure Grid of the LOFAR Two-metre Sky Survey: Data Release 2}",
      journal = {\mnras},
         year = 2023,
        month = mar,
       volume = {519},
       number = {4},
        pages = {5723-5742},
          doi = {10.1093/mnras/stac3820},
archivePrefix = {arXiv},
       eprint = {2301.07697},
 primaryClass = {astro-ph.CO},
       adsurl = {https://ui.adsabs.harvard.edu/abs/2023MNRAS.519.5723O}
}

@ARTICLE{2023A&A...675L...6V,
       author = {{Vedantham}, H.~K. and {Dupuy}, T.~J. and {Evans}, E.~L. and {Sanghi}, A. and {Callingham}, J.~R. and {Shimwell}, T.~W. and {Best}, W.~M.~J. and {Liu}, M.~C. and {Zarka}, P.},
        title = "{Polarised radio pulsations from a new T-dwarf binary}",
      journal = {\aap},
         year = 2023,
        month = jul,
       volume = {675},
          eid = {L6},
        pages = {L6},
          doi = {10.1051/0004-6361/202244965},
archivePrefix = {arXiv},
       eprint = {2301.01003},
 primaryClass = {astro-ph.SR},
       adsurl = {https://ui.adsabs.harvard.edu/abs/2023A&A...675L...6V}
}

@ARTICLE{2007ApJ...663..258B,
       author = {{Brown}, J.~C. and {Haverkorn}, M. and {Gaensler}, B.~M. and {Taylor}, A.~R. and {Bizunok}, N.~S. and {McClure-Griffiths}, N.~M. and {Dickey}, J.~M. and {Green}, A.~J.},
        title = "{Rotation Measures of Extragalactic Sources behind the Southern Galactic Plane: New Insights into the Large-Scale Magnetic Field of the Inner Milky Way}",
      journal = {\apj},
         year = 2007,
        month = jul,
       volume = {663},
       number = {1},
        pages = {258-266},
          doi = {10.1086/518499},
archivePrefix = {arXiv},
       eprint = {0704.0458},
 primaryClass = {astro-ph},
       adsurl = {https://ui.adsabs.harvard.edu/abs/2007ApJ...663..258B}
}

@ARTICLE{2024A&A...690A.314H,
       author = {{Hutschenreuter}, Sebastian and {Haverkorn}, Marijke and {Frank}, Philipp and {Raycheva}, Nergis C. and {En{\ss}lin}, Torsten A.},
        title = "{Disentangling the Faraday rotation sky}",
      journal = {\aap},
         year = 2024,
        month = oct,
       volume = {690},
          eid = {A314},
        pages = {A314},
          doi = {10.1051/0004-6361/202346740},
archivePrefix = {arXiv},
       eprint = {2304.12350},
 primaryClass = {astro-ph.GA},
       adsurl = {https://ui.adsabs.harvard.edu/abs/2024A&A...690A.314H}
}

@software{2020ascl.soft05003P,
       author = {{Purcell}, C.~R. and {Van Eck}, C.~L. and {West}, J. and {Sun}, X.~H. and {Gaensler}, B.~M.},
        title = "{RM-Tools: Rotation measure (RM) synthesis and Stokes QU-fitting}",
 howpublished = {Astrophysics Source Code Library, record ascl:2005.003},
         year = 2020,
        month = may,
          eid = {ascl:2005.003},
archivePrefix = {ascl},
       eprint = {2005.003},
       adsurl = {https://ui.adsabs.harvard.edu/abs/2020ascl.soft05003P}
}

@ARTICLE{2022A&A...657A..43H,
       author = {{Hutschenreuter}, S. and {Anderson}, C.~S. and {Betti}, S. and {Bower}, G.~C. and {Brown}, J. -A. and {Br{\"u}ggen}, M. and {Carretti}, E. and {Clarke}, T. and {Clegg}, A. and {Costa}, A. and {Croft}, S. and {Van Eck}, C. and {Gaensler}, B.~M. and {de Gasperin}, F. and {Haverkorn}, M. and {Heald}, G. and {Hull}, C.~L.~H. and {Inoue}, M. and {Johnston-Hollitt}, M. and {Kaczmarek}, J. and {Law}, C. and {Ma}, Y.~K. and {MacMahon}, D. and {Mao}, S.~A. and {Riseley}, C. and {Roy}, S. and {Shanahan}, R. and {Shimwell}, T. and {Stil}, J. and {Sobey}, C. and {O'Sullivan}, S.~P. and {Tasse}, C. and {Vacca}, V. and {Vernstrom}, T. and {Williams}, P.~K.~G. and {Wright}, M. and {En{\ss}lin}, T.~A.},
        title = "{The Galactic Faraday rotation sky 2020}",
      journal = {\aap},
         year = 2022,
        month = jan,
       volume = {657},
          eid = {A43},
        pages = {A43},
          doi = {10.1051/0004-6361/202140486},
archivePrefix = {arXiv},
       eprint = {2102.01709},
 primaryClass = {astro-ph.GA},
       adsurl = {https://ui.adsabs.harvard.edu/abs/2022A&A...657A..43H}
}

@ARTICLE{2024ApJ...970...95U,
       author = {{Unger}, Michael and {Farrar}, Glennys R.},
        title = "{The Coherent Magnetic Field of the Milky Way}",
      journal = {\apj},
         year = 2024,
        month = jul,
       volume = {970},
       number = {1},
          eid = {95},
        pages = {95},
          doi = {10.3847/1538-4357/ad4a54},
archivePrefix = {arXiv},
       eprint = {2311.12120},
 primaryClass = {astro-ph.GA},
       adsurl = {https://ui.adsabs.harvard.edu/abs/2024ApJ...970...95U}
}

@ARTICLE{2005A&A...441.1217B,
       author = {{Brentjens}, M.~A. and {de Bruyn}, A.~G.},
        title = "{Faraday rotation measure synthesis}",
      journal = {\aap},
         year = 2005,
        month = oct,
       volume = {441},
       number = {3},
        pages = {1217-1228},
          doi = {10.1051/0004-6361:20052990},
archivePrefix = {arXiv},
       eprint = {astro-ph/0507349},
 primaryClass = {astro-ph},
       adsurl = {https://ui.adsabs.harvard.edu/abs/2005A&A...441.1217B}
}

@ARTICLE{1966MNRAS.133...67B,
       author = {{Burn}, B.~J.},
        title = "{On the depolarization of discrete radio sources by Faraday dispersion}",
      journal = {\mnras},
         year = 1966,
        month = jan,
       volume = {133},
        pages = {67},
          doi = {10.1093/mnras/133.1.67},
       adsurl = {https://ui.adsabs.harvard.edu/abs/1966MNRAS.133...67B}
}

@INPROCEEDINGS{2018ASSL..426..159B,
       author = {{Brentjens}, Michiel A.},
        title = "{Polarization Imaging with LOFAR}",
    booktitle = {Astrophysics and Space Science Library},
         year = 2018,
       series = {Astrophysics and Space Science Library},
       volume = {426},
        month = jan,
        pages = {159},
          doi = {10.1007/978-3-319-23434-2_10},
       adsurl = {https://ui.adsabs.harvard.edu/abs/2018ASSL..426..159B}
}




\appendix

\section{Rotation measure map of the southern lobe of 7C\,1604+5447}
\label{sec:rmmap}
\begin{figure*}
    \includegraphics[width=0.49\textwidth]{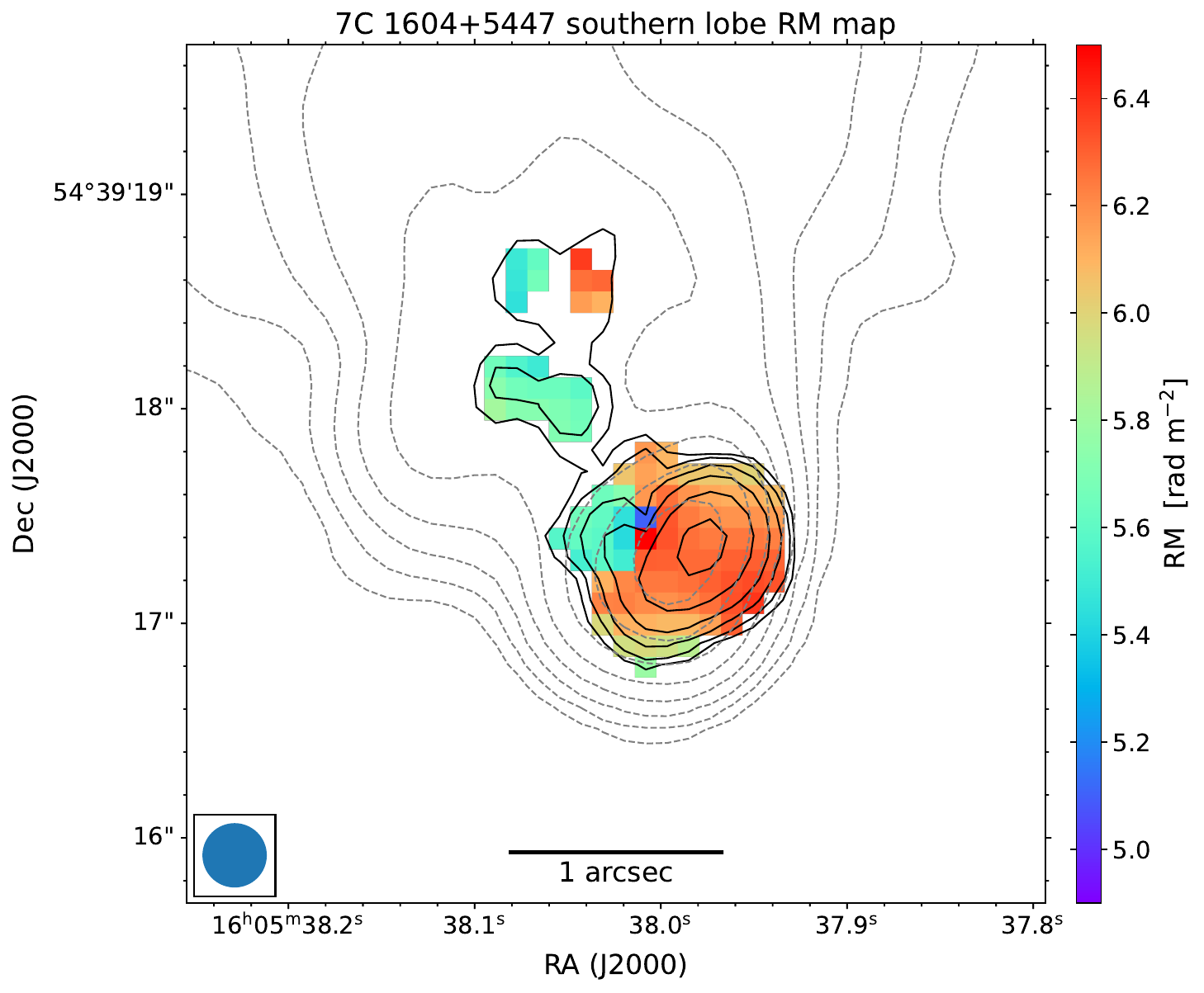}
    \includegraphics[width=0.49\textwidth]{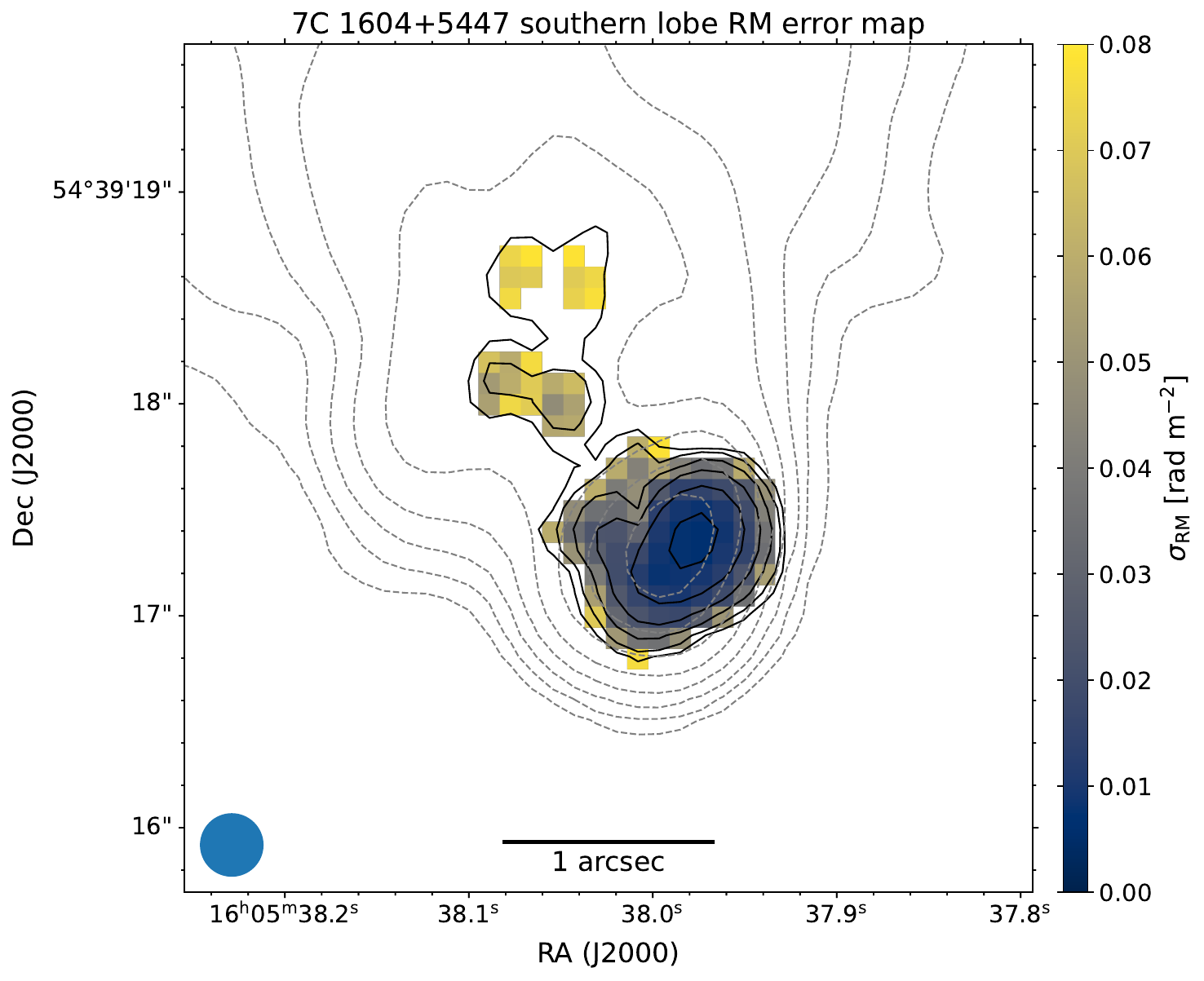}
    \caption{Left panel: RM map of the southern lobe of 7C\,1604+5447. The beam size is indicated in the bottom-left corner. Solid contours are the same as in the left panel of Fig.~\ref{fig:7C1604+5447}, taken from the linearly polarised intensity image and drawn at the same levels. The dashed grey contours are from the Stokes~$I$ image and are drawn at levels of $20 \times \sigma_{\rm rms}\,[1,2,4,8,\ldots]$, where $\sigma_{\rm rms} = 20.8$~\textmu Jy~beam$^{-1}$ is the rms noise in the Stokes~$I$ image. Right panel: The RM uncertainty map corresponding to the left panel.}
    \label{fig:rmmap}
\end{figure*}

We used the \texttt{rmtools\_peakfitcube} function from the \texttt{RM-tools} package to create an RM map of the southern lobe of 7C\,1604+5447 by fitting the peak of the Faraday spectrum in each pixel of the cleaned Faraday dispersion cube. The results are shown in Fig.~\ref{fig:rmmap} (left panel), with the corresponding uncertainty map ($\sigma_{\rm RM}$) shown in the right panel. Pixels with $\sigma_{\rm RM} > 0.08$~rad~m$^{-2}$ were blanked. The RM map shows a somewhat bimodal distribution of RM values, centred around 6.3~rad~m$^{-2}$ for the brightest part of the lobe and around 5.7~rad~m$^{-2}$ for most of the emission to the northeast.

\section{Faraday dispersion functions}    

\begin{figure}
    \includegraphics[width=0.49\textwidth]{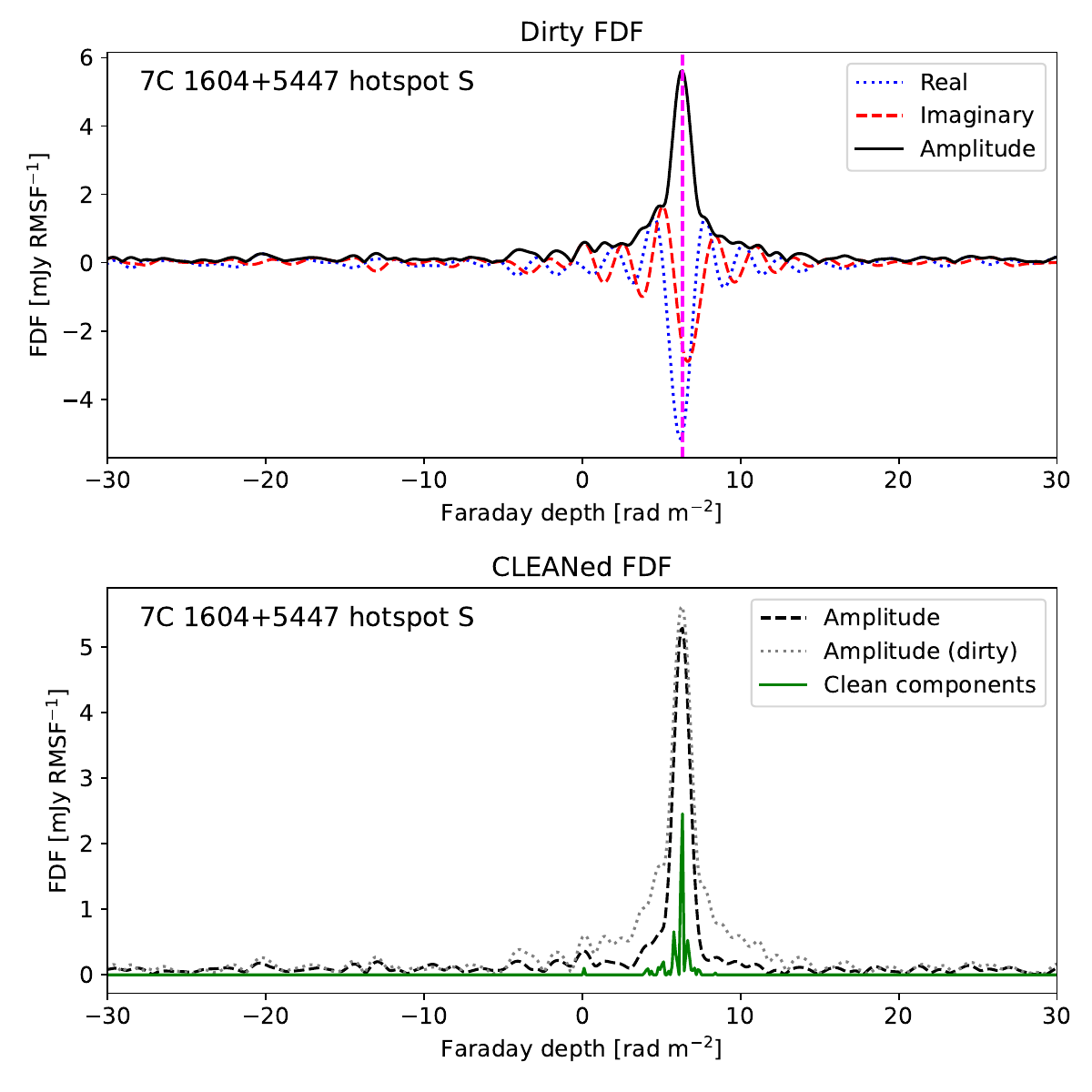}
    \caption{Top panel: Dirty Faraday dispersion function of the southern hotspot of 7C\,1604+5447. The real, imaginary, and amplitude components are shown in blue, red, and black, respectively. The dashed vertical magenta line indicates the peak of the Faraday dispersion function.  Bottom panel: RM-CLEANed Faraday spectrum corresponding to the top panel. The cleaned amplitude, dirty amplitude, and the amplitude of the clean components are shown in black, grey, and green, respectively.}
    \label{fig:APpolcalibratorhotspotS}
\end{figure}

\begin{figure}
    \includegraphics[width=0.49\textwidth]{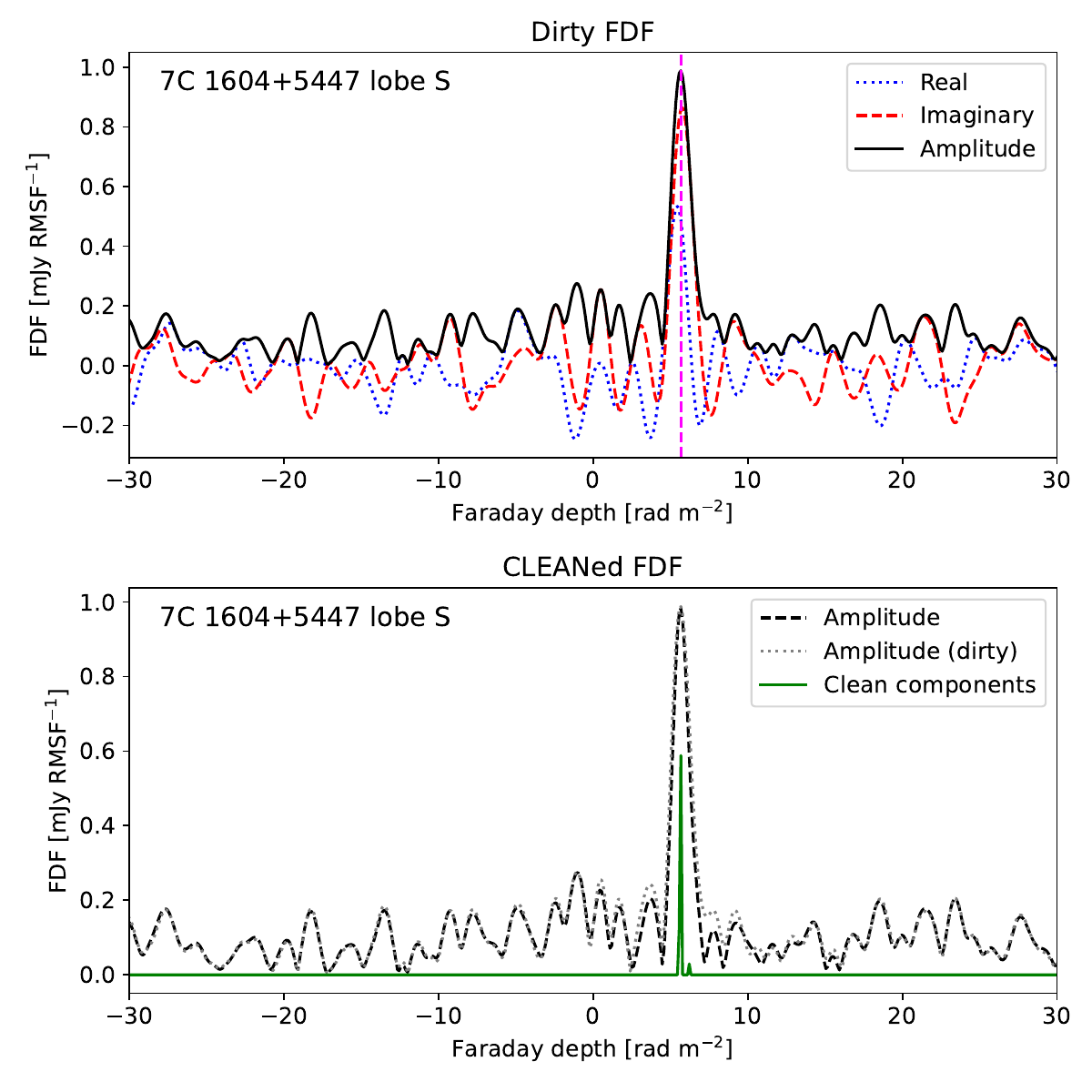}
    \caption{The same as for Fig.~\ref{fig:APpolcalibratorhotspotS} but for lobe S of 7C\,1604+5447}
    \label{fig:APpolcalibratorlobeS}
\end{figure}

\begin{figure}
    \includegraphics[width=0.49\textwidth]{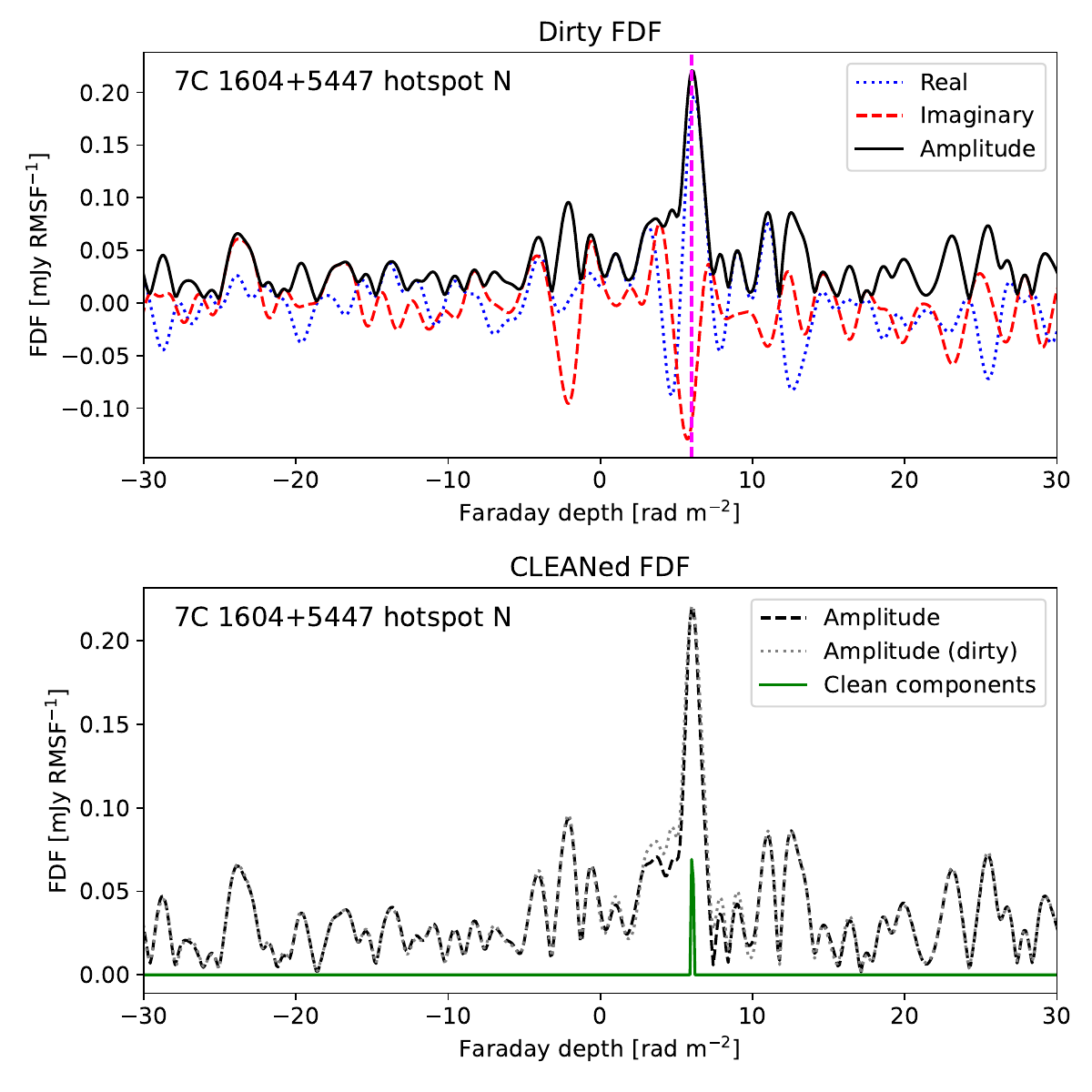}
    \caption{The same as for Fig.~\ref{fig:APpolcalibratorhotspotS} but for hotspot N of 7C\,1604+5447}
    \label{fig:APpolcalibratorhotspotN}
\end{figure}

\begin{figure}
    \includegraphics[width=0.49\textwidth]{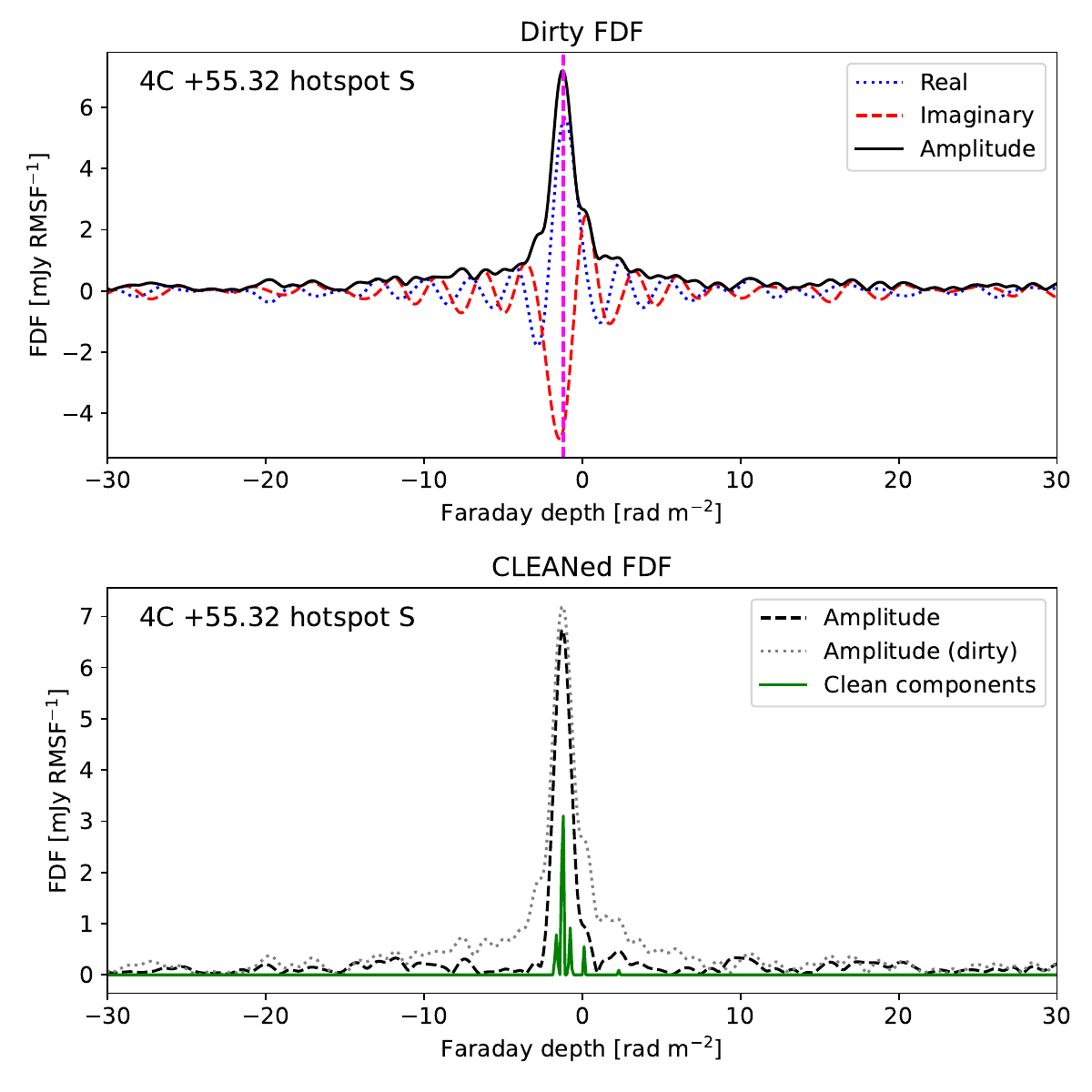}
    \caption{The same as for Fig.~\ref{fig:APpolcalibratorhotspotS} but for hotspot~S of 4C\,+55.32}
    \label{fig:APbrightnewhotspotS}
\end{figure}

\begin{figure}
    \includegraphics[width=0.49\textwidth]{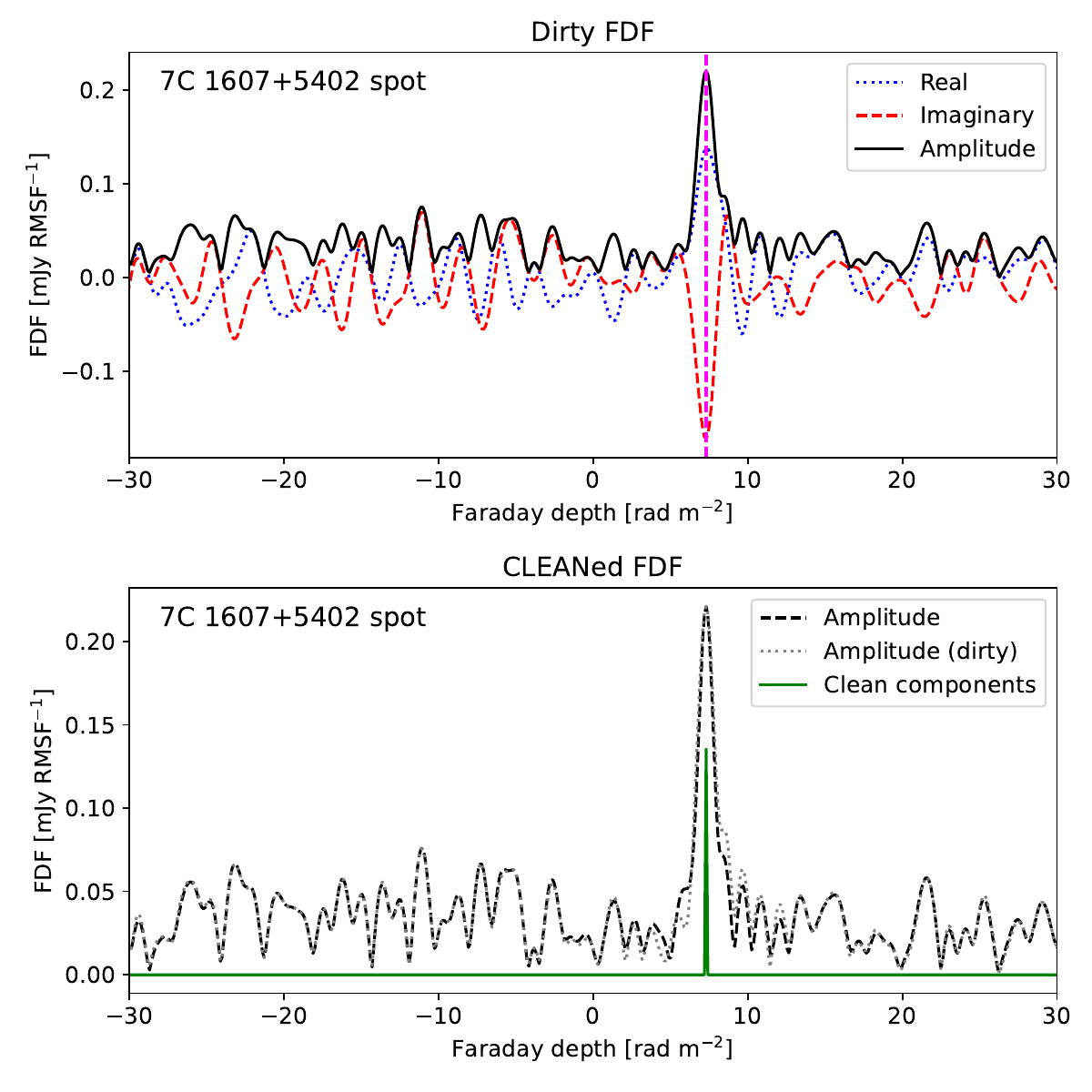}
    \caption{The same as for Fig.~\ref{fig:APpolcalibratorhotspotS} but for the spot of 7C\,1607+5402}
    \label{fig:APspecialsourcespot}
\end{figure}

\begin{figure}
    \includegraphics[width=0.49\textwidth]{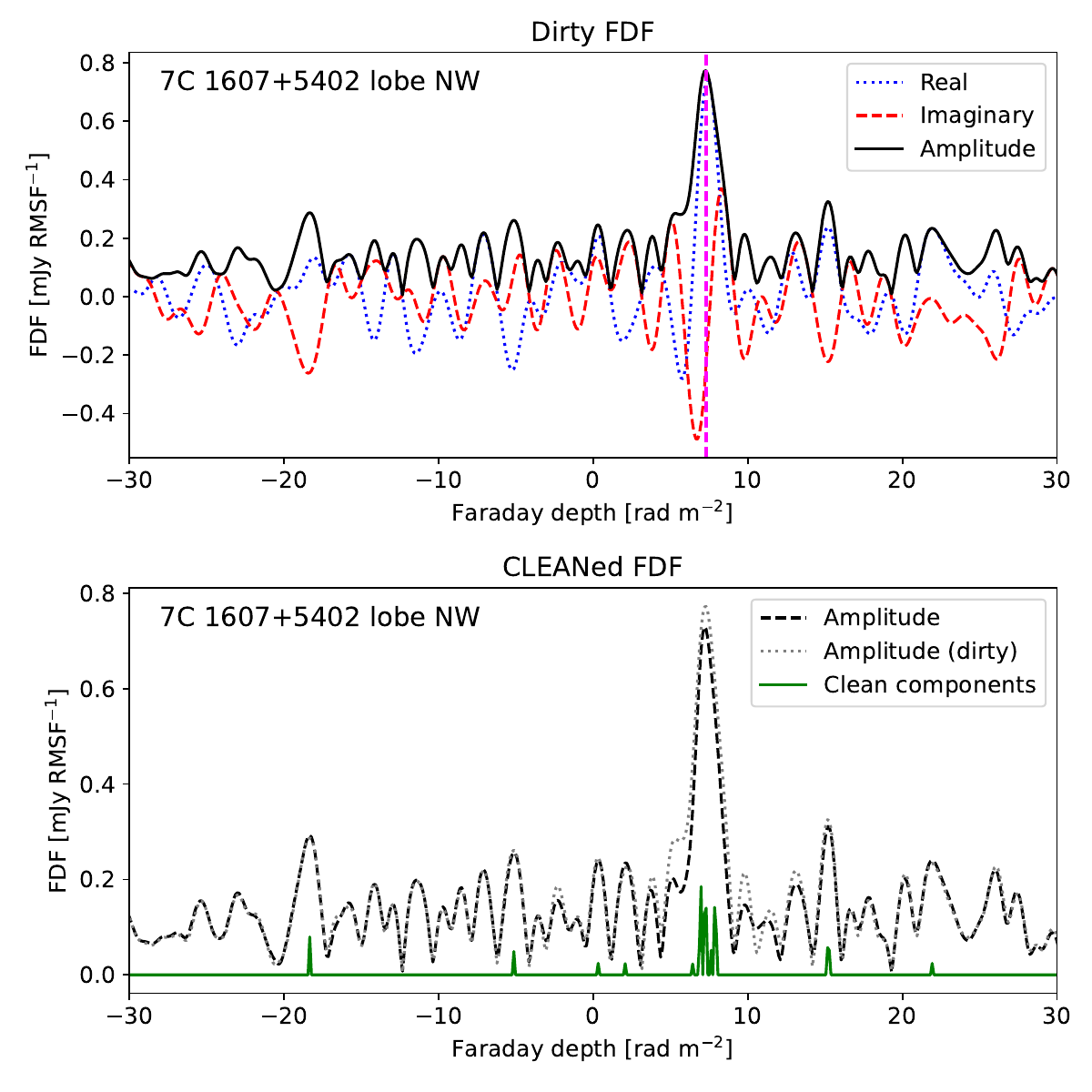}
    \caption{The same as for Fig.~\ref{fig:APpolcalibratorhotspotS} but for the NW lobe of 7C\,1607+5402}
    \label{fig:APspecialsourcelobeedge}
\end{figure}

In Figs.~\ref{fig:APpolcalibratorhotspotS}--\ref{fig:APspecialsourcelobeedge}, we present the dirty and RM-CLEANed Faraday dispersion functions, including the clean components, for the detected polarised source components in 7C\,1604+5447, 4C\,+55.32, and 7C\,1607+5402. The locations and names of these components within each source are shown in Figs.~\ref{fig:7C1604+5447}, \ref{fig:4C55.32}, and \ref{fig:7C1607+5402}, respectively.

\bsp	
\label{lastpage}
\end{document}